\documentclass[prd,amsmath,aps,floats,amssymb, floatfix,
  superscriptaddress,nofootinbib,twocolumn,preprintnumbers]{revtex4-2}
\usepackage{tabularx}
\usepackage{amsmath,amssymb,natbib,latexsym,times}

\usepackage{booktabs}
\usepackage{multirow}

\usepackage{graphicx}
\usepackage{dcolumn}
\usepackage{bm}

\usepackage{url}
\usepackage{color}
\usepackage{comment}
\usepackage{mathrsfs}
\usepackage{hyperref}
\usepackage[dvipsnames]{xcolor}
\usepackage{orcidlink}

\newcommand{\mnras}{Mon.~Not.~Roy.~Astron.~Soc.}
\newcommand{\physrep}{Phys.~Rep.}

\newcommand{\pasj}{Publ.~Astron.~Soc.~Japan}
\newcommand{\apjl}{\apj~Lett.}
\newcommand{\apjs}{\apj~Suppl.}
\newcommand{\aap}{Astronomy~\&~Astrophysics}
\newcommand{\aj}{The~Astronomical~J.}

\newcommand{\avrg}[1]{\left\langle#1\right\rangle}
\newcommand{\dSigma}{\Delta\!\Sigma}
\newcommand{\wproj}{w_{\rm p}}
\newcommand{\mpch}{h^{-1}{\rm Mpc}}
\newcommand{\sqdeg}{deg$^2$}

\newcommand{\dnnz}{{\sc DNNz}~}
\newcommand{\mizuki}{{\sc Mizuki}~}
\newcommand{\dempz}{{\sc DEmPz}~}

\begin{document}

\title{Hyper Suprime-Cam Year 3 Results: 
Cosmology from  Galaxy Clustering and Weak Lensing with HSC and SDSS using the Minimal Bias Model}

\author{Sunao~Sugiyama\orcidlink{0000-0003-1153-6735}}
\email{sunao.sugiyama@ipmu.jp}
\affiliation{Kavli Institute for the Physics and Mathematics of the Universe (WPI), The University of Tokyo Institutes for Advanced Study (UTIAS), The University of Tokyo, Chiba 277-8583, Japan}
\affiliation{Department of Physics, The University of Tokyo, Bunkyo, Tokyo 113-0031, Japan}

\author{Hironao~Miyatake\orcidlink{0000-0001-7964-9766}}
\affiliation{Kobayashi-Maskawa Institute for the Origin of Particles and the Universe (KMI),
Nagoya University, Nagoya, 464-8602, Japan}
\affiliation{Institute for Advanced Research, Nagoya University, Nagoya 464-8601, Japan}
\affiliation{Kavli Institute for the Physics and Mathematics of the Universe (WPI), The University of Tokyo Institutes for Advanced Study (UTIAS), The University of Tokyo, Chiba 277-8583, Japan}

\author{Surhud~More\orcidlink{0000-0002-2986-2371}}
\affiliation{The Inter-University Centre for Astronomy and Astrophysics, Post bag 4, Ganeshkhind, Pune 411007, India}
\affiliation{Kavli Institute for the Physics and Mathematics of the Universe
(WPI), The University of Tokyo Institutes for Advanced Study (UTIAS),
The University of Tokyo, Chiba 277-8583, Japan}

\author{Xiangchong~Li\orcidlink{0000-0003-2880-5102}}
\affiliation{McWilliams Center for Cosmology, Department of Physics, Carnegie Mellon University, Pittsburgh, PA 15213, USA}
\affiliation{Kavli Institute for the Physics and Mathematics of the Universe
(WPI), The University of Tokyo Institutes for Advanced Study (UTIAS),
The University of Tokyo, Chiba 277-8583, Japan}

\author{Masato~Shirasaki\orcidlink{0000-0002-1706-5797}}
\affiliation{National Astronomical Observatory of Japan, Mitaka, Tokyo 181-8588, Japan}
\affiliation{The Institute of Statistical Mathematics,
Tachikawa, Tokyo 190-8562, Japan}

\author{Masahiro~Takada\orcidlink{0000-0002-5578-6472}}
\affiliation{Kavli Institute for the Physics and Mathematics of the Universe (WPI), The University of Tokyo Institutes for Advanced Study (UTIAS), The University of Tokyo, Chiba 277-8583, Japan}

\author{Yosuke~Kobayashi\orcidlink{0000-0002-6633-5036}}
\affiliation{Department of Astronomy/Steward Observatory, University of Arizona, 933 North Cherry Avenue, Tucson, AZ 85721-0065, USA}
\affiliation{Kavli Institute for the Physics and Mathematics of the Universe
(WPI), The University of Tokyo Institutes for Advanced Study (UTIAS),
The University of Tokyo, Chiba 277-8583, Japan}

\author{Ryuichi~Takahashi}
\affiliation{Faculty of Science and Technology, Hirosaki University, 3 Bunkyo-cho, Hirosaki, Aomori 036-8561, Japan}

\author{Takahiro~Nishimichi\orcidlink{0000-0002-9664-0760}}
\affiliation{Center for Gravitational Physics and Quantum Information, Yukawa Institute for Theoretical Physics, Kyoto University, Kyoto 606-8502, Japan}
\affiliation{Kavli Institute for the Physics and Mathematics of the Universe
(WPI), The University of Tokyo Institutes for Advanced Study (UTIAS),
The University of Tokyo, Chiba 277-8583, Japan}
\affiliation{Department of Astrophysics and Atmospheric Sciences, Faculty of Science, Kyoto Sangyo University, Motoyama, Kamigamo, Kita-ku, Kyoto 603-8555, Japan}

\author{Atsushi~J.~Nishizawa\orcidlink{0000-0002-6109-2397}}
\affiliation{Gifu Shotoku Gakuen University, Gifu 501-6194, Japan}
\affiliation{Institute for Advanced Research/Kobayashi Maskawa Institute, Nagoya University, Nagoya 464-8602, Japan}

\author{Markus~M.~Rau\orcidlink{0000-0003-3709-1324}}
\affiliation{McWilliams Center for Cosmology, Department of Physics, Carnegie Mellon University, Pittsburgh, PA 15213, USA}
\affiliation{High Energy Physics Division, Argonne National Laboratory, Lemont, IL 60439, USA}

\author{Tianqing~Zhang\orcidlink{0000-0002-5596-198X}}
\affiliation{McWilliams Center for Cosmology, Department of Physics, Carnegie Mellon University, Pittsburgh, PA 15213, USA}

\author{Roohi~Dalal\orcidlink{0000-0002-7998-9899}}
\affiliation{Department of Astrophysical Sciences, Princeton University, Princeton, NJ 08544, USA}

\author{Rachel~Mandelbaum\orcidlink{0000-0003-2271-1527}}
\affiliation{McWilliams Center for Cosmology, Department of Physics, Carnegie Mellon University, Pittsburgh, PA 15213, USA}

\author{Michael~A.~Strauss\orcidlink{0000-0002-0106-7755}}
\affiliation{Department of Astrophysical Sciences, Princeton University, Princeton, NJ 08544, USA}

\author{Takashi~Hamana}
\affiliation{National Astronomical Observatory of Japan, National Institutes of Natural Sciences, Mitaka, Tokyo 181-8588, Japan}

\author{Masamune~Oguri\orcidlink{0000-0003-3484-399X}}
\affiliation{Center for Frontier Science, Chiba University, 1-33 Yayoi-cho, Inage-ku, Chiba 263-8522, Japan}
\affiliation{Research Center for the Early Universe, The University of Tokyo, Bunkyo, Tokyo 113-0031, Japan}
\affiliation{Department of Physics, The University of Tokyo, Bunkyo, Tokyo 113-0031, Japan}
\affiliation{Kavli Institute for the Physics and Mathematics of the Universe (WPI), The University of Tokyo Institutes for Advanced Study (UTIAS), The University of Tokyo, Chiba 277-8583, Japan}

\author{Ken~Osato\orcidlink{0000-0002-7934-2569}}
\affiliation{Center for Frontier Science, Chiba University, Chiba 263-8522, Japan}
\affiliation{Department of Physics, Graduate School of Science, Chiba University, Chiba 263-8522, Japan}

\author{Arun~Kannawadi\orcidlink{0000-0001-8783-6529}}
\affiliation{Department of Astrophysical Sciences, Princeton University, Princeton, NJ 08544, USA}

\author{Bau-Ching Hsieh\orcidlink{0000-0001-5615-4904}}
\affiliation{Academia Sinica Institute of Astronomy and Astrophysics, No. 1, Sec. 4, Roosevelt Rd., Taipei 10617, Taiwan}

\author{Wentao Luo\orcidlink{0000-0001-6579-2190}}
\affiliation{School of Physical Sciences, University of Science and Technology of China, Hefei, Anhui 230026, China}
\affiliation{CAS Key Laboratory for Researches in Galaxies and Cosmology/Department of Astronomy, School of Astronomy and Space Science, University of Science and Technology of China, Hefei, Anhui 230026, China}

\author{Robert~Armstrong}
\affiliation{Lawrence Livermore National Laboratory, Livermore, CA 94551, USA}

\author{James~Bosch\orcidlink{0000-0003-2759-5764}}
\affiliation{Department of Astrophysical Sciences, Princeton University, Princeton, NJ 08544, USA}

\author{Yutaka~Komiyama\orcidlink{0000-0002-3852-6329}}
\affiliation{Department of Advanced Sciences, Faculty of Science and Engineering, Hosei University, 3-7-2 Kajino-cho, Koganei-shi, Tokyo 184-8584, Japan}

\author{Robert~H.~Lupton\orcidlink{0000-0003-1666-0962}}
\affiliation{Department of Astrophysical Sciences, Princeton University, Princeton, NJ 08544, USA}

\author{Nate~B.~Lust\orcidlink{0000-0002-4122-9384}}
\affiliation{Department of Astrophysical Sciences, Princeton University, Princeton, NJ 08544, USA}

\author{Satoshi Miyazaki\orcidlink{0000-0002-1962-904X}}
\affiliation{Subaru Telescope,  National Astronomical Observatory of Japan, 650 N Aohoku Place Hilo HI 96720 USA}

\author{Hitoshi~Murayama\orcidlink{0000-0001-5769-9471}}
\affiliation{Berkeley Center for Theoretical Physics, University of California, Berkeley, CA 94720, USA}
\affiliation{Theory Group, Lawrence Berkeley National Laboratory, Berkeley, CA 94720, USA}
\affiliation{Kavli Institute for the Physics and Mathematics of the Universe (WPI), The University of Tokyo Institutes for Advanced Study (UTIAS), The University of Tokyo, Chiba 277-8583, Japan}

\author{Yuki~Okura\orcidlink{0000-0001-6623-4190}}
\affiliation{National Astronomical Observatory of Japan, National Institutes of Natural Sciences, Mitaka, Tokyo 181-8588, Japan}

\author{Paul~A.~Price\orcidlink{0000-0003-0511-0228}}
\affiliation{Department of Astrophysical Sciences, Princeton University, Princeton, NJ 08544, USA}

\author{Philip~J.~Tait}
\affiliation{Subaru Telescope,  National Astronomical Observatory of Japan, 650 N Aohoku Place Hilo HI 96720 USA}

\author{Masayuki~Tanaka}
\affiliation{National Astronomical Observatory of Japan, National Institutes of Natural Sciences, Mitaka, Tokyo 181-8588, Japan}

\author{Shiang-Yu~Wang}
\affiliation{Academia Sinica Institute of Astronomy and Astrophysics, No. 1, Sec. 4, Roosevelt Rd., Taipei 10617, Taiwan}

\date{\today}

\begin{abstract}
We present cosmological parameter constraints from a blind joint analysis of three two-point correlation functions measured from the Year 3 Hyper Suprime-Cam (HSC-Y3) imaging data, covering about 416~\sqdeg,
and the SDSS DR11 spectroscopic galaxies spanning the redshift range $[0.15, 0.70]$. We subdivide the SDSS galaxies into three 
luminosity-cut, and therefore nearly volume-limited
samples separated in redshift, each of which acts as a large-scale structure tracer characterized by the measurement of the projected correlation function, $\wproj(R)$. We also use the measurements of the galaxy-galaxy weak lensing signal $\dSigma(R)$ for each of these SDSS samples which act as lenses for a secure sample of source galaxies selected from the HSC-Y3 shape catalog based on their photometric redshifts. We combine these measurements with the cosmic shear correlation functions, $\xi_{\pm}(\vartheta)$ measured for our HSC source sample. We model these observables with the minimal bias model of the galaxy clustering observables in the context of a 
flat
$\Lambda$CDM cosmology. We use conservative scale cuts, $R>12$ and $8~h^{-1}$Mpc for $\dSigma$ and $\wproj$, respectively, 
where the minimal bias model is valid, in addition to conservative prior on the residual bias in the mean redshift of the HSC photometric source galaxies. We present various validation tests of our model as well as analysis methods. Our baseline analysis yields $S_8=0.775^{+0.043}_{-0.038}$ (68\% C.I.) for the $\Lambda$CDM model, after marginalizing over uncertainties in other parameters. Our value of $S_8$ is consistent with that from the {\it Planck} 2018 data, but the credible interval of our result is still relatively large. We show that various internal consistency tests based on  different splits of the data are passed. 
Our results are statistically consistent with those of a companion paper, which 
extends this analysis to smaller scales with an emulator-based halo model, using $\dSigma(R)$ and $\wproj(R)$ down to $R>3$ and $2~h^{-1}$Mpc, respectively. 
\end{abstract}

\maketitle

\section{Introduction}
\label{sec:intro}

Wide-area imaging surveys are powerful tools for constraining the composition 
of the Universe and 
the growth history of cosmic structure. Motivated by this fact, the international Subaru Hyper Suprime-Cam (HSC) collaboration consisting of scientists mainly from Japan, Taiwan and Princeton University has carried out a wide-area, multi-band imaging survey with HSC, covering about 1,100~\sqdeg 
\citep{2018PASJ...70S...1M,HSCoverview:17}. In particular, comparing the  
measurements of weak lensing effects due to large-scale structure in the Universe with predictions from cosmological models allows us to obtain precise estimates of cosmological parameters \citep{2019PASJ...71...43H,2020PASJ...72...16H,Miyatake:2021sdd,Sugiyama:2021,Porredon:2021,Heymansetal:2021}. Interestingly, however, the majority of large-scale structure probes infer a lower value of $\sigma_8$ or $S_8\equiv \sigma_8(\Omega_{\rm m}/0.3)^{0.5}$, a parameter that characterizes the amplitude of the matter clustering on scales of $8~h^{-1}$Mpc in the Universe today, compared to that inferred from the {\it Planck}
cosmic microwave background data \citep{2020A&A...641A...6P}, albeit at low statistical significance. 
If established at high statistical significance, and after ruling out any systematic biases as an origin,
this so-called $\sigma_8$ or $S_8$ tension \citep[see][for a review]{2022arXiv220911726A} might be a consequence of new physics beyond the standard $\Lambda$CDM model of the Universe. 

Establishing or ruling out the $S_8$ tension with cosmological datasets is, therefore, one of the most important contemporary problems in modern cosmology. To do this, we require high-precision, robust cosmology experiments. Here by ``high-precision'', we mean experiments yielding small credible intervals (small error bars) on the cosmological parameter of interest, here $S_8$, and by ``robust'', we mean experiments that can provide an unbiased estimate of the underlying true value of $S_8$. This is the direction the cosmology community is heading in the coming decade. 

In this paper, we model measurements from the Year 3 galaxy shape catalog of Subaru HSC (hereafter HSC-Y3) and the spectroscopic SDSS DR11 galaxy catalog, to perform a joint cosmological analysis of galaxy clustering, galaxy-galaxy clustering and cosmic shear  -- a 3$\times$2pt analysis. This paper is an extension of \citet{Sugiyama:2021} which used the Year 1 HSC (HSC-Y1) data of about 140~\sqdeg
to perform a joint cosmological analysis of galaxy clustering and galaxy-galaxy weak lensing (2$\times$2pt analysis). In this paper, we use the HSC-Y3 catalog covering about 416~\sqdeg,
which is approximately three times larger area than in HSC-Y1, and supplement 
the 2$\times$2pt measurements with the cosmic shear correlation functions to perform a joint analysis. 

The main challenge in the use of galaxy clustering for cosmological analyses lies in the relation between the distribution of galaxies and that of matter (mainly dark matter) in the large-scale structure -- the so-called galaxy bias uncertainty \citep{Kaiser:1984} \citep[also see][for a review]{Desjacquesetal:16}. Observationally, galaxy-galaxy weak lensing, measured from the cross-correlation of the positions of lens galaxies with shapes of background galaxies, can be used 
to measure the average matter distribution around lens galaxies, which, in turn, can be used to infer the galaxy bias 
when combined with the auto-correlation of galaxies in the lens sample
\citep{Sugiyama:2020, 2021arXiv210100113M, Miyatake:2021sdd}.

In order to carry out a cosmological analysis, we need a theoretical model to describe our observables. However, it is still very challenging to accurately model the galaxy bias and its scale dependence from first principles due to complex physical processes that are inherent to the formation and evolution of galaxies. Cosmological perturbation theory \citep{Bernardeauetal:02,Desjacquesetal:16} provides an accurate modeling framework to describe the clustering properties of galaxies. In this paper, we utilize the {\it minimal} bias model to describe the galaxy clustering observables. In this model, we model the auto-correlation function of galaxies as $\xi_{\rm gg}(r)=b_1^2 \xi^{\rm NL}_{\rm mm}(r)$ and the cross-correlation function of matter and galaxies as $\xi_{\rm gm}(r)=b_1\xi_{\rm mm}^{\rm NL}(r)$, where $b_1$ is a linear bias parameter and $\xi_{\rm mm}^{\rm NL}(r)$ is the nonlinear correlation function of matter. The galaxy-galaxy weak lensing and the projected galaxy clustering correlation function probe the line-of-sight projection of $\xi_{\rm gm}(r)$ and $\xi_{\rm gg}(r)$, respectively. Their combination allows us to break degeneracies between $b_1$ and $\xi_{\rm mm}^{\rm NL}$, the latter of which allows us to extract the cosmological information. On sufficiently large length scales, where gravity is the driving dominant force for structure formation and local baryonic physics do not affect the observables, the minimal bias model serves as a phenomenologically accurate theoretical framework that can be applied to any galaxy type. However, the model breaks down on smaller scales where a complex, scale-dependent galaxy bias appears. 

The purpose of this paper is to obtain robust cosmological constraints from the 3$\times$2pt cosmological analysis using the minimal galaxy bias model and flat  $\Lambda$CDM cosmological model. We use the galaxy-galaxy weak lensing and the projected correlation function of galaxies measured on conservatively-chosen scales, where the minimal bias model is safely valid. Furthermore, we will show that the addition of the cosmic shear correlation function can improve the cosmological constraints when combined with the 
2$\times$2pt measurements. We will present 
various
validation tests of the minimal bias model using mock 2$\times$2pt signals that include different galaxy bias models and other physical systematic effects. 
In addition, we employ a nuisance parameter to model a possible residual ensemble photometric redshift (hereafter photo-$z$) error in the HSC source galaxies that are used for the measurements and model of galaxy-galaxy weak lensing and cosmic shear signals, because ensemble photo-$z$ errors are one of the most important systematic effects in weak lensing measurements.
A companion paper, \citet{miyatake2023}, indicates a non-zero residual photo-$z$ error from the analysis of the same weak lensing data as is used in this paper. We will also perform various internal consistency tests to check that our results remain consistent across various splits of the data.

This paper is a companion paper to \citet{more2023} and \citet{miyatake2023}. The measurements, systematics and covariance estimates for the 3$\times$2pt signals 
are described in \citet{more2023} These signals are analysed using two different models in \citet{miyatake2023} and in this paper. In the former case, we model the measured signals down to smaller scales 
using an emulator-based halo model 
framework, where the galaxy bias is determined as an average over the bias of halos weighted by the halo occupation probability of galaxies. The additional degrees of freedom in the halo occupation distribution make it possible to model the signals on scales below those used in this paper \citep{2021arXiv210100113M,miyatake2023}. 
The results presented in this paper are thus complementary to those in the companion paper, \citet{miyatake2023}, which uses an emulator-based halo model to estimate the cosmological parameters from the same observables as those used in this paper, but down to smaller scales. We will show that the results using different theoretical models are consistent with each other. This consistency strengthens our confidence in the cosmological parameters inferred from our analyses, in particular that the results are robust against contamination from possible baryonic effects inherent in physical processes of galaxy formation and evolution (including the assembly bias effect).

In addition, there are two more companion papers, \citet{li2023} and \citet{dalal2023} that perform cosmological parameter analyses using the real- and Fourier-space tomographic cosmic shear analyses of the HSC-Y3 data, respectively. The two 3$\times$2pt analyses use the same blinded catalog (as they use the same measurements but on different scales), while the two cosmic shear analyses use entirely independent blinded catalogs. These cosmological analyses were led by different authors without comparing the cosmological parameter results until after each analysis was unblinded. We present the results without any change after unblinding; any results or conclusions derived post-unblinding are mentioned explicitly in the paper. For all these analyses, we employ a conservative prior on any residual errors on the true ensemble redshift distribution of HSC galaxies beyond $z\gtrsim 0.7$ to model possible residual photo-$z$ errors. All of these analysis choices were defined during the blind phase of the analysis. 

This paper is organized as follows. In Section~\ref{sec:data}, we briefly describe the data and measurement methods. In Section~\ref{sec:analysis-method}, we describe the theoretical model that we use to infer cosmological parameters for the
flat $\Lambda$CDM model, our choice of model for systematic effects, and the likelihood analysis method. 
In Section~\ref{sec:blind-internal-consistency}, we describe the strategy we adopt to perform a blind analysis. 
In Section~\ref{sec:model-validation}, we perform model validation using different types of mock data vectors.
In Section~\ref{sec:results} we show the cosmological parameters inferred from our 3$\times$2pt analysis. Section~\ref{sec:conclusion} is devoted to conclusions and a discussion of our results.

\section{Data and Measurement}
\label{sec:data}

In this section we briefly describe the data and the measurement methods. The details can be found in a companion paper, \citet{more2023}.

\subsection{HSC-Y3 data: source galaxies for 
weak lensing}
\label{sec:HSC-Y3}

HSC is a wide-field prime focus camera on the 8.2m Subaru Telescope \citep{2018PASJ...70S...1M,2018PASJ...70S...2K,2018PASJ...70S...3F,2018PASJ...70...66K}. The HSC Subaru Strategic Program (HSC SSP) survey started in 2014, and used 330 Subaru nights to conduct a five-band ($grizy$) wide-area imaging survey \citep{HSCoverview:17}. The combination of HSC's wide field-of-view (1.77~\sqdeg), superb image quality (a median $i$-band seeing FWHM of $0.6^{\prime\prime}$), and large photon-collecting power makes it one of the most powerful instruments for weak lensing measurements. The HSC SSP survey consists of three layers; Wide, Deep, and Ultradeep. The Wide layer, which is designed for weak lensing cosmology, covers about 1,100~\sqdeg\ of the sky with a $5\sigma$ depth of $i\sim26$ ($2^{\prime\prime}$ aperture for a point source). Since the $i$-band images are used for galaxy shape measurements in weak lensing analyses, they are preferentially taken under good seeing conditions.

In this paper, we use the HSC three-year (hereafter HSC-Y3) galaxy shape and photo-$z$ catalogs \citep{HSCDR1_shear:17,2018MNRAS.481.3170M} constructed from about 90~nights of HSC Wide data taken 
between March 2014 and April 2019. 
Both catalogs are based on the object catalog produced by the data reduction pipeline \cite{2018PASJ...70S...5B}. In the following subsections, we describe details of the shape and photo-$z$ catalogs.

\subsubsection{HSC-Y3 galaxy shape catalog}
\label{sec:HSC-Y3_shape}
In this paper, we use the shape catalog from the S19a internal data release
which was processed with hscPipe~v7 \citep{Li2021}. There were a number of
improvements to the PSF modelling, image warping kernel, background subtraction
and bright star masks, which have improved the quality of the shape catalog in
Year~3 compared to the Year~1 shape catalog  \citep{HSCDR1_shear:17,2018MNRAS.481.3170M}. The detailed selection of galaxies that
form the shape catalog is presented in \citet{Li2021}. Briefly, the shape
catalog consists of galaxies selected from the ``full depth full color region'' in
all five filters. Apart from some basic quality cuts related to pixel level
information, we select extended objects with an extinction corrected cmodel
magnitude $i<24.5$, $i$-band SNR$\ge 10$, resolution $>0.3$, $>5\sigma$ detection
in at least two bands other than $i$, a 1 arcsec diameter aperture magnitude
cut of $i<25.5$, and a blendedness cut in the $i$-band of $10^{-3.8}$.

The original shape catalog contains more than 35 million source galaxies covering 433~\sqdeg.
However, as described in detail in \citet{more2023}, \citet{dalal2023}, \citet{li2023},
we find 
a significant source of $B$-mode systematics in the cosmic shear correlation functions for 
a $\sim 20$~\sqdeg~
patch in the GAMA09H region, and we remove this problematic region from the following analysis. After removing this $\sim 20$~\sqdeg region, we have the HSC area of 416~\sqdeg, with
an effective weighted number density of 19.9 arcmin$^{-2}$. It is
divided into six disjoint regions: the XMM, VVDS, GAMA09H, WIDE12H, GAMA15H and
HECTOMAP fields \citep[see Fig.~2 in Ref.][]{Li2021}. The shape measurements in the catalog were calibrated using detailed
image simulations, such that the residual galaxy property-dependent multiplicative shear
estimation bias is less than $\sim 10^{-2}$. \citet{Li2021} also present a
number of systematic tests and null tests, and quantify the level of residual
systematics in the shape catalog that could affect the cosmological science
analyses carried out using the data. Given that \citet{Li2021} flag residual
additive biases due to PSF model shape residual correlations and star galaxy
shape correlations as systematics requiring special attention and
marginalization, we have included the effects of these systematics on
the cosmic shear measurements in our modelling scheme.

\subsubsection{Secure source galaxy sample definition}
\label{sec:source_galaxies}

The depth of the HSC-Y3 data enables us to define a conservative sample of source galaxies that are at  redshifts well beyond those of the lens galaxies, for weak lensing measurements. 
In this paper we select three distinct samples of lens galaxies from the database of spectroscopic SDSS galaxies up to $z_{\rm l, max}=0.7$. To select background galaxies, we use photometric redshift (hereafter photo-$z$) estimates for each HSC galaxy. 
The three year shape catalog is accompanied by a photometric redshift
catalog of galaxies based on three different methods \citep{2020arXiv200301511N}
\mizuki is a template fitting-based photo-$z$ estimation code. \dempz and \dnnz, on the other hand, provide machine learning-based estimates of the galaxy photo-$z$'s.
Each of these methods provides an estimate of the posterior distribution of redshift for each galaxy, denoted as $P(z_{\rm s})$. In this paper we use the \dempz photo-$z$ catalog as our fiducial choice. Photo-$z$ uncertainties are among the most important systematic effects in weak lensing cosmology, and can cause significant biases in the cosmological parameters if unknown residual systematic errors in photo-$z$ exist.

For the study of weak lensing, we define a sample of background galaxies whose redshifts are physically well beyond the maximum lens redshift $z_{\rm l, max}$. More specifically, we choose source galaxies  satisfying the following condition \citep{2014MNRAS.444..147O,2018PASJ...70...30M,2019ApJ...875...63M}:
\begin{equation}
\int_{z_{\rm l, max}+0.05}^{7} P_i(z_{\rm s}) \mathrm{d}z_{\rm s} \ge 0.99 \,,
\label{eq:source-selection}
\end{equation} 
where 
the
maximum redshift ($z_{\rm l,max}=0.7$) is that of the lens sample that we will use for the galaxy-galaxy
lensing measurements, and 
$P_i(z_{\rm s})$ is the posterior photo-$z$ distribution  for the $i$-th HSC galaxy.
Such cuts significantly reduce the contamination of
source galaxies that are physically associated with the lens galaxies.
With this additional cut, our weak lensing
sample includes  
about 24~percent of the galaxies in the original catalog, with an effective number density of 4.9~galaxies
per sq.~arcmin. The mean redshift of the sample is $\langle z_{\rm s}\rangle\simeq 1.3$. The resultant source redshift distribution is shown in Fig. 3 of \citet{more2023}.

\subsection{Lens galaxy sample}
\label{sec:lens_sample}

We use the large-scale structure sample compiled as part of the Data Release 11
(DR11) \footnote{\url{https://www.sdss.org/dr11/}} \cite{Alam:2015}
of the SDSS-III BOSS (Baryon Oscillation Spectroscopic Survey) project 
\citep{2013AJ....145...10D}
for
measurements of the clustering of galaxies and as lens galaxies for the galaxy-galaxy 
lensing signal measurements. The lens galaxy sample used in this paper is the
same as that used in the first year analysis of HSC data 
(\citet{2021arXiv210100113M,Sugiyama:2021}).
We describe
the resultant catalog here briefly.

The 
BOSS survey is a spectroscopic follow-up survey of galaxies and quasars
selected from the imaging data obtained by the SDSS-I/II, and covers an area of
approximately 11,000 \sqdeg\  \cite{2009ApJS..182..543A} using the dedicated 2.5m SDSS Telescope \cite{2006AJ....131.2332G}. 
 Imaging
data obtained in five photometric bands ($ugriz$) as part of the SDSS I/II
surveys \cite{1996AJ....111.1748F,2002AJ....123.2121S,2010AJ....139.1628D,2011AJ....142...72E,2012ApJS..203...21A,2013AJ....145...10D,2011ApJS..193...29A}. 
were augmented with an additional 3,000 deg$^2$ in SDSS DR9 to cover a
larger portion of the sky in the southern region \citep{2011AJ....142...72E,2012ApJS..203...21A,2013AJ....145...10D,2011ApJS..193...29A}. 
These data were processed by 
the 
SDSS 
photometric processing pipelines  \citep{2001ASPC..238..269L,2003AJ....125.1559P,2008ApJ...674.1217P}, 
and corrected for Galactic extinction  \citep{1998ApJ...500..525S} to obtain a 
reliable photometric catalog which serves as an input to select
targets
for spectroscopy \citep{2013AJ....145...10D}.
The resulting spectra were processed by an automated pipeline to perform redshift  determination and spectral classification \cite{2012AJ....144..144B}. The BOSS large-scale structure (LSS) samples are selected using algorithms focused on galaxies in different redshifts:  $0.15<z<0.35$ (LOWZ) and $0.43<z<0.7$ (CMASS). 

We use three galaxy subsamples in three redshift bins: ``LOWZ'' galaxies in the redshift range $z$ in $[0.15,0.35]$
and two subsamples of ``CMASS'' galaxies, hereafter called ``CMASS1'' and ``CMASS2'', respectively, which 
are obtained by subdividing CMASS galaxies into two redshift bins, $[0.43,0.55]$ and $[0.55,0.70]$, respectively. As shown in Fig.~1 of \citet{Miyatake:2021sdd}, we define each of the subsamples by selecting galaxies with $i$-band absolute magnitudes $M_i-5\log {\rm h}<-21.5$, $-21.9$ and $-22.2$ for the LOWZ, CMASS1 and CMASS2 samples, respectively. The comoving number densities of these samples for the {\it Planck} cosmological model \citep{2020A&A...641A...6P}
 are $\bar{n}_{\rm g}/[10^{-4}\,(h^{-1}{\rm Mpc})^{-3}] \simeq 1.8, 0.74$ and $0.45$, respectively. These are a few times smaller than the densities of the entire parent 
LOWZ and CMASS samples. 
The resultant lens redshift distributions are shown in Fig. 3 of \citet{more2023}.

As described in \citet{more2023} in detail, as our three clustering observables, 
we use the projected correlation functions for the three subsamples measured from the entire SDSS DR11 region 
of about $8,300$~\sqdeg,
and the galaxy-galaxy lensing signal and the cosmic shear correlations measured from the overlap 
416~\sqdeg~ area of the HSC-Y3 data. \citet{more2023} presented the results for 
various null and systematic tests, which are 
used to define the scale cuts used in this paper. 
The covariance matrices for our measurements were computed using a suite of mock catalogs, as described in that paper. We will use these  three two-point functions to constrain cosmological parameters.

\section{Analysis method}
\label{sec:analysis-method}

In this section, we describe the theoretical model and the analysis method we use in our cosmological analysis \citep[also see][for details]{Sugiyama:2020,Sugiyama:2021}. We also describe the blinding strategy we adopt for our cosmological analysis and the validation tests of the models/methods as well as the internal consistency tests that we performed before unblinding the results of our analysis. 

\subsection{Theoretical model}
\label{subsec:model}

\subsubsection{Projected correlation function: \texorpdfstring{$\wproj(R)$}{w(R)}}
\label{subsubsec:galaxy-clustering}

To model the projected auto-correlation, $\wproj$, and the galaxy-galaxy lensing
signal, $\dSigma$, which are related to 
the surrounding
matter distribution for the LOWZ, CMASS1 and CMASS2 galaxies, we adopt
the ``minimal bias'' model described in Ref.~\cite{Sugiyama:2020} \citep[also
see][]{Sugiyama:2021}. This model relates the number density fluctuation field of galaxies to 
the matter density fluctuation field via a linear galaxy bias parameter, 
$\delta_{\rm g}=b_{1}\!(z_{\rm l})\delta_{\rm m}$ for each galaxy sample 
at a representative redshift $z_{\rm l}$. \citet{Sugiyama:2020} 
demonstrated that the minimal bias model can serve as a sufficiently accurate model to
 recover 
the cosmological parameters without any significant systematic bias, 
as long as the model is applied to a sufficiently large scale: $R>8\mpch$ and $12\mpch$ 
for $\wproj$ and $\dSigma$, respectively. Given that these conclusions were based on the covariance matrix for HSC Y1 data, we will validate the minimal bias model 
using mock catalogs of galaxies in Section~\ref{sec:model-validation} using the covariance matrix for the HSC-Y3 and SDSS data.

The projected auto-correlation function of galaxies, $\wproj(R)$, is related
to the three-dimensional real-space correlation function $\xi_{\rm gg}$ such that
\begin{align}
	\wproj(R;z_{\rm l}) =& 2f_{\rm corr}^{\rm RSD}(R;z_{\rm l})
    \int_{0}^{\Pi_{\rm max}}\!{\rm d}\Pi~ \xi_{\rm gg}\left(\sqrt{R^2 + \Pi^2};z_{\rm l}\right),\label{eq:wproj}
\end{align}
where $\Pi_{\rm max}$ is the projection length along the line of sight, and throughout this paper we employ $\Pi_{\rm max}=100\,h^{-1}{\rm Mpc}$ in the measurement \cite[][]{vandenBosch:2013}. Typically, the redshift-space distortion 
(RSD) effects are expected to not contaminate the projected correlation function due to integration along the line-of-sight. But as we model the signal on scales $R$ approaching $ \Pi_{\rm max}$, we account for the residual RSDs using the Kaiser RSD factor \cite{Kaiser:1984}. This correction factor $f_{\rm corr}^{\rm RSD}(R;z_{\rm l})$ depends on the lens redshift and on cosmological parameters, especially $\Omega_{\rm m}$ \cite[see Eq.~(48) in Ref.][for the definition]{vandenBosch:2013} \citep[also see][]{2021arXiv210100113M}. Using the minimal bias model, we model the three-dimensional, real-space galaxy correlation function as 
\begin{align}
    \xi_{\rm gg}(r;z_{\rm l}) =b_1\!(z_{\rm l})^2 \int_0^\infty\!\frac{k^2\mathrm{d}k}{(2\pi^2)}~P^{\rm NL}_{\rm mm}(k;z_{\rm l})j_0(kr),\label{eq:xigg}
\end{align}
where $j_0(x)$ is the zeroth-order spherical Bessel function, 
and $b_1\!(z_{\rm l})$ is the linear galaxy bias parameter for each SDSS galaxy sample 
at the redshift $z_{\rm l}$. 
Throughout this paper, we model the non-linear matter power spectrum, $P^{\rm NL}_{\rm mm}$, using {\tt halofit} \citep{Smithetal:03} with the modification suggested by \cite{Takahashi_2012} for the assumed cosmological model.
While the galaxy clustering signal is also affected by magnification bias, we have checked that its contribution is at the sub-percent level compared to Eq.~(\ref{eq:wproj}), and hence neglect it in our model.

Each of the lens samples 
lies in a 
redshift bin with finite width, 
and the model signal must be evaluated by averaging the signal within the redshift bin. 
In this paper, we instead evaluate the model signal at a single representative redshift, defined as the mean redshift of the lens galaxies within each redshift bin.
The representative redshift for the LOWZ, CMASS1, and CMASS2 samples are $\bar{z}_{\rm l}\simeq 0.26, 0.51$ and $0.63$, respectively.
We have checked that the difference between the signal evaluated 
at the representative redshift and the signal averaged within the redshift bin is at most 4\% 
of
the statistical error in each $R$ bin, as long as we assume the linear galaxy bias does not evolve within the  redshift bin.

\subsubsection{Galaxy-galaxy weak lensing: \texorpdfstring{$\dSigma(R)$}{dSigma(R)}}
\label{subsubsec:model-galaxy-galaxy-lensing}

Our model for the galaxy-galaxy lensing signal has two contributions:
\begin{align}
    \dSigma(R)=\dSigma_{\rm gG}(R)+\dSigma_{\rm mag}(R). \label{eq:dsigma}
\end{align}
The first term represents the standard galaxy-galaxy weak lensing contribution 
that arises from the average projected matter density distribution around the lens galaxies.
Using the minimal bias model we model $\dSigma_{\rm gG}$ for each of 
the LOWZ, CMASS1 and LOWZ2 samples as 
\begin{align}
	\dSigma_{\rm gG}(R;z_{\rm l})=b_1\!(z_{\rm l})\bar{\rho}_{\rm m0}\int_0^\infty\!\frac{k\mathrm{d}k}{2\pi}~ 
	P^{\rm NL}_{\rm mm}(k; z_{\rm l})J_2(kR),\label{eq:dsigma-intrinsic}
\end{align}
where $J_{2}(x)$ is the second-order Bessel function, $\bar{\rho}_{\rm m0}$ is 
the mean matter density today, and the nonlinear matter power spectrum $P_{\rm mm}^{\rm NL}$ 
and sample galaxy bias $b_1$ are the same as in Eq.~(\ref{eq:xigg}).

The second term on the r.h.s. of Eq.~(\ref{eq:dsigma}) represents magnification bias, 
which arises from the cross-correlation between the lensing magnification effect 
in the observed number density field of lens galaxies and the lensing shear 
on the HSC source galaxy shapes due to the foreground matter density fluctuation 
along the same line-of-sight direction: 
\begin{align}
    \dSigma_{\rm mag}(R) = \int\!\!{\rm d}z_{\rm s}p_{\rm s}(z_{\rm s})
    \int\!\!{\rm d}z_{\rm l}p_{\rm l}(z_{\rm l})~ \widetilde{\dSigma}_{\rm mag}\!(R; z_{\rm l}, z_{\rm s}),
\end{align}
where $p_{\rm l}(z_{\rm l})$ and $p_{\rm s}(z_{\rm s})$ are the redshift distributions 
of lens and source galaxies, respectively, that are normalized as 
$\int\!\mathrm{d}z~p_i(z)=1$ ($i=$l or s). For $p_{\rm s}(z_{\rm s})$, 
we adopt the stacked photo-$z$ posterior distribution of source galaxies 
as our default choice and will discuss the impact of systematic redshift errors
on our cosmology analysis. For $p_{\rm l}(z_{\rm l})$ we can accurately evaluate 
the distribution using spectroscopic redshifts for lens galaxies. 
The integrand function, $\widetilde{\dSigma}_{\rm mag}(R;z_{\rm l},z_{\rm s})$ is defined as
\begin{align}
    \widetilde{\dSigma}_{\rm mag}(R;z_{\rm l}, z_{\rm s}) &\equiv 2(\alpha_{\rm mag, l}-1)\nonumber\\
    &\hspace{-2em}\times\int_0^\infty\!\!\frac{\ell{\rm d}\ell}{2\pi}~\Sigma_{\rm c}(z_{\rm l}, z_{\rm s})C_{\kappa}\left(\ell, z_{\rm l}, z_{\rm s}\right) J_2\left(\ell\frac{R}{\chi_{\rm l}}\right).
\end{align}
where $\chi_{\rm l}$ is the comoving distance at the representative redshift $z_{\rm l}$.
$\alpha_{\rm mag}$ is a parameter to model the power-law slope of the number counts
of the lens galaxies around a magnitude cut in each sample
\citep[see Eq.~10 and Fig.~2 in Ref.][for the estimated value and error]{Miyatake:2021sdd}, 
and $\Sigma_{\rm c}$ is the critical surface density defined as
\begin{align}
    \Sigma_{\rm c}(z_{\rm l}, z_{\rm s}) = \frac{1}{4\pi G}\frac{\chi(z_{\rm s})}{\chi(z_{\rm l})\chi(z_{\rm l}, z_{\rm s})(1+z_{\rm l})}.
\end{align}
Throughout this paper we adopt natural units with  $c=1$ for the speed of light.
The angular power spectrum of the lensing convergence field for galaxies at redshifts 
$z_{\rm l}$ and $z_{\rm s}$ is defined as
\begin{align}
    C_{\kappa}(\ell; z_{\rm l}, z_{\rm s}) = \int {\rm d}\chi \frac{W(\chi, \chi_{\rm l})W(\chi, \chi_{\rm s})}{\chi^2}P^{\rm NL}_{\rm mm}\left(\frac{l+1/2}{\chi}; z\right),
\end{align}
where the lensing efficiency kernel is defined using $\chi_s=\chi(z_{\rm s})$ as
\begin{align}
    W(\chi, \chi_s) \equiv \frac{3}{2}\Omega_{\rm m} H_0^2
    (1+z) \frac{\chi(\chi_{\rm s}-\chi)}{\chi_{\rm s}}.
\end{align}

\subsubsection{Cosmic shear correlation functions: \texorpdfstring{$\xi_{\pm}$}{xi}}
\label{subsubsec:model-cosmic-shear}

We model the two-point correlation functions of source galaxy shapes as a sum of the following three terms
\begin{align}
    \xi_\pm(\vartheta) = \xi_{{\rm GG},\pm}(\vartheta) + \xi_{{\rm GI},\pm}(\vartheta) + \xi_{{\rm II},\pm}(\vartheta). \label{eq:cosmic-shear}
\end{align}
The first term is the ``gravitational-gravitational'' term (i.e. cosmic shear, ``GG''), 
the second term is the ``gravitational-intrinsic'' correlation (``GI'') \citep{Hirata:2004} 
that arises in pairs of galaxies for which some common large-scale structure 
along the line of sight affects the intrinsic shapes of one of the galaxies and results in
a gravitational lensing shear on the other, while
the third term is the ``intrinsic-intrinsic'' (``II'') IA contribution \citep{Heavensetal:00}.

The GG term of Eq.~(\ref{eq:cosmic-shear}) is defined in terms of 
the cosmic shear power spectrum as 
\begin{align}
    \xi_{{\rm GG}, \pm}(\vartheta) = \int\frac{\ell{\rm d}\ell}{2\pi} C_\kappa(\ell) J_{0/4}(\ell\vartheta),\label{eq:cosmic-shear-GG}
\end{align}
where the zeroth and fourth order Bessel functions are for $\xi+$ and $\xi_-$, respectively. 
 $C_\kappa$ is the angular power spectrum of the lensing convergence, defined as 
\begin{align}
    C_\kappa(\ell) = \int{\rm d}\chi\frac{q^2(\chi)}{\chi^2}P_{\rm mm}^{\rm NL}\!\left(\frac{\ell+1/2}{\chi}; z \right),
\end{align}
where $q(\chi)$ is the lensing efficiency kernel averaged over the source redshift distribution 
defined as
\begin{align}
    q(\chi) = \int\!\!{\rm d}z_{\rm s}p_s(z_{\rm s})W(\chi, \chi_s).
\end{align}
In this paper, we use a single source sample, and hence we have no tomographic cosmic shear signals.

To model the GI and II terms of Eq.~(\ref{eq:cosmic-shear}), 
we employ the nonlinear alignment model (NLA) \citep{2007NJPh....9..444B}: 
\begin{align}
    C_{\rm GI}(\ell) &= 2\int{\rm d}\chi \frac{F(\chi)H(z)p_{\rm s}(z)q(\chi)}{\chi^2} P_{\rm mm}^{\rm NL}\left(\frac{\ell+1/2}{\chi}; z \right),\\
    C_{\rm II}(\ell) &= \int{\rm d}\chi \left[\frac{F(\chi)H(z)p_{\rm s}(z)}{\chi}\right]^2 P_{\rm mm}^{\rm NL}\left(\frac{\ell+1/2}{\chi}; z \right).
\end{align}
Here following the conventional method in the literature \citep[e.g.,][]{2019PASJ...71...43H}, 
we introduced the redshift- and cosmology-dependent factor $F(\chi)$ that relates 
the intrinsic galaxy ellipticity and the gravitational tidal field and is parametrized as
\begin{align}
    F(\chi) = -A_{\rm IA}C_{1}\rho_{\rm c}\frac{\Omega_{\rm m}}{D_+(z)},
\end{align}
where $A_{\rm IA}$ 
is
a free parameter that describes the amplitude of the NLA model, 
$C_1=5\times10^{-14}h^{-2}M_\odot^{-1}{\rm Mpc}^3$ is a normalization constant, 
$\rho_{\rm c}$ is the critical mass density at $z=0$, and $D_+(z)$ is the linear growth 
factor normalized to unity at $z=0$. 
Since we use the cosmic shear correlations for a single sample of the source galaxies, i.e., no lensing tomography, 
we employ a single parameter $A_{\rm IA}$ to model the IA contamination,  and do not include a parameter to model the redshift dependence of the IA effect.

\subsection{Modeling residual systematic errors}
\label{subsec:residual-systematics}

In this section, we present a method to model possible residual systematic effects 
in the measured signals. In our method, we include these effects in the model predictions
rather than in the measured signals to keep the data vector and the covariance invariant.

\subsubsection{Residual systematic redshift uncertainty: \texorpdfstring{$\Delta z_{\rm ph}$}{Delta z}}
\label{subsubsec:photo-z}

Residual systematic error in the mean redshift of the HSC source galaxies is one of the most important
systematic effects in weak lensing measurements, i.e., $\dSigma$ and $\xi_{\pm}(\vartheta)$ 
in our data vector. To study the impact of residual redshift error, we introduce 
a nuisance parameter to model the systematic error in the mean source redshift 
by shifting the posterior distribution of source redshifts, 
given as $z^{\rm est}=z^{\rm true} + \Delta z_{\rm ph}$ \citep{Hutereretal:06,OguriTakada:11,Miyatake:2021sdd}. 
Please see Section~IIIA2 in \citet{miyatake2023} and \citet{zhang:2022dvs}
for a justification of our parametrization ($\Delta z_{\rm ph}$) to model 
the impact of residual source redshift uncertainty on the weak lensing observables.
Therefore, we use the shifted $P(z)$ distribution to model the {\it mean} of the true redshift distribution as
\begin{align}
    p^{\rm true}_{\rm s}(z)=p^{\rm est}_{\rm s}(z+\Delta z_{\rm ph}).
    \label{eq:deltaz_def}
\end{align}
Thus, if $\Delta z_{\rm ph}>0$ or $<0$, the true mean redshift of our sources becomes lower or higher than what is anticipated from the photo-$z$
estimates, 
respectively. 

For $\dSigma$ (Eq.~\ref{eq:dsigma}), we first need to recompute the average lensing efficiency 
$\langle{\Sigma_{\rm cr}^{-1}}\rangle$ and the weight $w_{\rm ls}$ using 
the shifted redshift distribution: we define the correction factor as
\begin{align}
    f_{\dSigma}(\Delta z_{\rm ph}) \equiv  
    \frac{\sum_{\rm ls}w_{\rm ls}\langle\Sigma_{\rm c}^{-1}\rangle^{\rm true}_{\rm ls}/\langle\Sigma_{\rm c}^{-1}\rangle^{\rm est}_{\rm ls}}{\sum_{\rm ls}w_{\rm ls}}, \label{eq:photo-z-corr}
\end{align}
where the weight is given as $w_{\rm ls}=w_{\rm l}w_{\rm s}\avrg{\Sigma^{-1}_{\rm c}}^2_{\rm ls}$
and $w_{\rm l}$ and $w_{\rm s}$ 
are weights given in the HSC shape catalog and the BOSS catalog, 
respectively \citep[see Section~IIB in Ref.][for the definitions]{Miyatake:2021sdd}.
We compute the correction factor for each of the three lens samples, LOWZ, CMASS1, and CMASS2. 
In our method, we multiply the correction factor by the model template of $\dSigma$ as
\begin{align}
    \dSigma^{\rm corr}(R; z_{\rm l}, \Delta z_{\rm ph})
    = f_{\dSigma}(\Delta z_{\rm ph}; z_{\rm l})\dSigma(R|z_{\rm l})
\end{align}
Note that $\dSigma$ includes both the galaxy-galaxy weak lensing and the magnification term 
in Eq.~(\ref{eq:dsigma}): $\dSigma=\dSigma_{\rm gG}+\dSigma_{\rm mag}$, 
and we also use the shifted redshift distribution of source galaxies to compute 
the magnification term, $\dSigma_{\rm mag}$
\footnote{Note that the definition of $f_{\dSigma}$ is the inverse of 
the similar correction factor $f_{\rm ph}$ used in the HSC-Y1 papers 
\citep{Sugiyama:2021, Miyatake:2021sdd}.}.

Similarly, for nonzero $\Delta z_{\rm ph}$, we recompute the model prediction for the cosmic shear correlation functions 
$\xi_\pm(\vartheta)$ using the shifted redshift distribution of the source galaxies.

\subsubsection{Correction for the reference
cosmology used in our measurement}
\label{subsubsec:meascorr}

In the measurements of $\wproj$ and $\dSigma$, we need to assume a ``reference''
cosmology to convert the angular separation between galaxies in the pair to 
the projected separation $R$, and the redshift difference to the radial separation, 
$\Pi$. For $\dSigma$, we also need the reference cosmology to compute 
$\avrg{\Sigma_{\rm c}}^{-1}$, which is needed to convert the shear to $\dSigma$. 
In \citet{more2023}, where we present the measurements, we assume a reference cosmology 
with $\Omega_{\rm m}^{\rm ref}=0.279$, which is only the relevant parameter 
for a flat $\Lambda$CDM model. However, the reference cosmology generally 
differs from the underlying true cosmology, and we need to correct for the discrepancy 
in our cosmological parameter analysis. We follow Ref.~\citep{2015ApJ...806....2M} in order to perform these corrections. 
We denote the cosmological parameters in the parameter inference 
as $\mathbb{C}$ and the reference cosmological parameters for the measurements as $\mathbb{C}^{\rm ref}$. 
Similarly to Section~\ref{subsubsec:photo-z}, we can derive the correction factors, 
by keeping the observables invariant. The corrections for $R$ and $\Pi$ are obtained 
as follows
\begin{align}
    R &= \frac{\chi(z_{\rm l}; \mathbb{C})}{\chi(z_{\rm l}; \mathbb{C}^{\rm ref})}R^{\rm ref},\nonumber\\
    \Pi&= \frac{E(z_{\rm l};\mathbb{C}^{\rm ref})}{E(z_{\rm l};\mathbb{C})} \Pi^{\rm ref}, \label{eq:meascorr-pimax}
\end{align}
Here $E(z) \equiv H(z)/H_0$. Thus we include the measurement corrections 
in the theoretical templates of $\dSigma$ and $\wproj$ as
\begin{align}
    &\dSigma^{\rm ref}(R^{\rm ref},z_{\rm l}| \mathbb{C}, \Delta z_{\rm ph})=
    f_{\dSigma}(z_{\rm l}|\mathbb{C},\Delta z_{\rm ph})
    \dSigma(R, z_{\rm l}|\mathbb{C}), \nonumber\\
    &\wproj(R^{\rm ref},z_{\rm l}|\mathbb{C})= 2f^{\rm RSD}_{\rm corr}\!(R,z_{\rm l};\mathbb{C})
    \frac{E(z_{\rm l};\mathbb{C})}{E(z_{\rm l};\mathbb{C}^{\rm ref})}\nonumber\\
    &\hspace{5em}\times
    \int_0^{\Pi_{\rm max}}\!\mathrm{d}\Pi~\xi_{\rm gg}\!\left(\sqrt{R^2+\Pi^2},z_{\rm l}; \mathbb{C}\right),\label{eq:wp_varyingcosmology}
\end{align} 
where $R$ and $\Pi$ are given in terms of $R^{\rm ref}$ and $\Pi^{\rm ref}$
via the above Eq.~(\ref{eq:meascorr-pimax}) and evaluated at the sampling points 
of $R^{\rm ref}$ and $\Pi^{\rm ref}$ used in the measurements. 
Note that we adopt $\Pi_{\rm max}=[E(\mathbb{C}^{\rm fid})/E(\mathbb{C})]\Pi_{\rm max}^{\rm ref}=[E(\mathbb{C}^{\rm ref})/E(\mathbb{C})]\times 100\,h^{-1}{\rm Mpc}$, 
as we use the fixed $\Pi_{\rm max}^{\rm fid}=100\,h^{-1}{\rm Mpc}$ in the measurement. 
The overall correction factor for $\dSigma$ is defined as
\begin{align}
    f_{\dSigma}(z_{\rm l}|\mathbb{C},\Delta z_{\rm ph})\equiv 
    \frac{\sum_{\rm ls}w_{\rm ls}\langle\Sigma_{\rm c}^{-1}\rangle^{{\rm true}, \mathbb{C}}_{\rm ls}/\langle\Sigma_{\rm c}^{-1}\rangle^{{\rm est}, \mathbb{C}^{\rm ref}}_{\rm ls}}{\sum_{\rm ls}w_{\rm ls}}. \label{eq:photo-z-and-cosmology-corr}
\end{align}
Now this correction factor accounts for both the effects of residual photo-$z$ errors 
($\Delta z_{\rm ph}$) and the use of the reference cosmology. 

Note that the theoretical templates of $\xi_{\pm}(\vartheta)$ for cosmic shear 
correlation functions are not affected by the varying cosmological models, 
as $\xi_{\pm}$ is given as a function of the observed angular separation $\vartheta$.

\subsubsection{Residual multiplicative shear bias}
\label{subsubsec:mbias}

We account for possible residual biases on the weak lensing shear calibration, 
with a nuisance parameter describing the residual multiplicative bias $\Delta m$:
\begin{align}
    \dSigma(R; \Delta m) &= (1+\Delta m)\dSigma(R; \Delta m=0),\\
    \xi_\pm(\vartheta; \Delta m) &= (1+\Delta m)^2\xi_\pm(\vartheta; \Delta m=0).
\end{align}
Since we use the same source sample for both the galaxy-galaxy lensing and 
the cosmic shear measurements, we use the same residual multiplicative bias parameter
for $\dSigma$ and $\xi_{\pm}$. 

\subsubsection{Residual PSF modeling errors}
\label{subsubsec:psf-residual}

Systematic tests of the HSC-Y3 shear catalog presented in \citet{Li2021} indicate 
that there are small residual correlations between galaxy ellipticities and 
PSF ellipticities resulting from imperfect PSF correction. 
Such residual correlations
could produce artificial galaxy shape-shape  correlations and hence bias the cosmic shear measurements. 
Here we examine the impact of these systematics in our cosmic shear measurements, 
assuming that the measured galaxy shapes have an additional additive bias given by
\begin{align}
\epsilon^{(\rm sys)}=\alpha_{\rm psf}\epsilon^{\rm p}+\beta_{\rm psf}\epsilon^{\rm q}.
\end{align}
The first term, referred to as PSF leakage, represents the systematic error proportional to
the PSF model ellipticity $\epsilon^{\rm p}$ due to the imperfection in the method used to correct the galaxy shapes for the impact of the PSF. 
The second term represents the systematic error associated with the difference between 
the model PSF ellipticity, $\epsilon^{\rm p}$, and the true PSF ellipticity.  This difference is 
estimated from individual ``reserved'' stars $\epsilon^{\rm star}$, i.e. $\epsilon^{\rm q}\equiv \epsilon^{\rm p}-\epsilon^{\rm star}$ \citep{2018PhRvD..98d3528T}. 
A coherent residual PSF ellipticity $e^{\rm q} $ indicates an imperfect PSF estimate, 
which should propagate to shear estimates of galaxies. 

When the observed galaxy ellipticity is contaminated by $\epsilon^{(\rm sys)}$, 
these systematic terms cause an additional contamination to the measured cosmic shear 
correlation functions as 
\begin{align}
    \xi_{{\rm psf},\pm}(\vartheta)=\alpha_{\rm psf}^2\hat{\xi}^{\rm pp}_{\pm}(\vartheta)
    +2\alpha_{\rm psf}\beta_{\rm psf}\hat{\xi}^{\rm pq}_{\pm}(\vartheta)
    +\beta_{\rm psf}^2\hat{\xi}^{\rm qq}_{\pm}(\vartheta)\, ,
    \label{eq:psf_sys_cosmicshear}
\end{align}
where $\hat{\xi}^{\rm pp}_{\pm}$, $\hat{\xi}^{\rm qq}_{\pm}$ and $\hat{\xi}^{\rm pq}_{\pm}$ 
represent the auto-correlation of the model PSF ellipticity $\epsilon^{\rm p}_{\pm}$, 
the auto-correlation of the residual PSF ellipticity $\epsilon^{\rm q}_{\pm}$, 
and the cross-correlation of $\epsilon^{\rm p}_{\pm}$ and $\epsilon^{\rm q}_{\pm}$, 
respectively. The hat notation, ``$\hat{\hspace{1em}}$'', denotes the correlation function 
measured from the HSC data using the model PSF and the reserved stars 
(see \citet{more2023}).
The proportional coefficients $\alpha_{\rm psf}$ and $\beta_{\rm psf}$ are estimated 
by cross-correlating $\epsilon^{\rm p}_{\pm}$ and $\epsilon^{\rm q}_{\pm}$ 
with the observed galaxy ellipticities
 as
\begin{align}
    &\hat{\xi}^{\rm gp}_{\pm}(\vartheta)
    =\alpha_{\rm psf}\hat{\xi}^{\rm pp}_{\pm}(\vartheta)+
    \beta_{\rm psf}\hat{\xi}^{\rm pq}_{\pm}(\vartheta),\nonumber\\
    &\hat{\xi}^{\rm gq}_{\pm}(\vartheta)
    =\alpha_{\rm psf}\hat{\xi}^{\rm pq}(\vartheta)+\beta_{\rm psf}\hat{\xi}^{\rm qq}_{\pm}(\vartheta),
    \label{eq:galaxy_psf_correlation}
\end{align}
where $\hat{\xi}_{\pm}^{\rm gp}$ and $\hat{\xi}_{\pm}^{\rm gq}$ are 
the measured cross-correlations between galaxy ellipticities, used 
for the cosmic shear data vector, and $\epsilon^{\rm p}_{\pm}$ and $\epsilon^{\rm q}_{\pm}$. 
As shown in \citet{more2023}, we found $\alpha_{\rm psf}=-0.0292\pm 0.0129$ 
and $\beta_{\rm psf}=-2.59\pm 1.65$ for our fiducial source sample.

To take into account the impact of these additive shear residuals 
on parameter inference, 
we add the 
contamination term $\xi_{{\rm psf},\pm}$ (Eq.~\ref{eq:psf_sys_cosmicshear}) 
to the model cosmic shear correlation function $\xi_{\pm}$ in Eq.~(\ref{eq:cosmic-shear}) and then 
estimate parameters by varying the parameters $\alpha_{\rm psf}$ and $\beta_{\rm psf}$ 
with the Gaussian priors with widths inferred from the above errors.
Note that the above PSF systematics causes additive shear bias, and does not cause a bias in the galaxy-galaxy
weak lensing \citep{Mandelbaum:05b}.

We note that the PSF systematics model we adopted here is based on the second moments of PSF as done in \citet{2020PASJ...72...16H}, while the HSC-Y3 tomographic cosmic shear analyses  \cite{li2023,dalal2023} use a PSF systematics  model with additional terms including 
the fourth moments \citep{zhang:2022dvs}. Because 
the contamination from PSF systematics effects in the cosmic shear signal is relatively small for high-redshift HSC source galaxies compared to the signal at lower redshift, 
the second-moment-based PSF systematics model is sufficient for our analysis. We explicitly validate the use of our PSF systematics model 
in Appendix~\ref{sec:apdx-model-validation} by performing the cosmological parameter analysis on a synthetic data vector including the measured PSF systematics up to the fourth-moments in the synthetic cosmic shear data vector.

\subsection{Summary: theoretical template}
\label{sec:summary_theoretical_template}

For the convenience of the following discussion, here we summarize the theoretical templates, 
explicitly showing which parameters are used to model each of the theoretical templates:
\begin{widetext}
\begin{align}
    \dSigma^{\rm t}\!(R^{\rm ref},z_{\rm l}|\mathbb{C},b_1\!(z_{\rm l}),\Delta z_{\rm ph},\Delta m,
    \alpha_{\rm mag}(z_{\rm l})) &=(1+\Delta m)\dSigma^{\rm ref}(R^{\rm ref},z_{\rm l}|
    \mathbb{C},b_1\!(z_{\rm l}),\Delta z_{\rm ph},\alpha_{\rm mag}(z_{\rm l}))\, ,\nonumber\\
    \wproj^{\rm t}\!(R^{\rm ref},z_{\rm l}|\mathbb{C},b_1\!(z_{\rm l}))
    &: \hspace{1em} \mbox{Eqs.~(\ref{eq:wproj}) and (\ref{eq:wp_varyingcosmology})}\, ,\nonumber\\
    \xi_{\pm}^{\rm t}(\vartheta|\mathbb{C},\Delta z_{\rm ph},A_{\rm IA},\Delta m,\alpha_{\rm psf},
    \beta_{\rm psf})
    &=(1+\Delta m)^2
    \xi_{\pm}(\vartheta|\mathbb{C},\Delta z_{\rm ph}, A_{\rm IA})
    +\xi_{{\pm},{\rm psf}}(\vartheta|\alpha_{\rm psf},\beta_{\rm psf})\, .
    \label{eq:summary_theoretical_template}
\end{align}
\end{widetext}
For $\dSigma^{\rm ref}$ and $\wproj^{\rm t}$, 
we compute these model predictions 
at the sampling points of $R^{\rm ref}$ for each of the LOWZ,
CMASS1 and CMASS2 samples at their representative redshift. Here $\mathbb{C}$ denotes a 
cosmological model sampled in parameter inference, and characterized by 
5~cosmological parameters for the flat $\Lambda$CDM model, $(\Omega_{\rm de}, \ln(10^{10}A_{\rm s}), \omega_{\rm c}, \omega_{\rm b}, n_{\rm s})$.
$b_1\!(z_{\rm l})$ denotes the linear bias parameter for the LOWZ, CMASS1 or CMASS2 sample, 
and other parameters are nuisance parameters to model the residual systematic errors 
in photo-$z$'s, magnification bias, multiplicative shear bias, PSF modeling, and intrinsic alignment. 
For our baseline analysis, we have 16
parameters in total: 
$16=5~(\mathbb{C})+3\times 1~(b_1)+8~({\rm nuisance})$.

\subsection{Bayesian inference: Likelihood and Prior}
\label{subsec:bayes-likelihood-prior}

To infer parameters $\boldsymbol{\theta}$ from the measured clustering observables, we
compare a ``data vector'', denoted as $\hat{{\bm d}}$, to a ``theoretical model template'', 
denoted as ${\bm t}$. 
We define the data vector from the measured signals of $\wproj$, $\dSigma$, and $\xi_{\pm}$ as
\begin{align}
    {{\bm d}}\equiv \left\{\widehat{w}_{\rm p}\!(R^{\rm ref}_i|z_{\rm l}),~
    \widehat{\dSigma}\!(R^{\rm ref}_j|z_{\rm l}),~
    \widehat{\xi}_{\pm}(\vartheta_k)
    \right\}\, ,
    \label{eq:data_vector}
\end{align}
where $z_{\rm l}$ stands for the representative redshift of either LOWZ, CMASS1 or CMASS2 sample.
Here we emphasize that the measured signals are sampled at discrete values of
of separation, $R^{\rm ref}_i$, $R^{\rm ref}_j$ and $\vartheta_k$ for $\widehat{w}_{\rm p}$, $\widehat{\dSigma}$ and $\widehat{\xi}_{\pm}$, respectively, where $R^{\rm ref}_i$ and 
$R^{\rm ref}_j$ are estimated from the observed angular separations between galaxies 
in the pair assuming the reference cosmology as we described above. 

For $\wproj$, we use the signals in the range of $R^{\rm ref}=[8,80]~h^{-1}{\rm Mpc}$. 
The minimum scale cut is determined so that the minimum bias model fairly well 
describes the signals \citep{Sugiyama:2020,Sugiyama:2021}, without being so affected 
by the strongly nonlinear clustering. The maximum scale cut is determined 
such that our constraint purely comes from the large scale clustering amplitude, 
and does not include information from the baryonic acoustic oscillations.
The constraining power of cosmological parameters is mainly from scales around the minimum scale cut, where have 
higher signal-to-noise ratios. 
We take 14 logarithmically-spaced bins for each of the LOWZ, CMASS1 and CMASS2 samples. 
For $\dSigma$, we use the signals in the range of 
$R^{\rm ref}=[12,30]$, $[12,40]$ or $[12,80]~h^{-1}{\rm Mpc}$
for the LOWZ, CMASS1 or CMASS2 sample, respectively. The minimum scale cuts are 
determined by the same reason as the $\wproj$ case, while the maximum scale cuts 
are determined based on the systematic tests, i.e., where we do not find any significant signal 
of non-lensing $\dSigma_\times$ at scales below the maximum cuts compared to the statistical errors. The scale cuts give 4, 5, and 8 logarithmically-spaced radial bins
for the LOWZ, CMASS1 and CMASS2 samples, respectively. For cosmic shear 
we use 8 angular separation bins in the range of 
$\vartheta=[7.9,50.1]~{\rm arcmin}$ for 
$\xi_+(\vartheta)$, while 7 angular bins in
$\vartheta=[31.6,158]~{\rm arcmin}$ for $\xi_-(\vartheta)$.
Thus we use $74$~data points in total: $42(=3\times 14)$ for $\wproj$, $17(=4+5+8)$ 
for $\dSigma$ and $15(=7+8)$ for $\xi_{\pm}$, respectively.

For the theoretical template ${\bm t}$, we construct the ``model vector'' 
of the clustering correlation functions computed using a set 
of model parameters ${\boldsymbol{\theta}}$ within the flat $\Lambda$CDM framework:
\begin{align}
    {\bm t}({\boldsymbol{\theta}})\equiv \left\{w_{\rm p}^{\rm t}\!(R^{\rm ref}_i,z_{\rm l}|{\boldsymbol{\theta}}), \dSigma^{\rm t}\!(R^{\rm ref}_jz_{\rm l}|{\boldsymbol{\theta}}),
    \xi_{\pm}^{\rm t}(\vartheta_k|{\boldsymbol{\theta}})\right\},
    \label{eq:model_vactor}
\end{align}
where the theoretical templates of clustering observables (denoted by the superscript ``t'')
are computed at the representative redshift of each lens sample, 
$z_{\rm l}$,
using Eq.~(\ref{eq:summary_theoretical_template}). 
We use the publicly-available {\tt FFTLog} code developed in 
Ref.~\citep{Fang:2019xat}, which is a modified version from the original code \cite{Hamilton00}, 
to perform Hankel transforms in the model-prediction calculations. 
We also use {\tt FFTlog} to compute the average of model prediction 
within a finite radial bin width used in the measurements: 
$\Delta\ln R=0.169$ for $\wproj$, $\Delta\ln R=0.246$ for $\dSigma$ and $\Delta\ln\vartheta=0.242$ for $\xi_\pm$.

We assume that the likelihood of the data vector compared to the model vector 
follows a multivariate Gaussian distribution:
\begin{align}
    \ln {\cal L}({\bm d}|{\boldsymbol{\theta}})=-\frac{1}{2}\left[
    {{\bm d}}-{\bm t}(\boldsymbol{\theta})
    \right]^T {\bm C}^{-1}\left[
    {{\bm d}}-{\bm t}(\boldsymbol{\theta})
    \right],
    \label{eq:likelihood}
\end{align} 
where ${\bm C}$ is the covariance matrix of data vector [see \citet{more2023} for details], 
${\bm C}^{-1}$ is the inverse matrix.
The covariance matrix is estimated in \cite{more2023} from $1404(=$108$\times$13) realizations of the mock signals 
\citep{2019MNRAS.486...52S,Shirasakietal:17}. 
When we compute the inverse covariance in the likelihood, we multiply the factor 
$(108\times13-74-2)/(108\times13-1)=0.95$ to obtain 
the inverse covariance \citep{2007A&A...464..399H}\footnote{We mistakenly omitted one realization of the full sky simulation for the covariance estimation. Thus, in practice, we use $(107\times13-74-2)/(107\times13-1)=0.946$ for the Hartlap factor instead.}. 
The cross covariance between galaxy-galaxy lensing and cosmic shear is included 
because we use the same mock catalogs for clustering, galaxy-galaxy lensing, 
and cosmic shear measurements. Since the overlapping region 
between the HSC-Y3 and SDSS DR11 survey footprints, which has about 416~\sqdeg, is much smaller than the 
SDSS DR11 area (about 8,300~\sqdeg), we ignore the cross-covariance between the clustering 
($\wproj$) and galaxy-galaxy lensing ($\dSigma$). 
The mock catalogs used in the covariance matrix estimation 
are generated using full-sky simulations \citep{2017ApJ...850...24T}, 
and hence the covariance automatically includes the super-sample 
covariance contribution \cite{TakadaHu:13}. 
The additional covariance contribution due to the magnification bias effect on the lens galaxy distribution 
is analytically estimated and added onto the estimate from mock measurements \citep{Sugiyama:2021}. 
See \citet{more2023}
for more detail of the covariance matrix estimation.

We construct a posterior probability distribution for the parameters $\boldsymbol{\theta}$ given the data vector ${\bm d}$, 
denoted as ${\cal P}(\boldsymbol{\theta}|{\bm d})$,  
 by performing Bayesian inference: 
\begin{align}
    {\cal P}(\boldsymbol{\theta}|{\bm d})\propto {\cal L}({\bm d}|\boldsymbol{\theta})\Pi(\boldsymbol{\theta}),
    \label{eq:basyesian_posterior}
\end{align}
where $\Pi(\boldsymbol{\theta})$ is the prior distribution of $\boldsymbol{\theta}$.

In Table \ref{tab:parameters}, we summarize the model parameters and their priors. 
The first section summarizes the cosmological parameters: $\Omega_{\rm m}(=1-\Omega_{\rm de})$ 
and $\ln 10^{10}A_{\rm s}$ are the parameters to which our weak lensing and clustering 
observables are most sensitive, and we adopt uninformative uniform priors on these model parameters. 
On the other hand, the weak lensing and clustering analyses are not sensitive to 
$\omega_b=\Omega_{\rm b}h^2$ and $n_{\rm s}$, and hence we adopt informative priors 
using normal distributions: we use a BBN prior for $\omega_b$ \citep{Aver:2015iza,Cooke:2017cwo,Schoneberg:2019wmt}, and  a Planck prior on $n_{\rm s}$ \citep{2020A&A...641A...6P}. 
Note that we increased the uncertainty of the Planck prior on $n_{\rm s}$ by a factor of three 
to be conservative. The parameters $b_1\!(z_{\rm i})$ in the second section are the linear galaxy bias parameters 
for $i=$ LOWZ, CMASS1 and CMASS2. We use uninformative priors on each of these parameters, given that 
our samples could be affected by assembly bias \citep{Sugiyama:2020}. 
For the magnification bias parameter, $\alpha_{\rm mag}(z_{\rm l})$, we adopt a prior using a normal distribution: the central value is taken from the estimated slope of number counts at luminosity cut, while we adopt a relatively wide width, $\sigma(\alpha_{\rm mag})=0.5$,  for a conservative analysis.

The fourth section summarizes the residual redshift and the residual multiplicative bias parameters.
In the small-scale analysis by \citet{miyatake2023}, 
we find that the weak lensing signals have a
statistical power to calibrate the residual redshift error parameter 
($\Delta z_{\rm ph}$) to the precision of $\sigma(\Delta z_{\rm ph})\simeq 0.1$, based on the method in Ref.~\cite{OguriTakada:11}. 
On the other hand, as discussed in Section~\ref{sec:results}, we find that the statistical power of the large-scale
signals is not sufficient to calibrate $\Delta z_{\rm ph}$. 
Therefore, in this paper, we 
use a Gaussian prior on $\Delta z_{\rm ph}$, ${\cal N}(-0.05, 0.09)$, 
that is 
inferred from the mode and the credible interval of 
the posterior distribution of $\Delta z_{\rm ph}$ in the fiducial small-scale analysis by the companion work \citet{miyatake2023}.
That is, we employ the prior that is centered at $\Delta z_{\rm ph}=-0.05$, meaning that the source redshift distribution inferred from the photo-$z$ estimates is lower than the true distribution by $|\Delta z_{\rm ph}|=0.05$.

For the prior on the multiplicative shear bias, 
we use the Gaussian prior with zero mean and width of 
$\sigma=0.01$, which is estimated 
from HSC galaxy image simulations \citep{Li2021}.
The fifth section summarizes the PSF residual systematics modeling parameters. 
We use Gaussian prior for these model parameters. 
The center and width are estimated from the cross-correlation between star shapes 
and galaxy shapes \citep[see][for more detail]{more2023}. 
The sixth section summarizes the single intrinsic alignment parameter of the NLA model for this source sample, 
for which we use an uninformative uniform prior. 
The dimension of the fiducial model parameter vector is 16
in total, 
5 for the cosmological parameters and 
11
for the nuisance parameters.

We sample parameters from their posterior distribution given the data vector
using the Monte Carlo method in 
this high-dimensional parameter space. 
In particular, we utilize the nested sampling algorithm implemented in {\tt MultiNest} \cite{Feroz:2009}
from the python interface {\tt PyMultiNest} \citep{2014A&A...564A.125B}.
{\tt MultiNest} has 
two  hyper-parameters, 
the live points {\tt nlive} and the sampling efficiency rate {\tt efr}. 
We use {\tt nlive} $=600$ and {\tt efr} $=0.3$ as the fiducial setup. 
Another hyper-parameter, the tolerance {\tt tol}, is set to $0.1$, 
and replaced with a smaller value if necessary to check
for convergence.

In this paper, we report the inference result in the format of
\begin{align}
    \text{mode}^{+34\%~\text{upper}}_{-34\%~\text{lower}}~(\text{MAP}),
\end{align}
where the mode is the peak value of a parameter in the one-dimensional marginal posterior distribution, the 68\% credible interval is defined as the highest density interval of the 
posterior, and ``$+34\%$~upper'' and ``$-34\%$~lower'' are the upper and lower limit of the 68\% credible interval \citep[see Fig.~3 of ][for the illustration of the definitions of these statistics of marginalized posterior]{Sugiyama:2021}.
We also report the ``MAP'' value of the parameter that is the parameter value at 
the {\it maximum a posteriori} model which has
the highest posterior value in the chain. The mode value is defined with the marginal posterior and thus subject to the projection effect of the posterior distribution from the full-parameter space, while the MAP is not. Thus, a significant difference between the mode and MAP value may indicate the degree to which the mode value is affected due to lower dimensional
projection of the posterior distribution. However, we should note that the estimation of the MAP value can be noisy due to a finite number of samples in the chain, especially in the presence of severe parameter degeneracy(ies), resulting in a MAP that corresponds to 
a local minimum in the posterior surface.
Therefore, we will use the MAP value and its difference from the mode as an indicator of 
projection effects. 
In the summary table that gives 
the cosmological constraints for various setups and tests, 
we also report
the mean value as the third
point estimate so that one can easily compare our results with external results that also use the mean.

\begin{table}
\caption{Model parameters and priors used in our cosmological parameter inference. The label ${\cal U}(a,b)$ denotes a uniform (or equivalently flat) distribution with minimum $a$ and maximum $b$, while ${\cal N}(\mu, \sigma)$ denotes a normal distribution with mean $\mu$ and width $\sigma$. 
For the residual photo-$z$ error parameter, $\Delta z_{\rm ph}$, 
we employ the informative Gaussian prior 
${\cal N}(-0.05,0.09)$ in our baseline analysis, which is taken
from the companion analysis result in \citet{miyatake2023} that perform a parameter inference by comparing 
the halo model based predictions to 
exactly the same clustering observables (down to the smaller scale cuts for $\dSigma$ and $\wproj$).
}
\label{tab:parameters}
\setlength{\tabcolsep}{15pt}
\begin{ruledtabular}
\begin{center}
\begin{tabular}{ll}
Parameter & Prior \\ \hline
\multicolumn{2}{l}{\hspace{-1em}\bf Cosmological parameters}\\
$\Omega_{\rm de}$       & ${\cal U}(0.4594, 0.9094)$\\
$\ln(10^{10}A_{\rm s})$ & ${\cal U}(1.0,5.0)$\\
$\omega_{\rm c}$        & ${\cal U}(0.0998, 0.1398)$\\ 
$\omega_{\rm b}$        & ${\cal N}(0.02268,0.00038)$\\
$n_{\rm s}$             & ${\cal N}(0.9649,3\times0.0042)$\\ \hline
\multicolumn{2}{l}{\hspace{-1em}\bf galaxy bias parameters}\\
$b_1(z_{\rm LOWZ})$     & ${\cal U}(0.1,5.0)$\\ 
$b_1(z_{\rm CMASS1})$   & ${\cal U}(0.1,5.0)$\\ 
$b_1(z_{\rm CMASS2})$   & ${\cal U}(0.1,5.0)$\\ \hline
\multicolumn{2}{l}{\hspace{-1em}\bf magnification bias parameters}\\
$\alpha_{\rm mag}(z_{\rm LOWZ})$   & ${\cal N}(2.259, 0.5)$ \\
$\alpha_{\rm mag}(z_{\rm CMASS1})$ & ${\cal N}(3.563, 0.5)$ \\
$\alpha_{\rm mag}(z_{\rm CMASS2})$ & ${\cal N}(3.729, 0.5)$ \\\hline
\multicolumn{2}{l}{\hspace{-1em}\bf Photo-$z$ / Shear errors}\\
$\Delta z_{\rm ph}$     & ${\cal N}(-0.05, 0.09)$  \\
$\Delta m$       & ${\cal N}(0.0,0.01)$ \\ \hline
\multicolumn{2}{l}{\hspace{-1em}\bf PSF residuals}\\
$\alpha_{\rm psf}$      & ${\cal N}(-0.026, 0.010)$\\
$\beta_{\rm psf}$       & ${\cal N}(-1.656, 1.326)$\\
\hline
\multicolumn{2}{l}{\hspace{-1em}\bf Intrinsic Alignment parameters}\\
$A_{\rm IA}$            & ${\cal U}(-5, 5)$ \\
\end{tabular}\end{center}
\end{ruledtabular}
\end{table}

\section{Blinding scheme and internal consistency}
\label{sec:blind-internal-consistency}
To avoid confirmation bias, we perform our cosmological analysis in a blind fashion. 
The details of the blinding scheme can be also found in Section~IIB of \citet{more2023}.
We employ a two-tier blinding strategy to avoid unintentional
unblinding during the cosmological analysis. The two tiers are as follows:
\begin{itemize}
\item {\it Catalog level}: 
The analysis team performs the cosmological analysis using 
three different weak lensing shape catalogs. 
Only one is the true catalog and the other two catalogs have multiplicative biases which are different from the truth (see below for details). 
The analysis team members do not know which is the true catalog. 
\item {\it Analysis level}: When the analysis team makes plots 
comparing the measurements 
with  theoretical models, the $y$-axis values (e.g., the amplitudes of $\dSigma$) are hidden and the analysis team is not allowed to see the values of cosmological parameters used in the theoretical models. 
When the analysis team makes plots showing the credible intervals of 
cosmological parameters (i.e., the posterior distribution), the central value(s) of parameter(s) 
are shifted to zero, and only the range of the credible interval(s) can be seen. 
Finally, the analysis team does not compare the posterior for cosmological parameter(s) 
or the model predictions with external results such 
as the {\it Planck} CMB cosmological parameters prior to unblinding. 
\end{itemize}

See Section~IIB of \citet{more2023} \citep[also see][]{li2023}
for details of how the blinded catalogs were 
constructed in a manner that prevents accidental unblinding 
by the analysis team. 
The use of these catalogs means that the analysis team must perform three analyses, 
but this method avoids the need for reanalysis once the catalogs are unblinded.

The set of the three shape catalogs used in this paper is shared with the two companion papers, 
\citet{more2023} presenting details of the measurements of observables used in this paper
and \citet{miyatake2023}
presenting the cosmological parameter estimation from the same signals 
as those of this paper, but including the information down to smaller scales and 
using the halo model-based method.
We imposed the following criteria for deciding to unblind our results: 
\begin{itemize}
    \setlength{\itemsep}{0em}
    \item Analysis software is made available to collaboration members and 
    specific members have reviewed each part of the code.
    \item Validation tests of the cosmology analysis pipeline are performed using synthetic
    data vectors, some of which are generated using mock catalogs of galaxies. 
    In particular, the key cosmological parameter $S_8$ must be recovered 
    to within $0.5\sigma$ ($\sigma$ is the marginalized credible interval), for the fiducial mock catalogs (see below).
    \item Internal consistency tests are performed to check whether the estimate of 
    the key cosmological parameter is changed, 
    compared to the fiducial analysis method, using subsets of data vector and/or 
    different analysis setups. 
    Table~\ref{tab:internal-tests} summarizes the internal consistency tests that 
    we performed before unblinding.
    \item The goodness of fit of the best-fit model predictions to the data vector in each of the three blind catalogs is quantified.
\end{itemize}
The analysis team resolved that the results would be published regardless of the
outcome once the results are unblinded, without any changes or modifications to
the analysis method after unblinding. In the following text, we will explicitly indicate
any analysis or results that were obtained after unblinding.

\begin{table*}
\caption{Internal consistency tests carried out for the cosmological parameter analysis.  All of the analyses 
are performed {\it before} unblinding the results. 
In this paper, we use the prior $\Pi(\Delta z_{\rm ph})={\cal N}(-0.05, 0.09)$, which is 
obtained from the small-scale 3$\times$2pt analysis in \citet{miyatake2023},
for  the analysis setups that are denoted by 
the superscript 
``${}^\ast$'' in the label (see Table~\ref{tab:parameters} and main text explaining the table). 
We also consider the analyses using the different photo-$z$ prior centered at $\Delta z_{\rm ph}=0$, 
$\Pi(\Delta z_{\rm ph})={\cal N}(0,0.1)$, to study the impact of the different priors of the 
photo-$z$ error parameter on our results, as  
denoted by 
the superscript ``$^\dagger$'' in the label.
The third column denotes the dimension
of the sampled model parameters and the data vectors, $\mathcal{D}(\bm{\theta})$ and $\mathcal{D}(\bm{d})$,
respectively.}
\label{tab:internal-tests}
\renewcommand{\arraystretch}{1.2}
\setlength{\tabcolsep}{15pt}
\begin{center}
\begin{ruledtabular}
\begin{tabular}{llc}
setup label & description & $\mathcal{D}(\bm{\theta})$, $\mathcal{D}(\bm{d})$\\
\hline
3$\times$2pt${}^\ast$                                            & [{\it baseline analysis}] a joint analysis of $\dSigma$, $\wproj$ and $\xi_\pm$                        & 16, 74\\
2$\times$2pt${}^\ast$                                            & a joint analysis of $\dSigma$ and $\wproj$                                                             & 13, 59\\
cosmic shear${}^\ast$                                            & cosmic shear analysis using $\xi_\pm$                                                                  & 10, 15\\
\hline
3$\times$2pt, w/o LOWZ${}^\ast$                                  & 3$\times$2pt, without LOWZ sample for $\wproj$ and $\dSigma$                                           & 14, 56\\
3$\times$2pt, w/o CMASS1${}^\ast$                                & 3$\times$2pt, without CMASS1 sample for $\wproj$ and $\dSigma$                                         & 14, 55\\
3$\times$2pt, w/o CMASS2${}^\ast$                                & 3$\times$2pt, without CMASS2 sample for $\wproj$ and $\dSigma$                                         & 14, 52\\
\hline
2$\times$2pt, w/o LOWZ${}^\ast$                                  & 2$\times$2pt, without LOWZ sample for $\wproj$ and $\dSigma$                                           & 11, 41\\
2$\times$2pt, w/o CMASS1${}^\ast$                                & 2$\times$2pt, without CMASS1 sample for $\wproj$ and $\dSigma$                                         & 11, 40\\
2$\times$2pt, w/o CMASS2${}^\ast$                                & 2$\times$2pt, without CMASS2 sample for $\wproj$ and $\dSigma$                                         & 11, 37\\
\hline
no photo-$z$ error                                               & 3$\times$2pt, fixing $\Delta\!z_{\rm ph}=0$                                                            & 15, 74\\
no shear error${}^\ast$                                          & 3$\times$2pt, fixing $\Delta\!m=0$                                                                     & 15, 74\\
no magnification bias error${}^\ast$                             & 3$\times$2pt, fixing $\alpha_{\rm mag}=\mu$                                                            & 13, 74\\
no PSF error${}^\ast$                                            & 3$\times$2pt, fixing $\alpha_{\rm psf}=\beta_{\rm psf}=0$                                              & 14, 74\\
no IA${}^\ast$                                                   & 3$\times$2pt, fixing $A_{\rm IA}=0$                                                                    & 15, 74\\
extreme IA${}^\ast$                                              & 3$\times$2pt, fixing $A_{\rm IA}=5$                                                                    & 15, 74\\
\hline
$R_{\rm max}=30h^{-1}{\rm Mpc}$${}^\ast$                         & 3$\times$2pt, using the $\dSigma$ and $\wproj$ signals up to $R\leq R_{\rm max}=30h^{-1}{\rm Mpc}$     & 16, 51\\
\hline
2cosmo${}^\ast$                                                  & 3$\times$2pt, varying only two cosmological parameters, $\Omega_{\rm de}$ and $\ln(10^{10}A_{\rm s})$  & 13, 74\\
2cosmo, 2$\times$2pt${}^\ast$                                    & 2$\times$2pt, varying only two cosmological parameters, $\Omega_{\rm de}$ and $\ln(10^{10}A_{\rm s})$  & 10, 59\\
\hline
$\Delta z_{\rm ph}\sim{\cal U}(-1,1)$                            & 3$\times$2pt, with an uniform prior of $\Delta z_{\rm ph}\sim {\cal U}(-1,1)$                          & 16, 74\\
\hline
3$\times$2pt ${}^{\dagger}$                                      & 3$\times$2pt, with prior of $\Delta z_{\rm ph}\sim{\cal N}(0, 0.1)$                                    & 16, 74\\
2$\times$2pt ${}^{\dagger}$                                      & 2$\times$2pt, with prior of $\Delta z_{\rm ph}\sim{\cal N}(0, 0.1)$                                    & 16, 59\\
cosmic shear ${}^{\dagger}$                                      & cosmic shear, with prior of $\Delta z_{\rm ph}\sim{\cal N}(0, 0.1)$                                    & 16, 15\\
\hline
XMM      ($33.17{\rm deg}^2$)${}^\ast$                           & 3$\times$2pt, but using only XMM field for $\dSigma$ and $\xi_\pm$                                     & 16, 74\\
GAMA15H  ($40.87{\rm deg}^2$)${}^\ast$                           & 3$\times$2pt, but using only GAMA15H field for $\dSigma$ and $\xi_\pm$                                 & 16, 74\\
HECTOMAP ($43.09{\rm deg}^2$)${}^\ast$                           & 3$\times$2pt, but using only HECTOMAP field for $\dSigma$ and $\xi_\pm$                                & 16, 74\\
GAMA09H  ($78.85{\rm deg}^2$)${}^\ast$                           & 3$\times$2pt, but using only GAMA09H field for $\dSigma$ and $\xi_\pm$                                 & 16, 74\\
VVDS     ($96.18{\rm deg}^2$)${}^\ast$                           & 3$\times$2pt, but using only VVDS field for $\dSigma$ and $\xi_\pm$                                    & 16, 74\\
WIDE12H  ($121.32{\rm deg}^2$)${}^\ast$                          & 3$\times$2pt, but using only WIDE12H field for $\dSigma$ and $\xi_\pm$                                 & 16, 74\\
\hline
w/o star weight                                           & 3$\times$2pt, but without using star weight when computing $w_{\rm p}$ & 16, 74\\
\hline
\dempz\& WX${}^\ast$                                             & 3$\times$2pt, using \dempz \& WX for $p_{\rm s}(z_{\rm s})$                                            & 16, 74\\
\mizuki${}^\ast$                                                 & 3$\times$2pt, but using \mizuki for the source selection and the stacked $p_{\rm s}(z_{\rm s})$        & 16, 74\\
\dnnz${}^\ast$                                                   & 3$\times$2pt, but using \dnnz for the source selection and the stacked $p_{\rm s}(z_{\rm s})$          & 16, 74\end{tabular}
\end{ruledtabular}
\end{center}
\renewcommand{\arraystretch}{1}
\end{table*}

\section{Model validation}
\label{sec:model-validation}
In order to obtain unbiased cosmological parameters from the 
data, 
we need to validate our model. We adopt the minimal bias model for the galaxy bias, 
which was already validated using mock catalogs populated by galaxies using different models for the galaxy-halo connections
in \citet{Sugiyama:2020} for the hypothetical HSC-Y1 mock data.

In addition to the galaxy bias uncertainties, we also validate the use of 
{\tt halofit}
for the prediction of cosmic shear signals, neglecting  baryonic effects 
on the nonlinear matter power spectrum.
To validate this, we simulate the data vectors using {\tt HMCode} with various degrees of baryonic physics.

For the HSC-Y3 data,
we have greater statistical constraining power due to three times larger area coverage 
than the HSC-Y1 data, and also due to the inclusion of cosmic shear in the 3$\times$2pt analysis
compared to the HSC-Y1 2$\times$2pt analysis \citep{Sugiyama:2021}. 
Therefore, we subject our modeling and analysis methods
to validation tests using
the mock data vectors, but using the covariance matrix for the HSC-Y3 data.

In this paper, we use an informative prior on $\Delta z_{\rm ph}$ taken
from the result of the small-scale 3$\times$2pt
analysis, which can more precisely self-calibrate $\Delta z_{\rm ph}$
due to its higher signal-to-noise ratio.
We validate the use of the posterior from the small-scale analysis as a prior on $\Delta z_{\rm ph}$
by analyzing a mock data vector that includes the photo-$z$ bias effect. 

We describe 
the model validation results 
in Appendix~\ref{sec:model-validation} in detail.
In brief summary, the validation tests are passed for most of the synthetic data vectors; 
our method can recover the input $S_8$ with a bias smaller than $0.5\sigma$ ($\sigma$ is the marginalized error 
for the baseline 3$\times$2pt analysis), where we test our method using the mode value of $S_8$ in the posterior, rather than the MAP value, compared to the input value used in generating the synthetic data. 
However, given the fact that there is no established, accurate theory of the galaxy formation physics or galaxy-halo connection, we also consider the worst-case scenario in order for us to be ready for any unexpected result in the cosmological parameter estimation. As for the worst-case scenario, we consider the extreme mock data of SDSS galaxies, where we consider a non-standard prescription of the galaxy-halo connection when generating the mock SDSS catalogs from $N$-body simulations. 
Even for these extreme mocks, the minimal bias model can recover the input $S_8$ value with a bias $\lesssim 1\sigma$, while the halo model method suffers from a greater bias. For these worst-case cases, we have a diagnostic to identify a signature of the strong nonlinearities in the galaxy-halo connection. Since changes in the galaxy-halo connection cause a stronger modification in the clustering observables at smaller scales around and below virial radii of massive halos (Mpc scales), while the clustering observables have the linear theory limit on sufficiently large scales, which is captured by the minimal bias model. Hence, if the actual SDSS galaxies are affected by such extreme galaxy-halo connection, the cosmological parameters display systematic shifts with changing the scales cuts from small to large scales in the cosmology inference. We carefully study these behaviors using the different mocks. For the actual cosmology analysis of the HSC-Y3 and SDSS data, we did not observe such a systematic change in the $S_8$ values for the different scales cuts or from the small- and large-scale 3$\times$2pt analyses. These tests gave us a justification that we can unblind the results. 

\section{Results}
\label{sec:results}

\subsection{HSC-Y3 \texorpdfstring{$\Lambda$CDM}{LambdaCDM} constraint}
\label{subsec:hsc-y3-result}

In this section, we present the main results from the cosmological parameter inference using 
the HSC-Y3 lensing and SDSS clustering measurements for
the flat $\Lambda$CDM model.  All of the analyses in this section were done before
unblinding, and the results are presented without any change after unblinding.

\begin{figure*}
    \includegraphics[width=1.8\columnwidth]{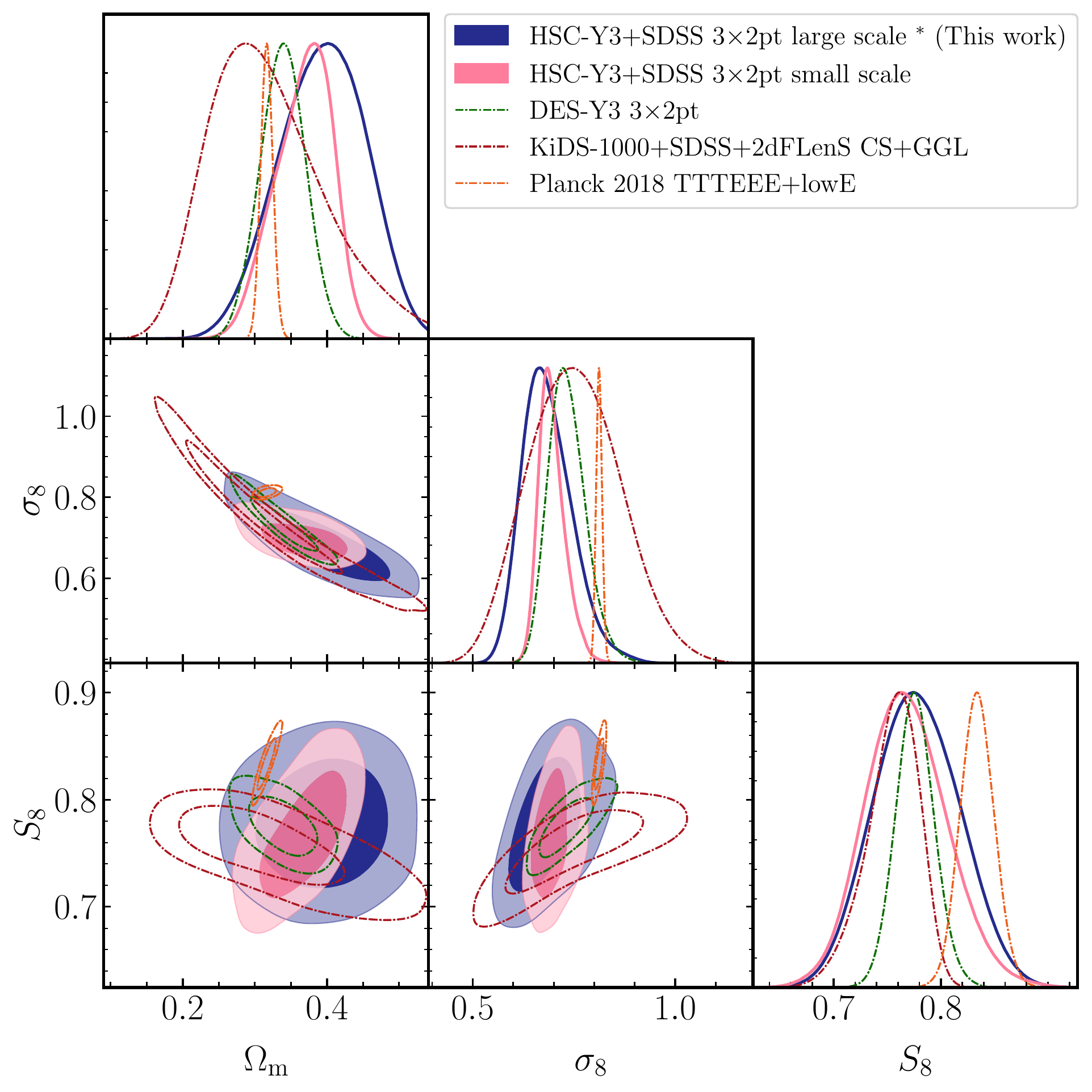}
    \caption{The cosmological constraint from the HSC Y3 data with the large-scale 3$\times$2pt analysis carried out in this work, along with the HSC Y3 small-scale 3$\times$2pt analysis in \citet{miyatake2023} and  external experiments: 
    {\it Planck} 2018 \cite{2020A&A...641A...6P}, 
    DES Y3 3$\times$2pt \citep{DES-Y3}, 
    and KiDS-1000  \citep{Heymansetal:2021}. 
    Here the marginal posterior distributions in one- or two-dimensional parameter space are shown for the main cosmological parameters constrained in this work,  $\Omega_{\rm m}$, $\sigma_8$, and $S_8\equiv\sigma_8(\Omega_{\rm m}/0.3)^{0.5}$.}
    \label{fig:hscy3main}
\end{figure*}

Fig.~\ref{fig:hscy3main} shows the cosmological constraints from the large-scale
3$\times$2pt 
analysis carried out in this paper. The central values and associated errors are
\begin{align}
\begin{aligned}
\Omega_{\rm m}  &= 0.401^{+0.056}_{-0.064} (0.394)\\
\sigma_8        &= 0.666^{+0.069}_{-0.051} (0.705)\\
S_8             &= 0.775^{+0.043}_{-0.038} (0.808)
\end{aligned}
\label{eq:3x2pt-cosmology-constraint}
\end{align}
The HSC-Y3 large-scale 3$\times$2pt analysis achieves $\sim5$\% fractional accuracy on 
$S_8$. The improvement in the statistical precision of $S_8$ 
compared to the HSC-Y1 2$\times$2pt analysis is a factor of $\sim2$, due to the 
increase in the HSC survey area and 
the 
inclusion of the 
cosmic shear data.

\begin{figure}
    \includegraphics[width=\columnwidth]{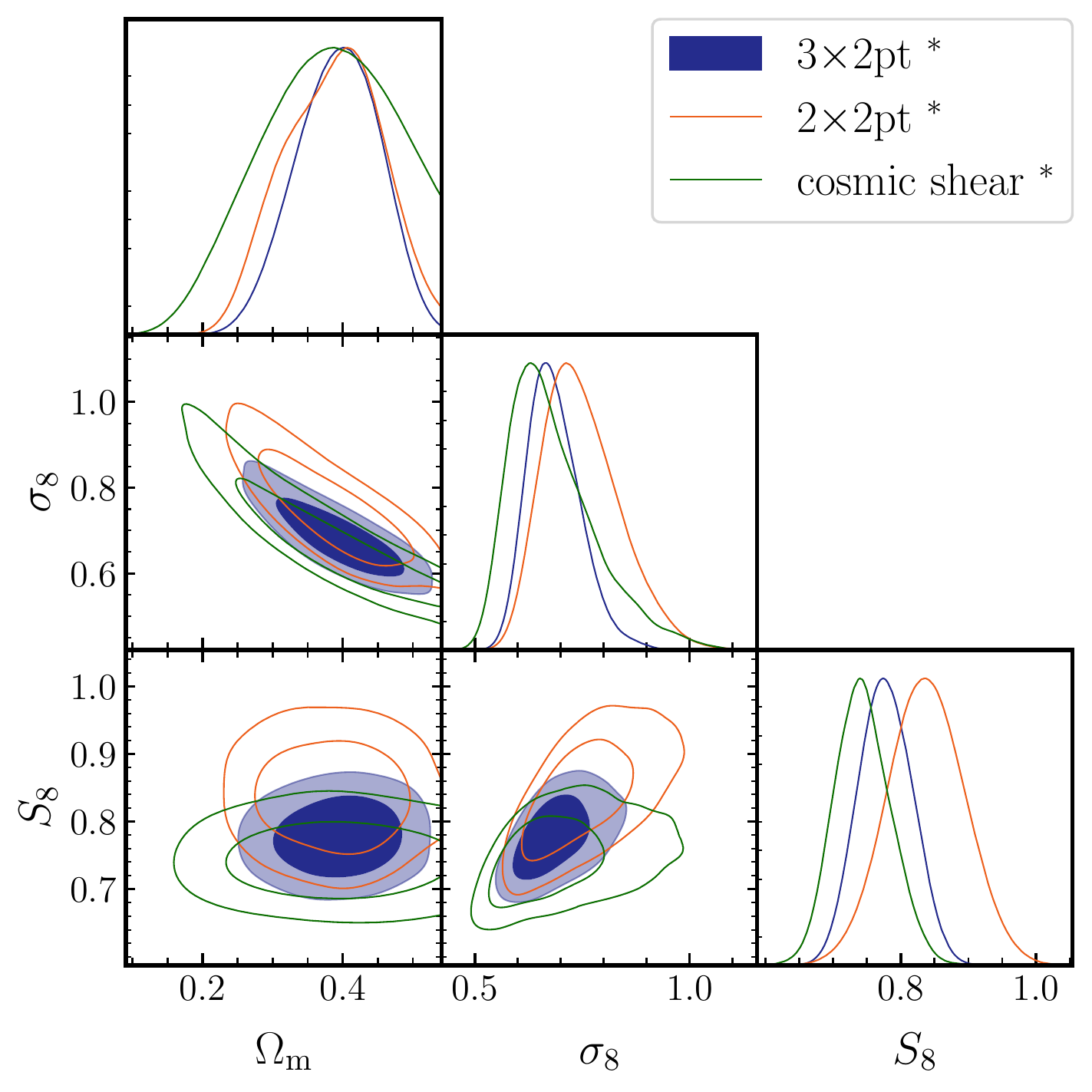}
    \caption{The cosmological parameter constraints for the baseline 3$\times$2pt, 2$\times$2pt, and cosmic shear analyses. Here, every analysis uses the informative prior on $\Delta z_{\rm ph}$
    from the HSC-Y3 small-scale analysis by \citet{miyatake2023}.}
    \label{fig:corner-3x2pt-2x2pt-cs}
\end{figure}

Fig.~\ref{fig:corner-3x2pt-2x2pt-cs} shows how the cosmological parameter constraints are improved by combining the 2$\times$2pt and cosmic shear analysis in 3$\times$2pt.
It is clear that the 2$\times$2pt analysis and cosmic shear are complementary to each other for constraining the cosmological parameters; combining them improves the $S_8$ constraint. 
Note that the cosmic shear analysis alone cannot constrain
$\Omega_{\rm m}$, 
because the cosmic shear in this paper does not include tomographic 
information and therefore is not sensitive to
the growth history of matter clustering, which depends on $\Omega_{\rm m}$.

\begin{figure}
    \includegraphics[width=\columnwidth]{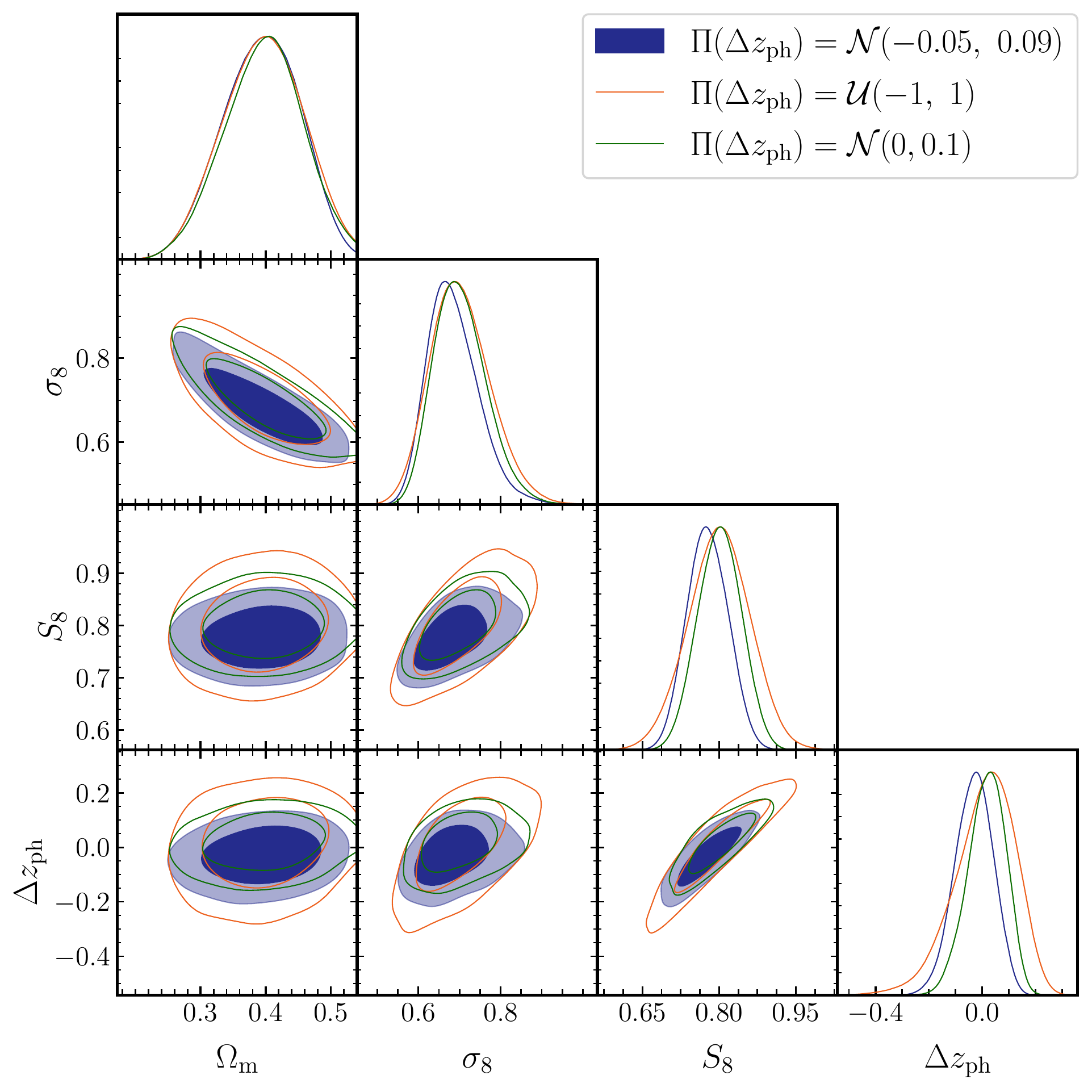}
    \caption{Cosmological parameter constraints for HSC-Y3 3$\times$2pt analyses with three different $\Delta z_{\rm ph}$ prior choices: the informative prior taken from the fiducial small-scale analysis result, the uninformative prior ${\cal U}(-1, 1)$, and the informative prior $\Pi(\Delta z_{\rm ph})={\cal N}(0, 0.1)$. 
    }
    \label{fig:hscy3main-dzp-self-calibrated}
\end{figure}

Fig.~\ref{fig:hscy3main-dzp-self-calibrated} shows the result of the 3$\times$2pt analyses for three cosmological parameters and the  $\Delta z_{\rm ph}$ parameter, using three different $\Delta z_{\rm ph}$ priors: the informative prior taken from the baseline small-scale analysis result, the uninformative prior, $\Pi(\Delta z_{\rm ph})={\cal U}(-1, 1)$, and the informative Gaussian prior 
with little room for a large shift from zero, 
$\Pi(\Delta z_{\rm ph})={\cal N}(0, 0.1)$. 
In the baseline analysis of this paper, we use the first prior.
Comparing the results with ${\cal N}(-0.05,0.09)$ and 
${\cal N}(0,0.1)$ tells us the impact of the central value of the Gaussian prior on the cosmological parameters. 
When the uninformative prior is used, 
the large-scale 3$\times$2pt analysis has insufficient information to constrain  $\Delta z_{\rm ph}$, although the posterior of $\Delta z_{\rm ph}$ prefers 
a negative value of $\Delta z_{\rm p}$ as found in the small-scale 
3$\times$2pt analysis. 
As a result, the $S_8$ constraint   is significantly degraded.
This was the main reason that 
we decided to use the informative prior obtained from the baseline small-scale 3$\times$2pt analysis in \citet{miyatake2023}
as shown in the blue contour. 
This prior choice was made during the blind analysis phase, and therefore the baseline result is free of confirmation bias.

From Fig.~\ref{fig:hscy3main-dzp-self-calibrated}, we find that adopting the informative Gaussian prior around 
$\Delta z_{\rm ph}=0$ results in a larger $S_8$ value.
The shift in 
$S_8$ between 
the baseline analysis and the Gaussian prior of $\Delta z_{\rm ph}$ in this case
corresponds to a shift of $0.7\sigma$, a non-negligible amount. 
In Appendix~\ref{sec:apdx-model-validation}, we validated the use of the posterior obtained from the small-scale 
3$\times$2pt analysis
as the prior on $\Delta z_{\rm ph}$
 using the mock analysis. 
We found that the posterior of $\Delta z_{\rm ph}$
from the small-scale analysis can be safely used as the prior in the large-scale 
3$\times 2$pt analysis, i.e., the analysis in this paper,  
to recover the input 
$S_8$ value. 
Thus, we decided {\it prior to unblinding} to adopt the informative prior of ${\cal N}(-0.05,0.09)$ for our baseline analysis.

\begin{figure*}
    \includegraphics[width=2\columnwidth]{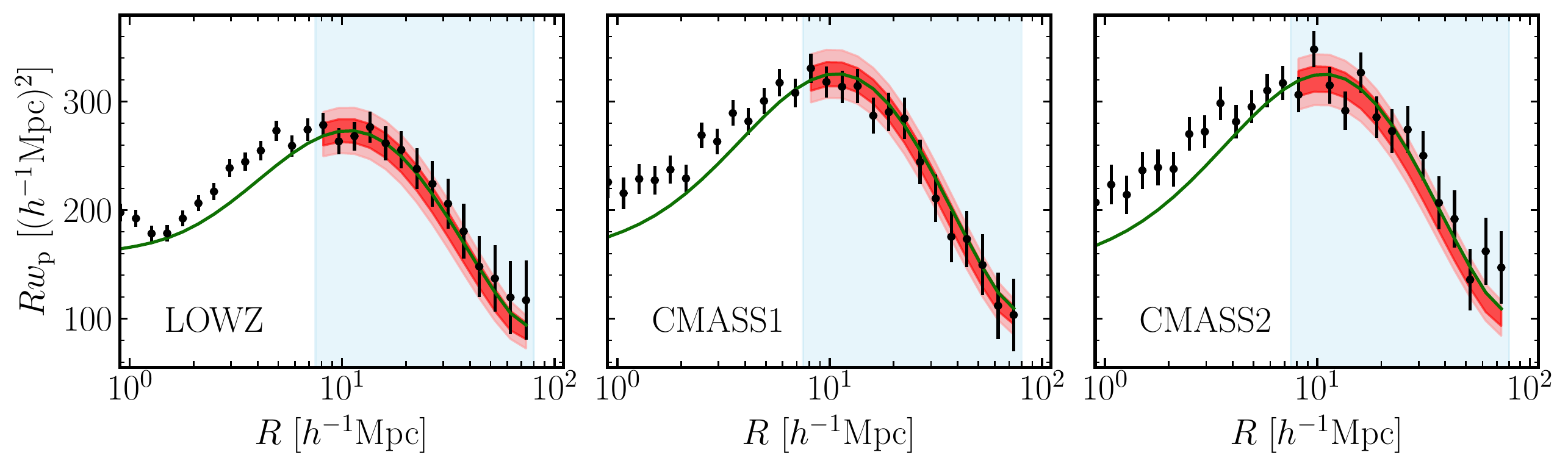}
    \includegraphics[width=2\columnwidth]{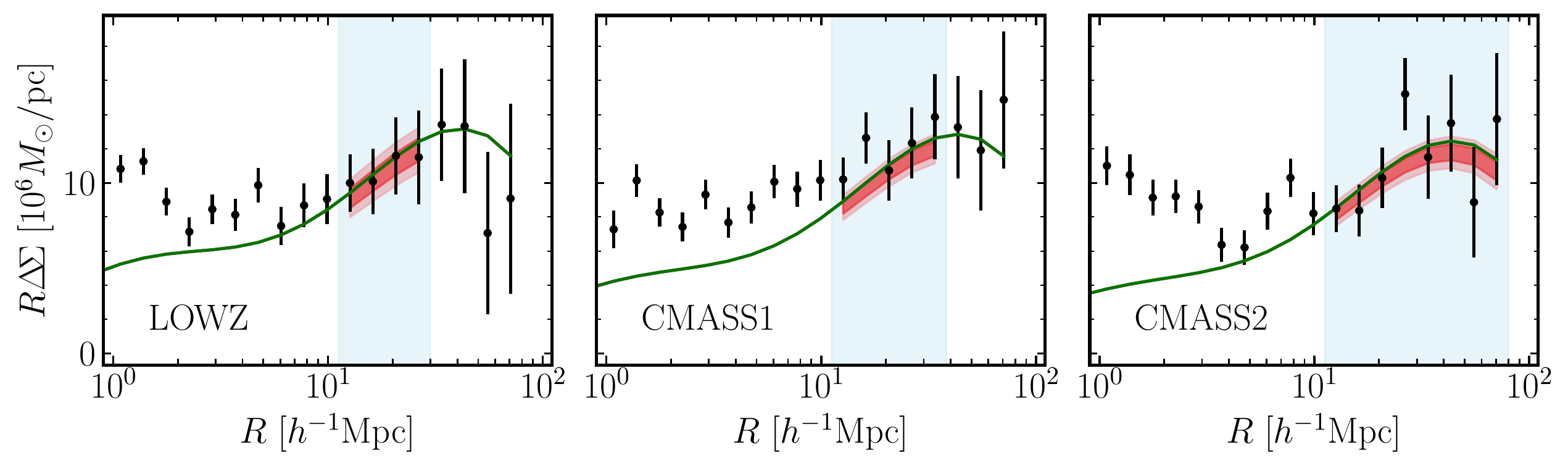}
    \includegraphics[width=1.4\columnwidth]{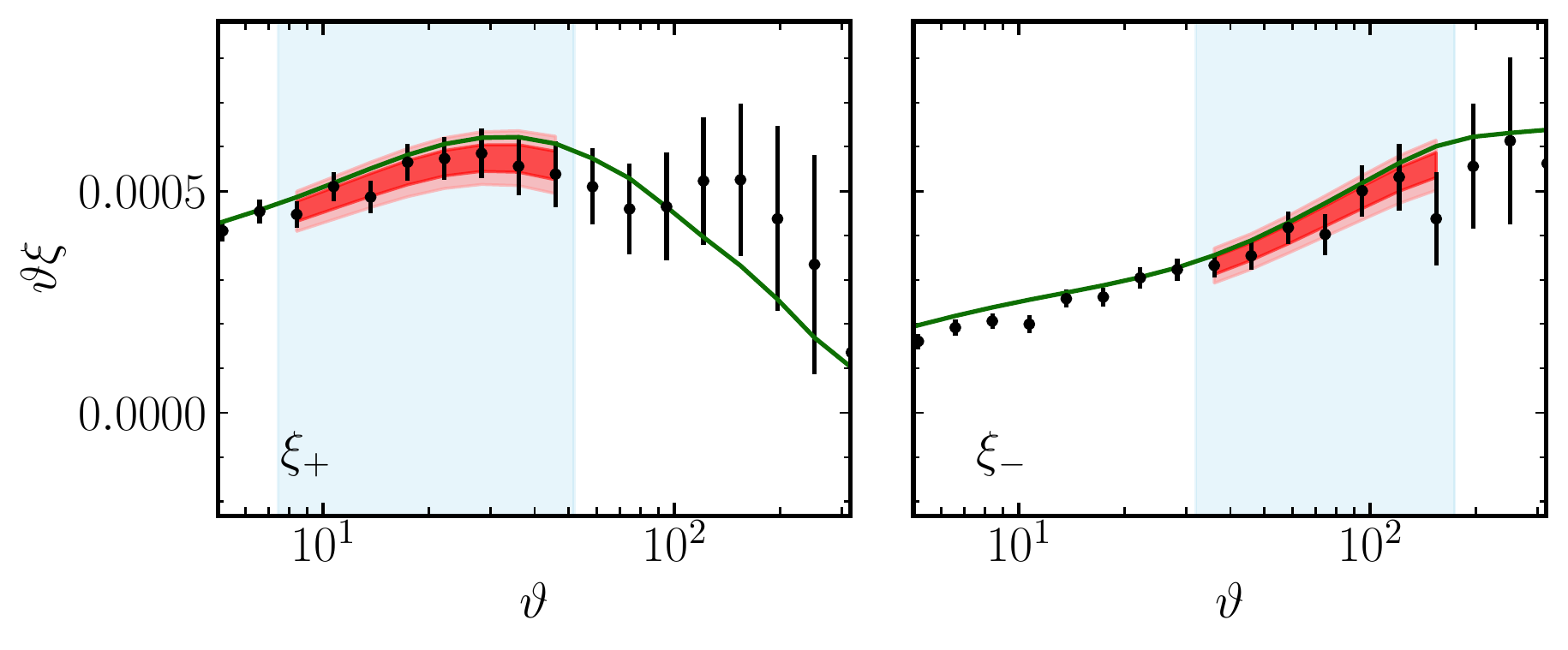}
    \caption{Comparison between the measured signals and
     the best-fit model predictions for the baseline large-scale 3$\times$2pt analysis.
    From the top to the bottom panels, we show the comparison for 
    $\wproj(R)$ and $\dSigma(R)$ for the three SDSS samples (LOWZ, CMASS1 and CMASS2), and the cosmic shear correlation functions, $\xi_{\pm}(\vartheta)$, respectively. 
In each panel, the black points with error bars denote the measured signals in each $R$ or $\vartheta$ bin, where the error bar is computed from the diagonal component of the covariance matrix. The solid line denotes 
    the model prediction at the MAP ({\it maximum a posteriori} model of the chain), while
    the red-shaded regions show the 68\% and 95\% credible intervals of the
    model predictions in each bin.
    The blue shaded region in each panel indicates the range of $R$ or $\vartheta$ that is used for the cosmological parameter inference in this paper. 
    See Section~\ref{sec:analysis-method} for details of the model predictions.
    Note that the $x$-axes in the top and middle panels are the comoving distance in the reference cosmology (see the main text), $R^{\rm ref}$, but we omit the superscript ``${}^{\rm ref}$'' for simplicity.
    }
    \label{fig:signal-fitting}
\end{figure*}

Fig.~\ref{fig:signal-fitting} compares the model predictions at the MAP ({\it maximum a posteriori}) model of the chain
with the measured signals.
The best-fit model fairly well reproduces the measured signals over the range of separations used 
for the cosmological analysis. The figure also shows that the best-fit model fails to reproduce
the measured $\wproj$ and $\dSigma$ on scales below the fitting range. This is expected, as the simple minimum bias model is invalid on such small scales.

\begin{figure}
    \includegraphics[width=\columnwidth]{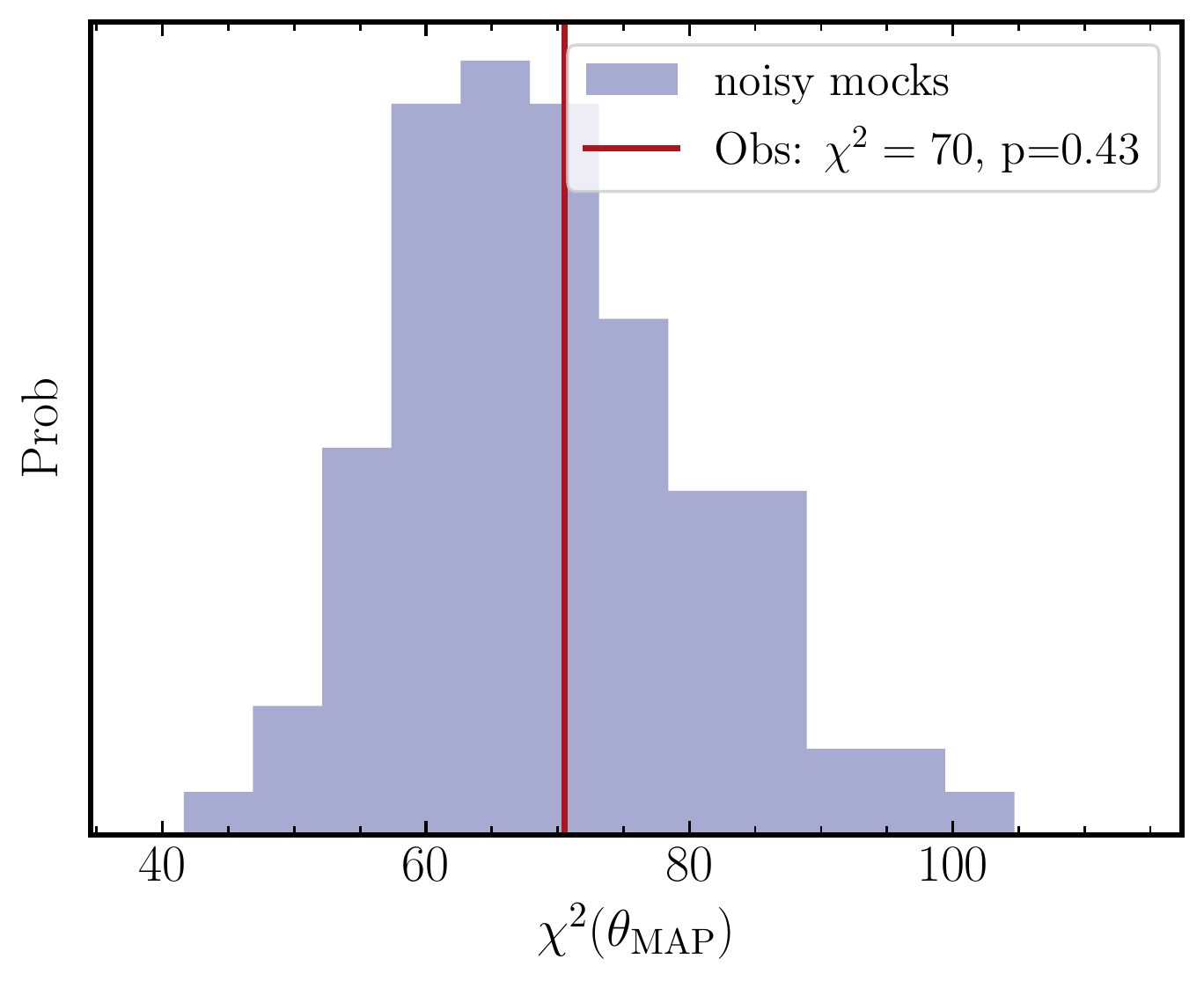}
    \caption{The evaluation of the goodness-of-fit with the $\chi^2(\theta_{\rm MAP})$ value at the {\it maximum a posteriori} (MAP) model. 
    The reference distribution (blue histogram) is obtained by analyzing 100 noisy mock data vectors. See the main text for how the noisy mock data vectors are generated. The vertical solid line denotes the observed 
    $\chi^2$ value for the cosmology analysis of the actual HSC-Y3 and SDSS data. The probability of finding the $\chi^2$ value larger than the observed value ($p$ value) is about 43\%. 
    }
    \label{fig:goodness-of-fit}
\end{figure}

Fig.~\ref{fig:goodness-of-fit} shows the goodness-of-fit test of the 3$\times$2pt analysis. To quantitatively evaluate the goodness-of-fit, we follow the same method as for the S16A analysis \citep{Sugiyama:2021}: we simulate 100 noisy mock data vectors, apply the same analysis to each mock data vector as for 
the real data, and obtain the distribution of the $\chi^2$ values at the MAP model for each mock. 
Note that here we generate the noisy data vector using 
the covariance matrix with a multiplicative shear bias parameter of
$m=0$ as described 
in the Appendix \ref{sec:model-validation}. 
We find that the probability to exceed the observed value, 
$\chi^2\simeq 70$, by chance
is $p=0.43$.
We have also checked that the dependence of the reference $\chi^2$ distribution on the 
assumed multiplicative shear bias value used when generating the noisy mock data  
is weak and does not change our conclusion. 
We therefore conclude that within the statistical constraining power of our data, the model is able to describe the data with no signs of model mis-specification.

\subsection{Internal consistency}
\label{subsec:internal-consistency}

\begin{figure*}
    \includegraphics[width=2\columnwidth]{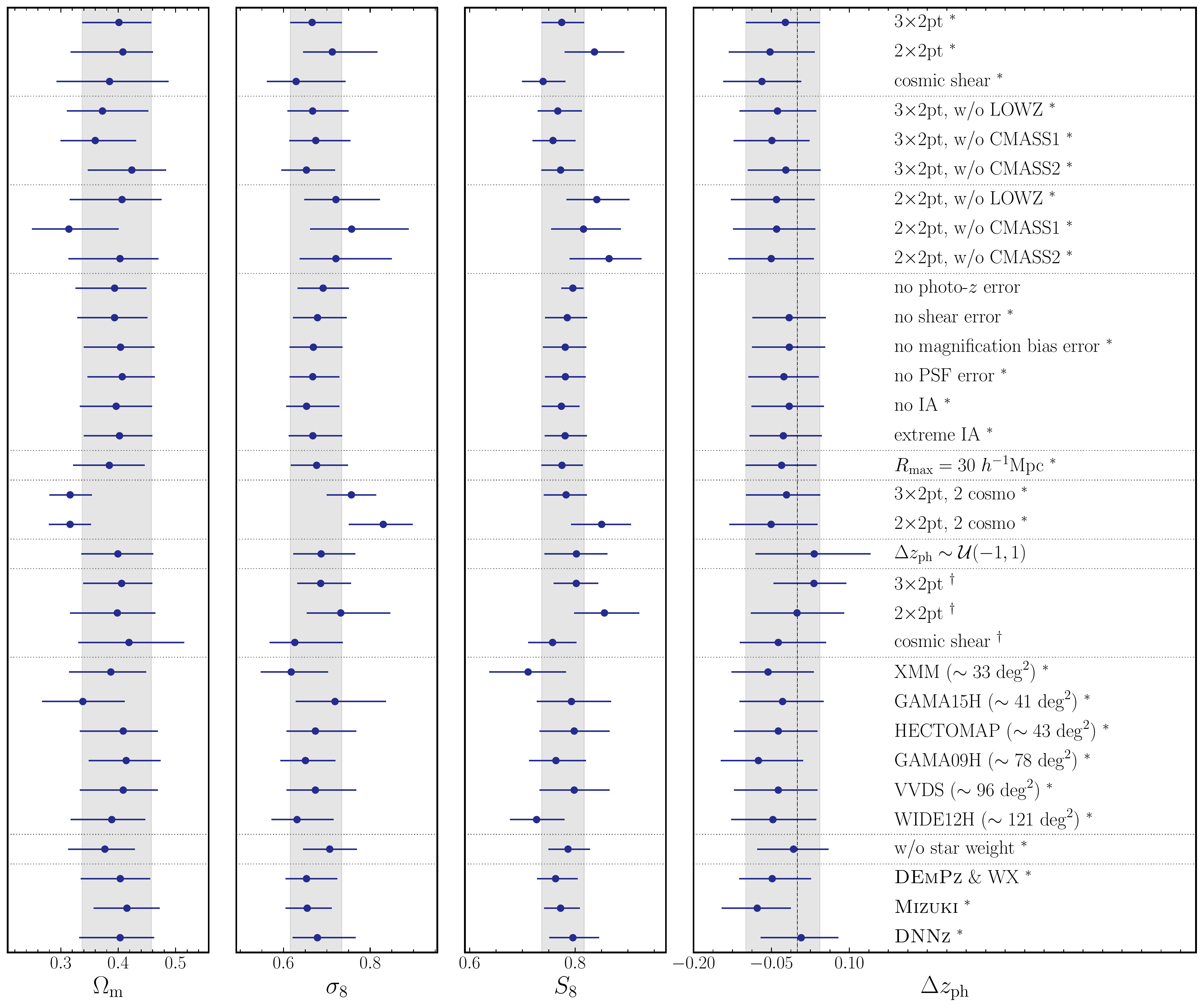}
    \caption{Summary of internal consistency tests. The estimates of cosmological parameters, $\Omega_{\rm m}$, $\sigma_8$, and $S_8$, and the photo-$z$ parameter, $\Delta z_{\rm ph}$, are summarized for each of the analysis setups 
    in Table~\ref{tab:internal-tests}.
    Here, the central point is the mode and the error bar is the 68\% highest density interval estimated from the 
    one-dimensional posterior distribution of each parameter. For comparison, the shaded band is the constraint from the baseline analysis. The vertical black dotted line in the $\Delta z_{\rm ph}$ panel denotes $\Delta z_{\rm ph}=0$.
    }
    \label{fig:summary-internal-tests}
\end{figure*}

In this section, we present the results of internal consistency tests of our analysis.
The analysis setups for 
the internal consistency tests are summarized in Table~\ref{tab:internal-tests}. Fig.~\ref{fig:summary-internal-tests} summarizes the result of the internal consistency tests, and Table~\ref{tab:cosmo-constraint} in Appendix~\ref{sec:apdx-internal-consistency-test} summarizes the central values and credible intervals for each parameter. 
In short,
we did not find any significant shift in each of the cosmological parameters compared to the expected statistical scatter. 

\begin{figure*}
    \includegraphics[width=\columnwidth]{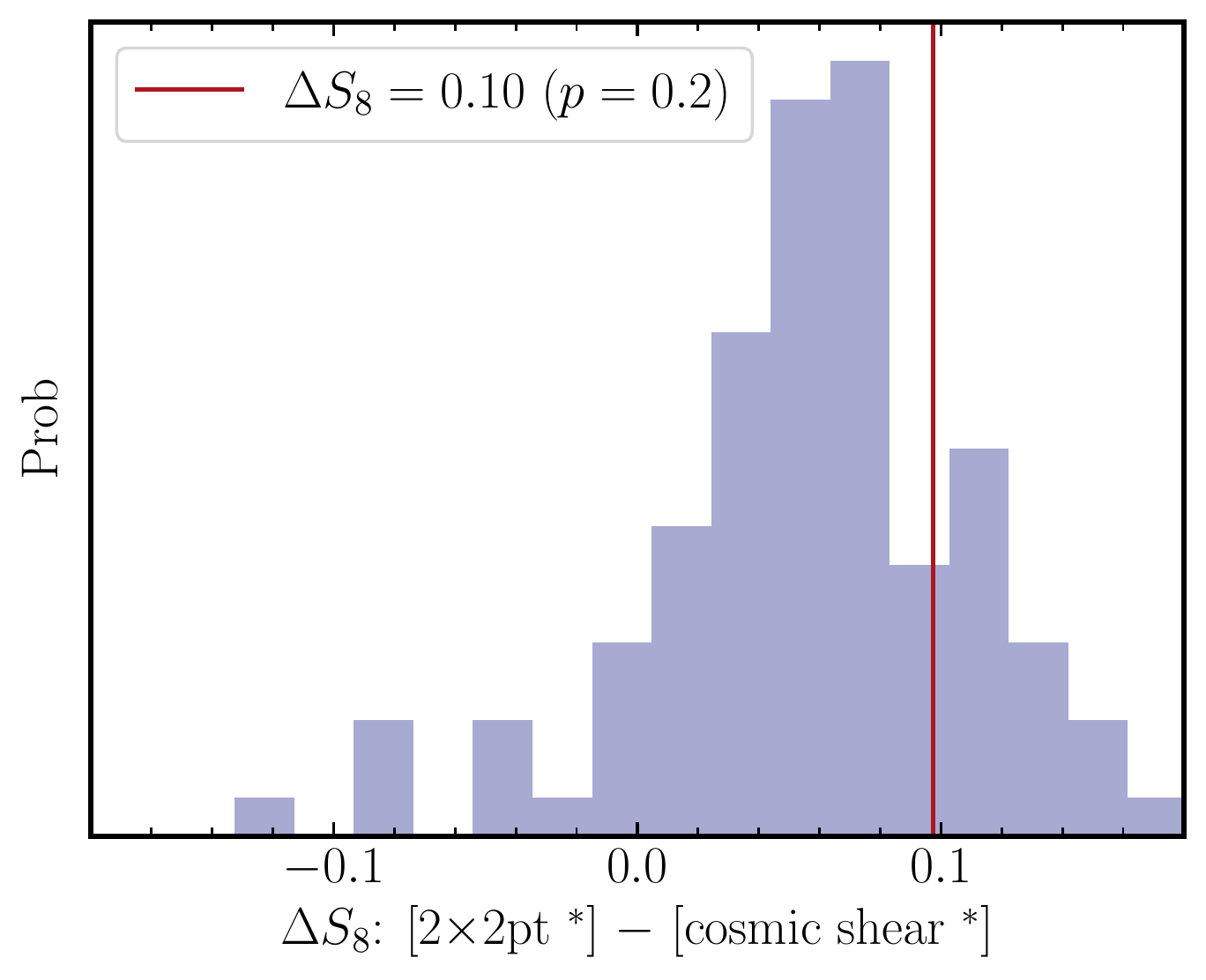}
    \includegraphics[width=\columnwidth]{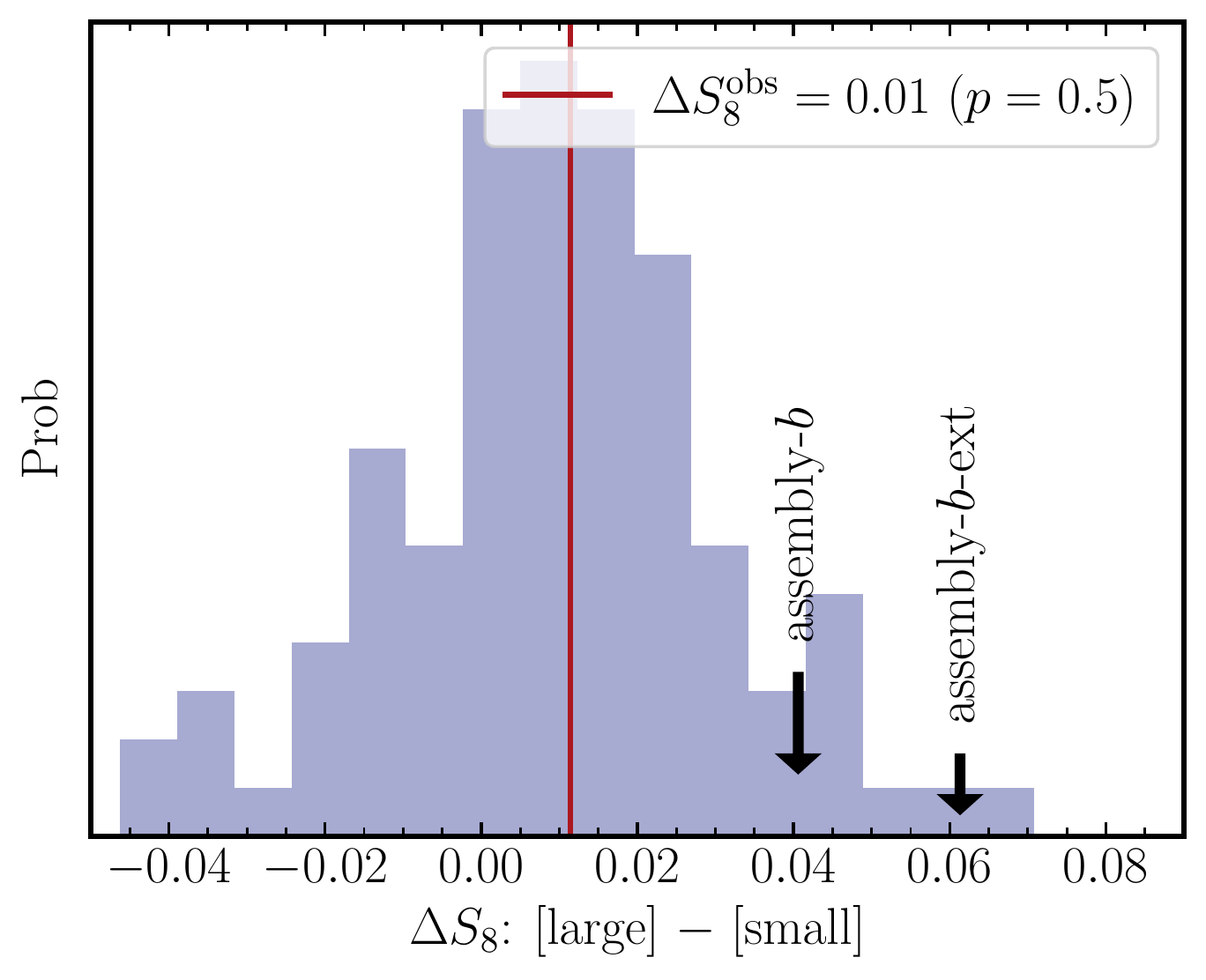}
    \caption{A statistical significance 
    of the difference between the $S_8$ values from the two analyses, where 
    we define $\Delta S_8$ by the difference between the $S_8$ modes values in the 1D posteriors of the two analyses. {\it Left panel}: The result for $\Delta S_8$ between 
    the large-scale 2$\times$2pt and  cosmic shear analyses. The blue histogram denotes the distribution of $\Delta S_8$ that is obtained by carrying out these analyses on each of 100 realizations of the noisy mock data vector 
    (see text for details). The red solid line is the observed $\Delta S_8=0.10$ in the real data. 
    The 
    $p$-value
    is $p=0.2$. {\it Right panel}: The result for $\Delta S_8$
    between the large-scale 3$\times$2 analysis in this paper and the small-scale 3$\times$2pt analysis in the companion paper by \citet{miyatake2023}. The observed difference is $\Delta S_8=0.01$ and the probability to exceed this value by chance is $p=0.5$. 
    The two arrows indicated by ``assembly-$b$'' and 
    ``assembly-$b$-ext'' denote the expected difference values of $S_8$ obtained from the simulated synthetic data, where the assembly bias effects with different amplitudes are included.}
    \label{fig:S8-diff}
\end{figure*}

The largest variation is 
the difference between the $S_8$ values obtained from the 
2$\times$2pt analysis versus from the cosmic shear analysis alone. First,  the trends in $S_8$ -- i.e., the larger and smaller $S_8$ values for the two analyses
than the $S_8$ value from the 3$\times$2pt analysis -- are also found in the model validation tests as shown 
in Fig.~\ref{fig:summary-mock-validation}.
To make a  quantitative estimate of
the statistical significance of the $S_8$ difference, $\Delta S_8$, 
we use the same 100 noisy mock realizations of the vector as those 
used in the goodness-of-fit analysis. We run the 2$\times$2pt and cosmic shear analyses for each noisy mock realization and then assess how often the measured difference in $S_8$ from the real data occurs in the distribution 
of the $S_8$ difference measured in the 100 noisy mock realizations.
The left panel of Fig.~\ref{fig:S8-diff} shows that
the probability to exceed the observed value 
$\Delta S_8=0.10$ by chance is $p=0.2$.

As another internal consistency test, we compare our result with that from the HSC-Y3 small-scale 3$\times$2pt analysis in \citet{miyatake2023} as shown in Fig.~\ref{fig:hscy3main}.  The two 3$\times$2pt analyses use the same observables albeit on different ranges of scales, but use different theoretical model templates for the cosmology analysis: the minimal bias model in this paper and the emulator-based halo model in \citet{miyatake2023}.
The figure shows that our result is in good agreement with that in \citet{miyatake2023}.
In order to assess the consistency of these two analyses, we again used 
the 100 mocks, which allow us to
account for the correlations between the cosmological parameters from the large- and 
small-scale 3$\times$2pt analyses. 
As shown in the right panel of Fig.~\ref{fig:S8-diff}, 
the probability to observe an $S_8$ difference larger than what we obtain $\Delta S_8\simeq 0.01$, is $p=0.5$.
Thus we 
conclude that the large- and small-scale 3$\times$2pt
analyses 
are consistent with each other.
The two vertical arrows in the right panel of Fig.~\ref{fig:S8-diff} indicate the expected size of $\Delta S_8$ in the presence of assembly bias effects that are simulated using the mock catalogs in \citet{2021arXiv210100113M}. The {\tt assembly-$b$-ext} is the worst-case scenario of the assembly bias effect and can be flagged at a $2\sigma$ level if such effect really exists (see Section~\ref{sec:apdx-model-validation} for the detail of the mocks).
The agreement between the results from the large-scale and small-scale analyses also indicates that the SDSS galaxies in our sample are not largely affected by the possible 
assemnly bias effect \cite{2021arXiv210100113M,Miyatake:2021sdd}.

\subsection{Comparison with external data and \texorpdfstring{$S_8$}{S8} tension}

In 
Fig.~\ref{fig:hscy3main} we also compare our result with
external experiments. 
For the CMB, we consider the {\it Planck} 2018 \citep{2020A&A...641A...6P} cosmological constraints -- in particular, those derived from primary CMB information, referred to as ``TT, EE, TE+lowE'' in their paper
\footnote{We use the Planck 2018 public chain of ``base/plikHM\_TTTEEE\_lowl\_lowE/base\_plikHM\_TTTEEE\_lowl\_lowE'' downloaded from their wiki \url{https://pla.esac.esa.int/pla/aio/product-action?COSMOLOGY.FILE_ID=COM_CosmoParams_fullGrid_R3.01.zip}}. 
For the lensing experiments, we use the cosmological constraints from DES-Y3 \citep{DES-Y3} and KiDS-1000 \citep{Heymansetal:2021}. 
In particular, we use the cosmological constraint from a 3$\times$2pt analysis with the MagLim sample from DES-Y3 data\footnote{We use the DES-Y3 public chain of ``chain\_3x2pt\_lcdm\_SR\_maglim.txt'' downloaded from DES Data Management: \url{https://des.ncsa.illinois.edu/releases/y3a2/Y3key-products}}. 
The fiducial KiDS-1000 3$\times$2pt analysis included
the angular diameter distance from the measured BAO scale in addition to the clustering information, which well constrains $\Omega_{\rm m}$ for the flat $\Lambda$CDM model.
Hence, we instead compare with the result from the cosmic shear and galaxy-galaxy lensing (CS+GGL) analysis\footnote{We use the KiDS-1000 public chain of ``samples\_multinest\_blindC\_EE\_nE\_w.txt'' downloaded from their website
\url{https://kids.strw.leidenuniv.nl/DR4/KiDS-1000_3x2pt_Cosmology.php}}.

Our result is generally in good agreement with both the DES-Y3 and KiDS-1000 results.
For our result, the degeneracy direction in cosmological parameter sub-spaces such as the $\Omega_{\rm m}$-$S_8$ plane
is slightly different from those of the DES-Y3 and KiDS-1000 results.

When comparing our result to the {\it Planck} 2018 result, we did not find any significant tension in 
the cosmological parameters. More quantitatively, 
we compare the cosmological parameter constraints from this paper and the {\it Planck} 2018 
    using the eigen tension metric \citep{Park:2020}. We first identify the eigenmodes of the cosmological parameters by diagonalizing the posterior covariance. We found that the first two eigenmodes, $e_1\equiv\sigma_8(\Omega_{\rm m})^{0.52}$ and $e_2\equiv\Omega_{\rm m}(\sigma_8)^{-0.52}$ are well constrained compared to the prior distribution.
For this reason we 
use these eigenmodes for tension assessment. As an independent criterion of principle eigenmodes, we also computed the effective number of  cosmological parameters constrained by the large-scale analysis using the Gaussian linear model \citep{2019PhRvD..99d3506R}. Focusing only on the cosmological parameters, we find that the effective number of constrained cosmological model parameters is 1.99, which supports the choice to use the first two eigenmodes for tension assessment. For these eigenmodes, we estimate the parameter difference distribution, $P(\Delta\bm{e})$, from the MCMC of HSC-Y3 and Planck 2018, where $\Delta\bm{e}\equiv\bm{e}_{\rm HSC-Y3}-\bm{e}_{\rm Planck}$. We then compute the $p$-value of the null hypothesis, i.e., the case that the {\it Planck 2018} and HSC-Y3 results are in perfect agreement with each other:
\begin{align}
    p = \int_{P(\Delta\bm{e})<P(\bm{0})}{\rm d}(\Delta\bm{e})~P(\Delta \bm{e}).
\end{align}
We find $p=0.846$, corresponding to at most a
$1.4\sigma$-level difference between our HSC-Y3 result and the {\it Planck} result. 
Therefore we conclude that our result is consistent with the {\it Planck} CMB results.
Although at face value this seems different from the $2.5\sigma$ tension
between {\it Planck} and the small-scale HSC-Y3 result presented in \citet{miyatake2023},
this difference is entirely due to the lesser statistical constraining power of our result, as reflected in the larger credible intervals.

\section{Conclusion}
\label{sec:conclusion}

In this paper, we have presented cosmological constraints from a joint analysis of 
galaxy clustering ($\wproj$), galaxy-galaxy lensing ($\dSigma$), and the cosmic shear correlation ($\xi_\pm$),
measured from the HSC-Y3 shape catalog and the 
SDSS DR11 spectroscopic galaxy catalog.
We have adopted a conservative analysis strategy: we employed the ``minimal bias model'' 
as a theoretical template to model $\dSigma$ and $\wproj$, and strict scale cuts to 
ensure its validity.
Using mock data vectors, 
we showed that the minimal bias model can recover the input 
cosmological parameter to within the statistical error for the HSC-Y3 data, as long as the analysis 
is restricted to large scales, $R>8$ and $12~h^{-1}{\rm Mpc}$ for $\wproj$ and $\dSigma$, respectively. This is because 
structure formation on such large scales is governed by gravity alone whereas the nonlinear galaxy bias
and baryonic effects are confined to smaller  scales. 
In addition, we employed a conservative prior on the nuisance parameter, $\Delta z_{\rm ph}$, to model a residual systematic effect in the mean redshift of HSC source galaxies used in the weak lensing measurements.
We adopted a Gaussian prior of $\Delta z_{\rm ph}$ given by  ${\cal N}(-0.09,0.05)$, based on 
a similar 3$\times$2pt analysis
in the companion paper \citet{miyatake2023} using 
the same data vector down to smaller scales.
Another key feature of this analysis is that we do not include the tomographic information in weak lensing signals, but rather adopt a 
a single conservatively-selected source sample to make our results robust against the residual photo-$z$ error 
following the method in \citet{OguriTakada:11}.

Our cosmological parameter constraint for the flat $\Lambda$CDM model is $S_8=0.775^{+0.043}_{-0.038}$, with $\sim5$\% precision. From the comparison with {\it Planck} 2018 \citep{2020A&A...641A...6P}, we found that our result is consistent with {\it Planck}, indicating no significant tension. 
We found that the cosmic shear correlation function not only improves the cosmological parameter estimation, but also helps to calibrate the residual photo-$z$ error $\Delta z_{\rm ph}$.
If we employ a prior on $\Delta z_{\rm ph}$ centered at no bias value, 
${\cal N}(0,0.1)$, the $S_8$ value is shifted to a higher value by $\sim 1\sigma$. Hence we concluded that a treatment of $\Delta z_{\rm ph}$ is important for our cosmological analysis. 
We emphasize that, using various validation tests, we defined 
the analysis setups and methods during the blinding analysis stage. 
Using the 100 noisy mock analyses, we confirmed that the main result of the large-scale analysis in this paper is statistically consistent with the small-scale analysis in \citet{miyatake2023}.

The constraining power of the large-scale 3$\times$2pt analysis in this paper is weaker than the {\it Planck} 2018 result. Increasing the statistical constraining power 
is important for a stringent test of the $S_8$ tension or, more generally, 
the $\Lambda$CDM model. There are several ways to improve the statistical constraints from the large-scale analysis. 
First, in future, we will use the full data of the HSC survey covering about 
1,100~\sqdeg\ 
of sky to carry out a similar 3$\times$2pt analysis. 
Second, we can 
include tomographic cosmic shear tomography information in the 3$\times$2pt analysis. As shown in \citet{li2023} and \citet{dalal2023}, the cosmic shear tomography can self-calibrate the residual photo-$z$ error; therefore, we expect that adding cosmic shear tomography to the large-scale 
2$\times$2pt signals can improve the cosmological constraints as well as the residual photo-$z$ error calibration. 
Third, 
to improve the cosmological constraint, we could  push the scale cuts down to smaller scales. In order to use the smaller scale signals of $\wproj$ or $\dSigma$, we would need to  account for the nonlinear physics in the $R\lesssim10h^{-1}{\rm Mpc}$ regime, which would require more complicated modeling of galaxy bias than the minimal bias model. We will leave these improvements to our future studies.

\begin{acknowledgments}

This work was supported in part by World Premier International Research Center Initiative (WPI Initiative), MEXT, Japan, and JSPS KAKENHI Grant Numbers JP18H04350, JP18H04358, JP19H00677, JP19K14767, JP20H00181, JP20H01932, JP20H04723, JP20H05850, JP20H05855, JP20H05856, JP20H05861, JP21J00011, JP21H05456, JP21J10314, JP21H01081, JP21H05456,  JP22K03634, JP22K03655, JP22K21349, JP23H00108 and JP23H04005 by Japan Science and Technology Agency (JST) CREST JPMHCR1414, by JST AIP Acceleration Research Grant Number JP20317829, Japan, and by Basic Research Grant (Super AI) of Institute for AI and Beyond of the University of Tokyo. SS was supported in part by International Graduate Program for Excellence in Earth-Space Science (IGPEES), WINGS Program, the University of Tokyo. 
YK is supported in part by the David and Lucile Packard foundation. 
RD acknowledges support from the NSF Graduate Research Fellowship Program under Grant No.\ DGE-2039656.
RM is supported by a grant from the Simons Foundation (Simons Investigator in Astrophysics, Award ID 620789).

The Hyper Suprime-Cam (HSC) collaboration includes the astronomical
communities of Japan and Taiwan, and Princeton University. The HSC
instrumentation and software were developed by the National Astronomical
Observatory of Japan (NAOJ), the Kavli Institute for the Physics and
Mathematics of the Universe (Kavli IPMU), the University of Tokyo, the
High Energy Accelerator Research Organization (KEK), the Academia Sinica
Institute for Astronomy and Astrophysics in Taiwan (ASIAA), and
Princeton University. Funding was contributed by the FIRST program from
Japanese Cabinet Office, the Ministry of Education, Culture, Sports,
Science and Technology (MEXT), the Japan Society for the Promotion of
Science (JSPS), Japan Science and Technology Agency (JST), the Toray
Science Foundation, NAOJ, Kavli IPMU, KEK, ASIAA, and Princeton
University.
This paper makes use of software developed for the Vera C. Rubin Observatory. We thank the Rubin Observatory for making their code available as
free software at \url{http://dm.lsst.org}.

The Pan-STARRS1 Surveys (PS1) have been made possible through
contributions of the Institute for Astronomy, the University of Hawaii,
the Pan-STARRS Project Office, the Max-Planck Society and its
participating institutes, the Max Planck Institute for Astronomy,
Heidelberg and the Max Planck Institute for Extraterrestrial Physics,
Garching, The Johns Hopkins University, Durham University, the
University of Edinburgh, Queen's University Belfast, the
Harvard-Smithsonian Center for Astrophysics, the Las Cumbres Observatory
Global Telescope Network Incorporated, the National Central University
of Taiwan, the Space Telescope Science Institute, the National
Aeronautics and Space Administration under Grant No. NNX08AR22G issued
through the Planetary Science Division of the NASA Science Mission
Directorate, the National Science Foundation under Grant
No. AST-1238877, the University of Maryland, and Eotvos Lorand
University (ELTE) and the Los Alamos National Laboratory.

Based in part on data collected at the Subaru Telescope and retrieved
from the HSC data archive system, which is operated by Subaru Telescope
and Astronomy Data Center at National Astronomical Observatory of Japan.
\end{acknowledgments}

\appendix
\section{Model validation}
\label{sec:apdx-model-validation}

\begin{table*}
\caption{A summary of mock signals used for the validation tests.
Please see \citet{2021arXiv210100113M} for the details of each mock catalogs
which are used to simulate the synthetic data of $\dSigma$ and $\wproj$.
}
\label{tab:mock-validation}
\setlength{\tabcolsep}{20pt}
\begin{center}
\begin{tabular}{ll}
\hline\hline
setup label & description\\
\hline
\multicolumn{2}{l}{\hspace{-1em}{\bf Fiducial mock analyses}}\\
3$\times$2pt                                                         & 3$\times$2pt analysis using clustering, galaxy-galaxy lensing and cosmic shear\\ 
2$\times$2pt                                                         & 2$\times$2pt analysis using clustering and galaxy-galaxy lensing\\
cosmic shear                                                         & cosmic shear analysis\\ 
\hline
\multicolumn{2}{l}{\hspace{-1em}{\bf Effects of galaxy bias uncertainties on $\wproj$ and $\dSigma$}}\\
sat-mod, sat-DM, sat-sub
        & Use different ways to populate satellite galaxies into host halos\\
off centering 1, 2, 3, 4                                             & Mocks include the off-centering effects of central galaxies \\ 
baryon                                                               & Mock includes baryonic feedback effect \\
assembly-$b$-ext
      & Mock includes the extremely 
        large assembly bias effect \\
assembly-$b$ 
                                                & Mock includes the large assembly bias effect \\
cent-incomp                                                               & Mock includes the incompleteness effect of central galaxies \\
fof                                                                  & Mock
uses fof halos to populate galaxies \\ 
\hline
\multicolumn{2}{l}{\hspace{-1em}{\bf Baryonic effect on cosmic shear $\xi_\pm$}}\\
HMCode v2015 (DM only)                                               & Mock cosmic shear is generated by HMCode v2015 with $A_{\rm bary}=3.13$\\ 
HMCode v2015 ($A_{\rm bary}$)                                        & Mock cosmic shear is generated by HMCode v2015 with $A_{\rm bary}\in[2.8, 1.6]$ \\ 
HMCode v2020 ($T_{\rm TGN}$)                                         & Mock cosmic shear is generated by HMCode v2020 with $T_{\rm AGN}\in[7.3, 8.3]$\\ 
\hline
\multicolumn{2}{l}{\hspace{-1em}{\bf Photo-$z$ bias on $\dSigma$ and $\xi_\pm$}}\\
$\Delta z_{\rm ph}^{\rm in}=-0.2$                                         & Mock signals include 
the effect of the worst-case photo-$z$ error: $\Delta z_{\rm ph}^{\rm in}=-0.2$\\
\hline
\multicolumn{2}{l}{\hspace{-1em}{\bf PSF systematics on $\xi_\pm$}}\\
PSF 4th                                                              & 
Mock 
signals include the measured PSF systematic effects up to the fourth moments of PSF\\
\hline\hline
\end{tabular}
\end{center}
\end{table*}

\begin{figure*}
    \includegraphics[width=2\columnwidth]{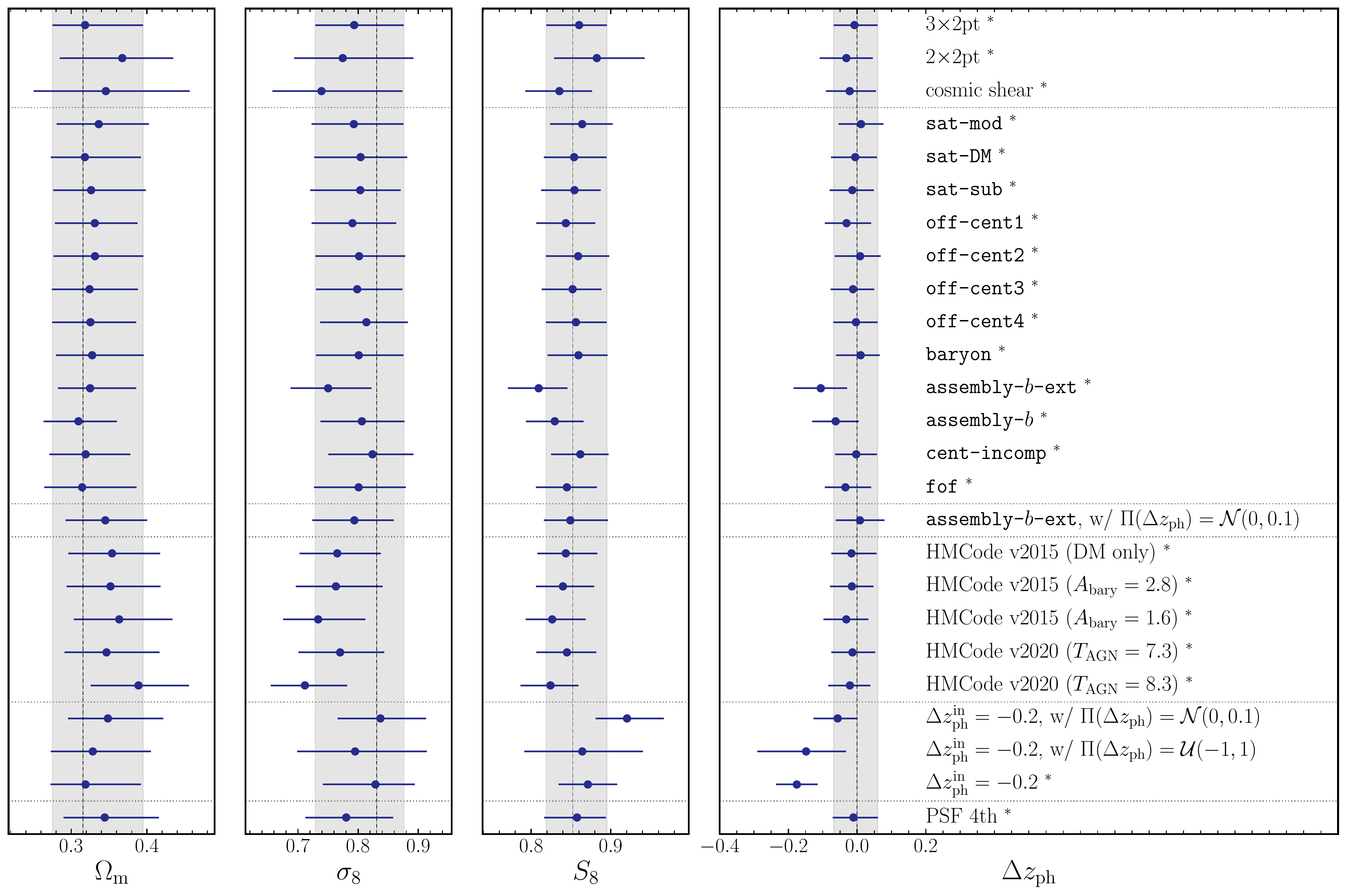}
    \caption{Summary of the validation tests of the model and method. 
    We apply the baseline analysis method to the synthetic data vectors in Table~\ref{tab:mock-validation}
    to obtain 
    constraints on the three cosmological parameters, $\Omega_{\rm m}$, $\sigma_8$ and $S_8$, and the photo-$z$ parameter $\Delta z_{\rm ph}$. 
    Similar types of validation tests are grouped by  
    horizontal dotted lines. 
    The superscript ``${}^\ast$'' denotes the analysis using 
    the 
    informative prior of $\Delta z_{\rm ph}$ that is
    taken from the posterior distribution of the small-scale 3$\times$2pt analysis in \citet{miyatake2023} on the {\it same} synthetic data vector.
    The top section shows the results of the baseline analysis on the fiducial mock, i.e. the data vector uncontaminated by systematic effects. 
    The second section shows the robustness of the minimum bias model against uncertainties in the galaxy bias or the galaxy-halo connection, where we use the different SDSS galaxy mock catalogs to simulate 
    $\dSigma$ and $\wproj$ affected by the different galaxy-halo connections. 
    The row ``assembly-$b$-ext w $\Pi(\Delta z_{\rm ph}={\cal N}(0.1)$'' shows the result obtained 
    using the synthetic data for the ``assembly-$b$-ext'' mock, but using the informative Gaussian 
    prior on $\Delta z_{\rm ph}$ given by $\Pi(\Delta z_{\rm ph})={\cal N}(0,0.1)$.
    The fourth section shows the validation of the model of cosmic shear signal against  
    baryonic effect contamination simulated by different versions of {\tt HMCode}. 
    The fifth section shows the results for the validation tests using the synthetic data that is affected by a systematic error in the mean source redshift by $|\Delta z_{\rm ph}^{\rm in}|=0.2$
    (see text for details). Here $|\Delta z_{\rm ph}^{\rm in}|=0.2$ gives the worst case scenario 
    for the effect of the unknown systematic photo-$z$ error, because the small-scale 3$\times$2pt has the precision of $\sigma(\Delta z_{\rm ph})\sim 0.1$ for the calibration of the photo-$z$ error parameter. 
    The analysis with ``$^{\ast}$'' uses the informative prior of $\Delta z_{\rm ph}$ taken from the
    small-scale 3$\times$2pt analysis on the same synthetic data (that is, this is our baseline analysis method). 
    The analysis with the labels ``$\Pi(\Delta z_{\rm ph}={\cal N}(0,0.1)$'' shows the result using the informative Gaussian prior around $\Delta z_{\rm ph}=0$, 
    and ``$\Pi(\Delta z_{\rm ph}={\cal U}(-1,1)$'' shows the result using the uninformative flat prior.
    The last section is the validation of our PSF systematics modeling in which the synthetic cosmic shear signals include the PSF systematics up to the fourth moment of PSF.
    }
    \label{fig:summary-mock-validation}
\end{figure*}

\begin{figure}
    \includegraphics[width=\columnwidth]{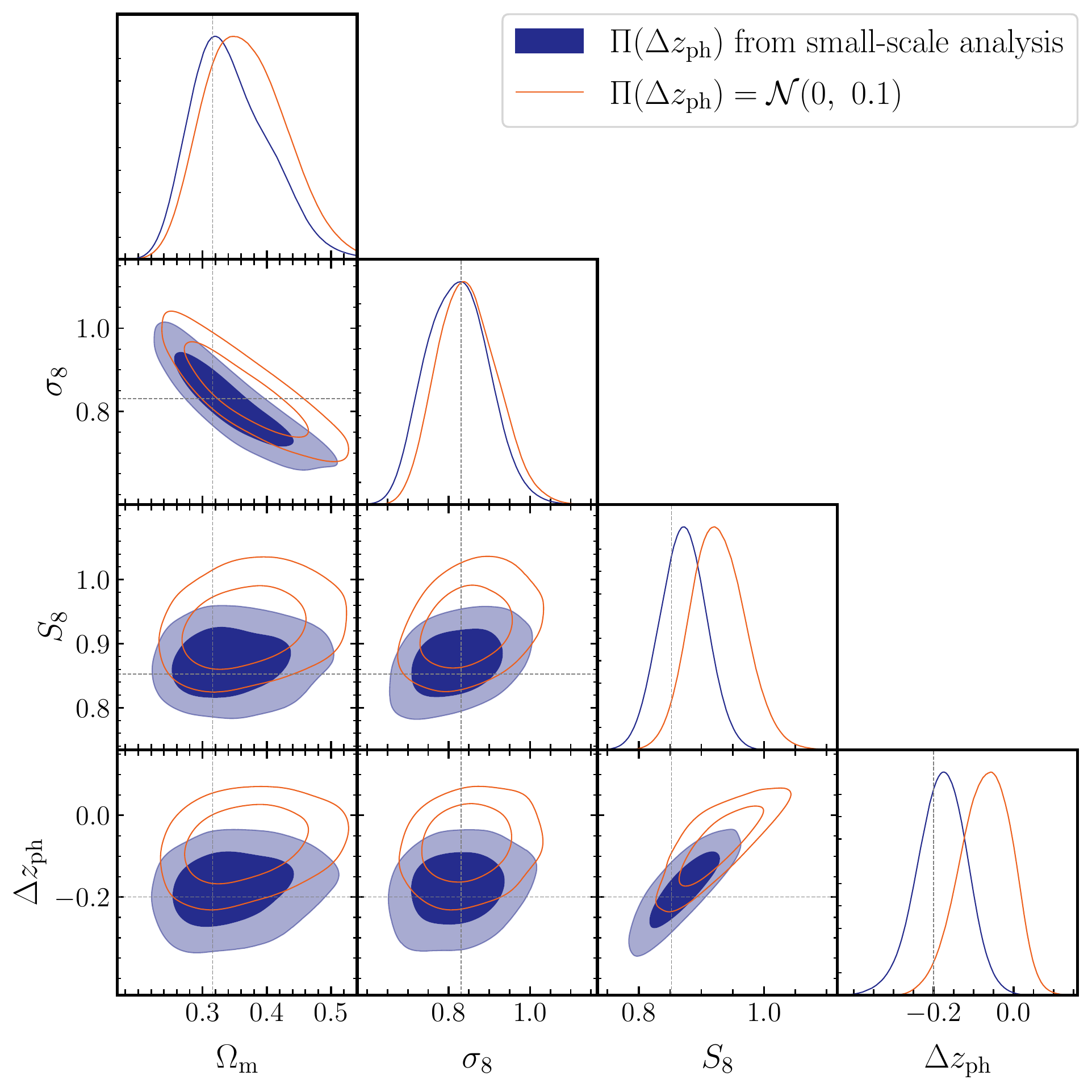}
    \caption{Validation of the use of the $\Delta z_{\rm ph}$ prior taken from the small-scale 3$\times$2pt 
    analysis, which is our baseline analysis method. 
    For this test, we use the synthetic data vector where we implemented the systematic error in the source redshift distribution modeled by 
    $\Delta z_{\rm ph}^{\rm in}=-0.2$ (see Section~\ref{subsec:validation-selfcalib-photo-z}). 
    The blue contours show the results obtained when analyzing 
    the synthetic data vector with a prior taken from small-scale 3$\times2$pt analysis method, 
    while the orange contours show the results with an informative (but wrong!) prior ${\cal N}(0,0.1)$.}
    \label{fig:photo-z-hod-prior}
\end{figure}

In this appendix, we validate the model and the analysis setup we adopted for the HSC-Y3 real data analysis. To validate the baseline choice of model and setup in this paper, we generate various kinds of data vectors which include systematic errors. We analyze each of the contaminated data vectors with the baseline model and setup; checking whether the cosmological parameter constraints are robust given the constraining power of HSC-Y3 data allows us to validate our modeling framewor. We first make a fiducial mock which does not include any systematic errors, and then add simulated systematic errors of various types to the fiducial mock. We categorize the simulated systematic errors on the data vectors into four groups, as  described in the following subsections. 

The galaxy-galaxy lensing and galaxy clustering signals in the fiducial mock is the same as used in the HSC-Y1 analysis of \citet{Sugiyama:2020,2021arXiv210100113M}. These are measured from the galaxy distribution populated by an HOD prescription on halos identified in $N$-body simulation data. The cosmic shear signal is generated from the {\tt halofit} \citep{Smithetal:03} code, updated by \citet{Takahashi_2012}. To simulate the data vector, we use the Planck 2015 cosmology \citep{PlanckCosmology:16}.

Table~\ref{tab:mock-validation} summarizes the validation tests done in the following subsections, and the results are summarized in Fig.~\ref{fig:summary-mock-validation}.

\subsection{Generation of contaminated mock data vectors}
\subsubsection{Galaxy bias uncertainties}
\label{subsec:validation-galaxy-bias}
Galaxies are biased tracers of the underlying matter field, and thus we can extract the cosmological information 
only after marginalizing over galaxy bias uncertainty.
At cosmological scales, gravity is the only force that drives structure formation, and hence the cosmological perturbation theory of structure formation and the galaxy bias expansion based on  perturbation theory should work well. However, at quasi-nonlinear scales, perturbation theory and the galaxy bias expansion break down due to nonlinear physics.

In this paper, we analyze the galaxy clustering signal $\wproj$ at $R>8\mpch$, and the galaxy-galaxy lensing signal $\dSigma$ at $R>12\mpch$, using the minimal bias model based on cosmological perturbation theory. We validated the use of the minimal bias model using the same scale cuts as those in \citet{Sugiyama:2020}, which checked that the minimal bias model can
recover 
cosmological parameters 
within the HSC-Y1 statistical error. In this HSC-Y3 3$\times$2pt analysis, we have higher statistical power than HSC-Y1 due to the larger area coverage of the HSC-Y3 shape catalog and the inclusion of the cosmic shear signal. Therefore, we repeat the validation of the minimal bias model as in \citet{Sugiyama:2020} using the HSC-Y3 covariance matrix.

\subsubsection{Baryonic effects on cosmic shear}
\label{subsec:validation-baryon}
For the evaluation of the cosmic shear signal, we use {\tt halofit} \citep{Smithetal:03} updated by \citet{Takahashi_2012} as the fiducial modeling method to compute the nonlinear matter power spectrum 
for an input model.

The calibration of the fitting formula, 
{\tt halofit},
was obtained using
$N$-body simulations for $\Lambda$CDM cosmologies.
However, baryonic effects inherent in galaxy formation physics alter the total matter power spectrum at nonlinear scales, $k\gtrsim 0.1~h{\rm Mpc}^{-1}$, as shown by
hydrodynamical simulations \citep[e.g.][]{2014Natur.509..177V}. 
To validate the use of {\tt halofit}, which does not include baryonic feedback effects, 
we generate mock data vectors in which we simulate the baryonic effects on the power spectrum for a given cosmological model using 
the 2015 and 2020 versions of {\tt HMCode} \citep{Mead:2020vgs}\footnote{In the tomographic cosmic shear analyses with HSC-Y3 data by \citet{li2023} and \citet{dalal2023}, we use the 2016 version of {\tt HMCode} instead of the 2015 version used in this paper. The 2016 version of {\tt HMCode} is an extension of the 2015 version of {\tt HMcode} to the beyond $\Lambda$CDM model, e.g. dark energy, neutrino mass, and modified gravity. Thus there is no difference between the two versions used in \citet{li2023} or \citet{dalal2023} and this paper as long as we focus on $\Lambda$CDM model.}.
In {\tt HMCode-2015} \citep{Mead:2015yca}, the baryonic effect is parameterized by the halo concentration parameter
$A_{\rm bary}$.  
$A_{\rm bary} = 3.13$ 
corresponds to the case of no baryonic effects, i.e. dark matter only case.
The smaller values of $A_{\rm bary}$ correspond to the case where the baryonic effects are greater. 
For our validation tests we consider  
$A_{\rm bary}=3.13,\, 2.8,\, 2.5,\, 2.2,\, 1.9, $ and $1.6$.
The most extreme value we assume, $A_{\rm bary}=1.6$, is designed to reproduce 
the 
OWLS simulation result \citep{2010mnras.402.1536s}.
In {\tt HMCode-2020} \citep{Mead:2020vgs}, the parameter
$\log_{10}(T_{\rm AGN}/{\rm K})$
is used to model
the AGN 
feedback effect on the matter power spectrum, and we use $\log_{10}(T_{\rm AGN}/{\rm K})=7.3,\, 7.5,\, 7.7,\, 7.9,\, 8.1$ and $8.3$ for the baryon-affected mocks. Here, 
$\log_{10}(T_{\rm AGN}/{\rm K})=7.6$ $(8.3)$ is designed to reproduce 
the AGN feedback effects in 
the {\tt BAHAMAS} \citep{2017MNRAS.465.2936M} ({\tt COSMO-OWLS} \citep{brun:2013yva}) simulations.

\subsubsection{A systematic error in the mean source redshift}
\label{subsec:validation-selfcalib-photo-z}

The small-scale 3$\times$2pt analysis by \citet{miyatake2023} uses a uniform prior to self-calibrate the 
residual systematic error in the source redshift, 
$\Delta z_{\rm ph}$. In this paper we use the same source sample, and use the mode and the credible interval 
of $\Delta z_{\rm ph}$ from \citet{miyatake2023} as an informative prior on 
$\Delta z_{\rm ph}$ in our fiducial analysis method.

To validate our analysis method, we perform the following test in a similar way to what we do in the actual analysis. 
In this test,  we keep the observed shear invariant, 
but assume that there is a bias in the mean redshift of source galaxies inferred from their photo-$z$'s, by an amount $\Delta z_{\rm ph}^{\rm in}$. We assume that the ``estimated'' redshift distribution, denoted as $p_s(z)$, is given by a shift of the true distribution, ${p}_s(z)$, as 
\begin{align}
p_s(z)={p}^{\rm true}_s(z-\Delta z_{\rm ph}^{\rm in}).
\end{align}
That is, $p_s(z+\Delta z_{\rm ph})={p}_s^{\rm true}(z)$, recovering the true distribution, if $\Delta z_{\rm ph}=\Delta z_{\rm ph}^{\rm in}$. 
This $\Delta z_{\rm ph}$ is our parametrization of the systematic photo-$z$ error (Eq.~\ref{eq:deltaz_def}). For our test we take $\Delta z_{\rm ph}^{\rm in}=-0.2$.  
Here $\Delta z_{\rm ph}^{\rm in}=-0.2$ is almost $2\sigma$ away from 
the Gaussian prior of $\Delta z_{\rm ph}$, ${\cal N}(-0.05,0.09)$ used in our analysis. 
Hence this test gives the worst-case scenario for the impact of the photo-$z$ bias error. 

For our weak lensing observables, a systematic error in the estimated redshift distribution, $p(z)$, causes a biased estimate of 
the excess surface mass density of
\begin{equation}
\widehat{\dSigma}=\frac{\left\langle\Sigma_{\rm cr}^{-1}\right\rangle^{-1}}{\left\langle\Sigma_{\rm cr}^{-1}\right\rangle^{-1}_{\rm true}}\dSigma_{\rm true}, 
\end{equation}
where $\left\langle\Sigma_{\rm cr}^{-1}\right\rangle$ is the ``estimated'' average of the critical surface mass density with $p_s(z)$
in the ensemble average sense (see Eq.~16 in \citet{more2023}), $\left\langle\Sigma_{\rm cr}^{-1}\right\rangle_{\rm true}$ is the true value computed 
with $p^{\rm true}_s(z)$, and $\dSigma_{\rm true}$ is the true excess surface mass density. 
In this way, we generate a synthetic data vector of $\dSigma$ including the effect of photo-$z$ errors. 
The cosmic shear correlation functions $\xi_{\pm}$ are invariant, and we do not change the synthetic data vector of $\xi_{\pm}$ in our test. 
However, the theoretical model of 
$\xi_{\pm}$ for a given cosmology is biased because the model assumes an input source redshift distribution, i.e. $p(z)$. 
Then we assess whether our analysis method 
can recover the input $S_8$ and other parameters 
including $\Delta z_{\rm ph}$; if our calibration method works perfectly, the best-fit model should give $\Delta z_{\rm ph}=\Delta z_{\rm ph}^{\rm in}$. 
To estimate the impact of source redshift error, we also study how the $S_8$ value is biased {\it if} we employ the 
informative, narrower Gaussian prior of $\Delta z_{\rm ph}$ around $\Delta z_{\rm ph}=0$, i.e. ${\cal N}(0,0.1)$.
Finally, and for completeness, we also perform the test using a flat prior, ${\cal U}(-1,1)$.

\subsubsection{PSF model}

To model the PSF systematic effects on cosmic shear correlations ($\xi_{\pm}$), we take into account the PSF modeling error and PSF leakage based on the second moments of the PSF
(Eq.~\ref{eq:psf_sys_cosmicshear}) in our model. 
The HSC-Y3 cosmic shear papers (\citet{li2023}
and \citet{dalal2023}) accompanying this paper, used a more sophisticated, accurate model of PSF systematics than we do; their model incorporates terms depending on the fourth moment of the PSF following \citet{zhang:2022dvs}, while we will only include the second-moment terms.
In this paper, we use HSC source galaxies at high redshifts, $z\gtrsim 0.75$, where the lensing efficiency is higher, and therefore the impact of PSF systematics on the cosmic shear signal should be smaller than for the lower-redshift source galaxies used in the cosmic shear tomography analysis.  
Nevertheless we validate our methodas follows. 

Using the method in \citet{zhang:2022dvs}, we measured the second and fourth moment PSF systematics terms  for 
the HSC source galaxy sample used in this paper. We then generated a mock data vector of cosmic shear signal 
($\xi_\pm$), contaminated with these PSF systematics, and then assessed whether our analysis method using
the second moment PSF model
can recover the input 
$S_8$ value.

\subsection{Results of model validation tests}

Fig.~\ref{fig:summary-mock-validation} shows the results of the validation tests of our model and method as outlined above and in Table~\ref{tab:mock-validation}.
Here we compare the mode values of the cosmological parameters and the photo-$z$ error parameter $\Delta z_{\rm ph}$ between our baseline analysis and the various analyses using subsets of our data and mock data contaminated with various kinds of systematics. 
For the analysis 
with superscript ${}^\ast$, we run the large-scale analysis to the synthetic data using 
the informative prior on $\Delta z_{\rm ph}$ taken from the posterior of the small-scale 3$\times$2pt analysis \citep{miyatake2023} on the {\it same} synthetic data vector. This is our baseline analysis method that we use for the actual 
HSC-Y3 and SDSS data. 

The top section of Fig.~\ref{fig:summary-mock-validation} shows the results obtained using subsets of the data vector with the baseline analysis method. 
If we use either 
the 2$\times$2pt data vector ($\dSigma$ and $\wproj$) or the cosmic shear, 
the constraints on the cosmological parameters are degraded compared to the baseline
3$\times$2pt analysis. 
In addition, this method yields a somewhat biased estimate of 
$S_8$. This might explain the larger value of $S_8$ in the HSC-Y1 large-scale 2$\times$2pt analysis in 
\citet{Sugiyama:2021}, compared to the $S_8$ value from the small-scale 2$\times$2pt or the HSC-Y3 3$\times$2pt analyses. 

The second section of Fig.~\ref{fig:summary-mock-validation} shows the results of the validation tests obtained by
applying the baseline analysis method to mock data vectors measured from different types of mock 
SDSS galaxies. Here we used the mock SDSS catalogs described by \citet{2021arXiv210100113M}, where mock galaxies 
are populated into halos in $N$-body simulations using different models of galaxy-halo connection.
Our baseline analysis 
recovers the input cosmological parameter, $\Omega_{\rm m}$ and $S_8$, within 
$0.5\sigma$, 
except for the {\tt assembly-$b$} and {\tt assembly-$b$-ext} mocks. 
Hence the results give validation of our analysis method for most of the mock catalogs, if the SDSS galaxies follow the galaxy-halo connection as that simulated by these mock catalogs. 

The assembly bias is one of the most important systematic effects in the galaxy-halo connection. From 
Fig.~\ref{fig:summary-mock-validation} one might conclude that the minimal bias model fails to pass the validation test using the assembly bias mocks.
However, this is not so simple as explained below. First of all, we would like to note that the assembly bias mocks we use in the tests assume the overwhelmingly large assembly bias effects and therefore give 
the worse-case scenario, where the {\tt assembly-$b$} and {\tt assembly-$b$-ext} 
mocks have the greater amplitudes in the 2-halo term of $\wproj$ by a factor of 1.3 and 1.5, even though the assembly bias has not yet been detected at a high significance from the SDSS galaxies.
The apparent failure of our method for the assembly bias mocks is due to the degeneracies between the photo-$z$ error 
parameter $\Delta z_{\rm ph}$ and the cosmological parameters. In our baseline method, 
we use the informative prior on $\Delta z_{\rm ph}$
taken from the small-scale 3$\times$2pt analysis of \citet{miyatake2023}, which uses the uninformative flat prior 
$\Pi(\Delta z_{\rm ph})={\cal U}(-1,1)$ to minimize the impact of the unknown source redshift uncertainty. 
To being with, the small-scale 3$\times$2pt analysis fails to reproduce both the input cosmological parameters (e.g. $S_8$) and the input photo-$z$ error parameter ($\Delta z_{\rm ph}=0$) for  these synthetic data from the assembly bias mocks, because the small-scale analysis is severely affected by the assembly bias effect due to a violation in the simple scaling relation of galaxy bias amplitude with the host halo masses. On top of this, the small-scale analysis suffers from 
the parameter degeneracies due to a {\it positive} correlation between $S_8$ and $\Delta z_{\rm ph}$. 
For these reasons, the prior information of $\Delta z_{\rm ph}$ delivered from the small-scale 3$\times$2pt analysis is {\it biased} in the first place. 
The row ``{\tt assembly-$b$-ext} w $\Pi(\Delta z_{\rm ph})={\cal N}(0,0.1)$'' shows the result obtained for our baseline method {\it if} 
we can use the informative Gaussian prior with mean around the true $\Delta z_{\rm ph}$
($\Delta z_{\rm ph}=0$), where the prior width $\sigma(\Delta z_{\rm ph})=0.1$ roughly matches the precision of $\Delta z_{\rm ph}$ for the small-scale 3$\times$2pt analysis. In this case, our method can nicely recover the input $S_8$ value, meaning that the minimal bias model can work even for the worst-case assembly bias scenario. 
However, this is not the case for the small-scale 3$\times$2pt case; if such informative prior of $\Pi(\Delta z_{\rm ph}={\cal N}(0,0.1)$ is employed for the small-scale analysis, $S_8$ is significantly biased due to the assembly bias effect. Thus comparing the inferred $S_8$ values between the small- and large-scale 3$\times$2pt analyses
with the informative prior of $\Pi(\Delta z_{\rm ph})={\cal N}(0,0.1)$ can be used as a flag of the possible assembly bias effect. For the actual HSC-Y3 and SDSS data, we did not find a significant shift in the $S_8$ values for the analyses 
using $\Pi(\Delta z_{\rm ph})={\cal N}(0,0.1)$.

There is another diagnostic to flag 
the assembly bias effect in an actual analysis.  
The large-scale 3$\times$2pt analysis is less affected by the assembly bias effect, even though the amount of the bias depends on the prior choice of 
$\Delta z_{\rm ph}$ as we discussed above. Hence, we can flag the significant assembly bias effect by comparing the $S_8$ values estimated from the large-scale and small-scale 3$\times$2pt analyses. The right panel of Fig.~\ref{fig:S8-diff}
shows this test. 
The histogram is from the noisy 100 mock data where we assume that the simulated data is not affected by the assembly bias effect (i.e. no assembly bias simulations). The two arrows denote the $S_8$ differences expected from the mock data that are simulated from the assembly bias mocks we used above. 
The measured $S_8$ difference is quite consistent with the noisy mock data. 
With the statistical power of the HSC-Y3 data, we cannot conclude that our results are not contaminated by the assembly bias effect that is 
as large as simulated in the {\tt assembly-$b$} mock, but the probability that our results are contaminated by the assembly bias effect as large as  
the {\tt assembly-$b$-ext} mock is quite unlikely (at a 2$\sigma$ level). 
Another rationale that we found after unblinding the HSC-Y3 cosmology results, although not objective, is a nice agreement between the $S_8$ values from the small-scale 3$\times$2pt analyses and the cosmic shear analyses
in \citet{li2023} and \citet{dalal2023}. If the SDSS galaxies are significantly contaminated by the assembly bias, this agreement is not guaranteed.

The third section in Fig.~\ref{fig:summary-mock-validation} tests
the {\tt halofit} model which we use to compute the 
cosmic shear prediction in our analysis method. 
The row ``HMCode v2015 (DM only)'' denotes the result for the mock cosmic shear data where baryonic effects are set to zero (i.e.,  
$A_{\rm bary}=3.13$) 
Hence, shifts in the cosmological parameters between our baseline analysis and 
``HMCode v2015 (DM only)'' are due to the difference in the nonlinear matter power spectra of {\tt halofit} and 
the HMCode. Although $\Omega_{\rm m}$ and $\sigma_8$ show sizable shifts, the $S_8$ value is essentially identical in the two cases. 
We carried out tests with mock data generated using 
$A_{\rm bary}=2.8$ and $1.6$ ({\tt HMCode 2015}), and with $\log_{10}(T_{\rm AGN}/{\rm K})=7.3$ or 8.3 ({\tt HMCode 2020}).
Our baseline method recovers the $S_8$ value within the $0.33\sigma$ uncertainties for the mocks with $A_{\rm bary}=2.8$ ({\tt HMCode 2015}) and $\log_{10}(T_{\rm AGN}/{\rm K})=7.3$ ({\tt HMCode 2020}), and within the $1\sigma$ uncertainties for the mocks with $A_{\rm bary}=1.6$ ({\tt HMCode 2015}) and $\log_{10}(T_{\rm AGN}/{\rm K})=8.3$ ({\tt HMCode 2020}).
Thus these results confirm that our results are robust, given the  cosmic shear scale cuts we have used
($\theta_{\rm min,+}=10^{0.8}{\rm arcmin}$ and $\theta_{\rm min,-}=10^{1.5}{\rm arcmin}$).

The validation tests for the residual photo-$z$ error parameter ($\Delta z_{\rm ph}$) shown in Fig.~\ref{fig:summary-mock-validation} are encouraging. 
Our baseline analysis method
using 
the informative prior on $\Delta z_{\rm ph}$ taken from the posterior distribution of the small-scale 3$\times$2pt analysis \citet{miyatake2023} nicely recovers the input $S_8$, even if the redshift source distribution inferred 
from the photo-$z$ estimates is wrong, with a systematic error by $|\Delta z_{\rm ph}^{\rm in}|=0.2$. 
On the other hand, if we employ the informative Gaussian prior with mean around the wrong value $\Delta z_{\rm ph}=0$, i.e. $\Pi(\Delta z_{\rm ph})={\cal N}(0,0.1)$, which is the prior used in the HSC-Y1 analysis, 
the estimate of $S_8$ is significantly biased compared to the input value,
as can be found from  the row ``$\Delta z_{\rm ph}^{\rm in}=-0.2$,~w/~$\Pi(\Delta_{\rm ph})={\cal N}(0, 0.1)$''. 
If we employ the uninformative flat prior of $\Delta z_{\rm ph}$ for our large-scale 3$\times$2pt analysis,  
we can recover the $S_8$ value, but the constraining power is significantly degraded,
as shown in the row of ``$\Delta z_{\rm ph}^{\rm in}=-0.2$,~w/~$\Pi(\Delta_{\rm ph})={\cal U}(-1, 1)$''.
Hence we conclude that our baseline method is valid in the sense that it 
can safely recover the value of $S_8$ even if 
a systematic error in the mean redshift of HSC source galaxies  
is as large as 0.2.
Fig.~\ref{fig:photo-z-hod-prior} gives 
 closer look at how the informative prior of the photo-$z$ parameter from the small-scale analysis works in the large-scale analysis. 
The figure shows that $S_8$ strongly correlates with $\Delta z_{\rm ph}$.  
We again stress that, when we adopt an informative prior from the small-scale analysis, which is {\it correctly} calibrated by the galaxy-galaxy lensing self-calibration \citep{OguriTakada:11}, we recover the input value of $S_8$ within the statistical error.

Finally, the row labeled ``PSF 4th'' in Fig.~\ref{fig:summary-mock-validation}
shows that our baseline analysis can recover the value of $S_8$ even if the cosmic shear signal is contaminated by fourth-moment PSF systematics.
Thus we conclude that our model limited to second-moment PSF systematics is sufficient to model the cosmic shear signal.

\section{A corner plot for all the model parameters}
\begin{figure*}
    \includegraphics[width=2\columnwidth]{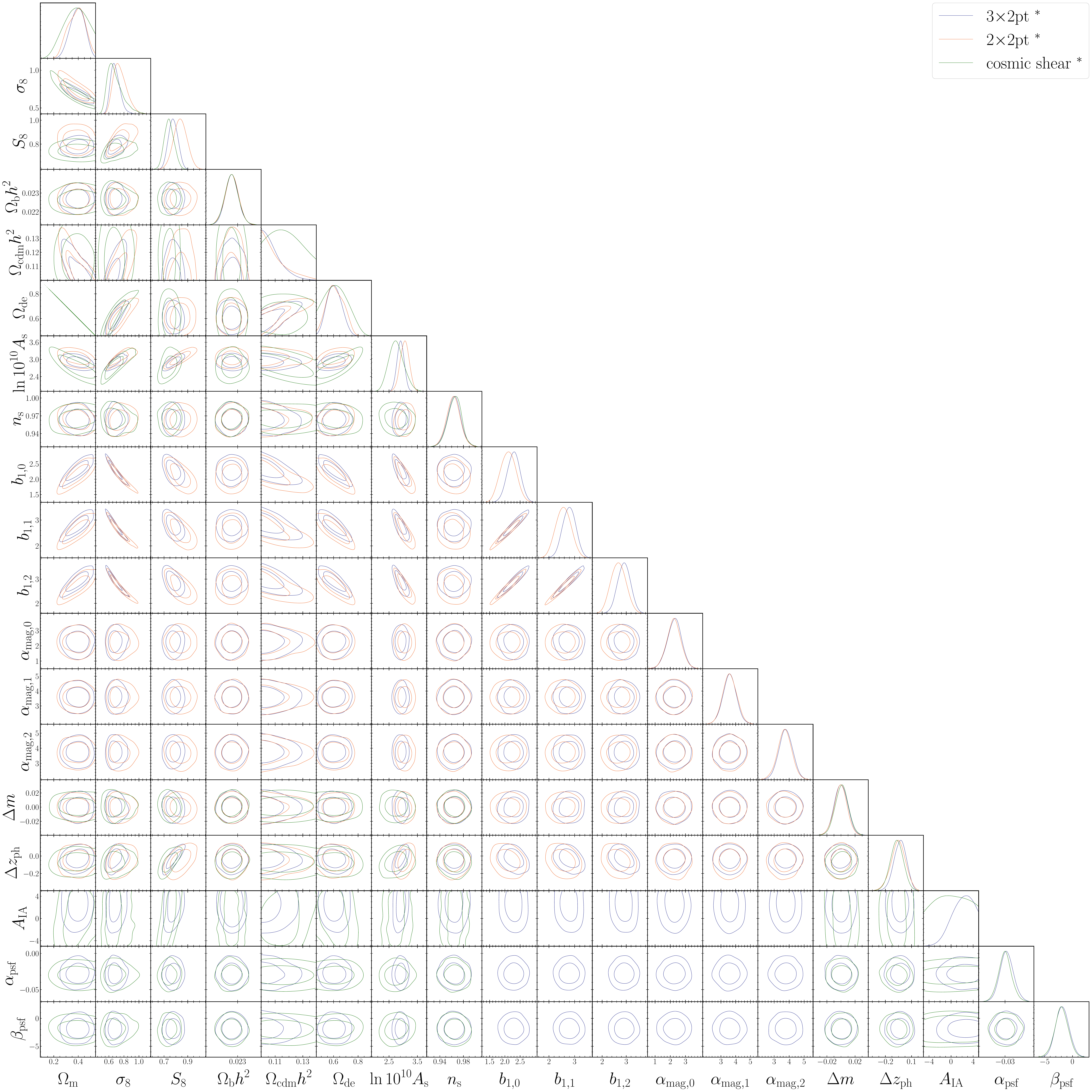}
    \caption{Marginalized posterior distribution for all the model parameters with derived parameters, $\Omega_{\rm m}$, $\sigma_8$, and $S_8$. Here the parameter constraints are obtained from the baseline 3$\times$2pt analysis (blue), the 2$\times$2pt-only analysis (orange), and the cosmic shear alone analysis (green). 
    }
    \label{fig:full-corner}
\end{figure*}

We present a corner plot for all the model parameters in Fig.~\ref{fig:full-corner}. 
This figure shows the correlations between different model parameters. This figure can be used to infer how we can improve the parameter constraints with prior information or external data set in future studies.
\label{sec:full-corner}

\section{Internal consistency tests}
\label{sec:apdx-internal-consistency-test}

\begin{table*}
    \centering
    \setlength{\tabcolsep}{12pt}
    \renewcommand{\arraystretch}{1.5}
    \caption{Summary of the 
    cosmological parameter constraints for $\Omega_{\rm m}$, $\sigma_8$, and $S_8$, obtained from our large-scale 3$\times$2pt analysis of the HSC-Y3 and SDSS data. 
    The estimates are presented in the format of $\text{mode}^{+34\%~\text{upper}}_{-34\%~\text{lower}}$ $(\text{MAP},~\text{mean})$, where the mode is the peak of the marginalized posterior distribution, the credible interval is defined as the 68\% highest density interval, the MAP is obtained from the MC chain with the highest posterior value, and the mean is defined as the parameter means with respect to the posterior. The analysis setup for each row is summarized in Table~\ref{tab:internal-tests}.
    The analyses with superscript ${}^\ast$ denote an analysis using 
    the informative prior on $\Delta z_{\rm ph}$, given by $\Pi(\Delta z_{\rm ph})={\cal N}(-0.05,0.09)$, 
    taken from the small-scale 3$\times$2pt analysis in \citet{miyatake2023}, and the analysis with superscript ${}^\dagger$ denotes the result using  the 
    informative Gaussian prior $\Pi(\Delta z_{\rm ph})={\cal N}(0, 0.1)$.
    }
\begin{tabular}{llll}
\toprule
{} &                          $\Omega_{\rm m}$ &                                $\sigma_8$ &                                     $S_8$ \\
\midrule
3$\times$2pt ${}^{\ast}$                        &  $0.401_{-0.064}^{+0.056} (0.394, 0.393)$ &  $0.666_{-0.051}^{+0.069} (0.705, 0.685)$ &  $0.775_{-0.038}^{+0.043} (0.808, 0.777)$ \\
2$\times$2pt ${}^{\ast}$                        &  $0.408_{-0.091}^{+0.053} (0.420, 0.385)$ &  $0.713_{-0.068}^{+0.105} (0.710, 0.749)$ &  $0.837_{-0.056}^{+0.057} (0.841, 0.838)$ \\
cosmic shear ${}^{\ast}$                        &  $0.385_{-0.092}^{+0.103} (0.411, 0.374)$ &  $0.629_{-0.068}^{+0.114} (0.671, 0.684)$ &  $0.739_{-0.040}^{+0.043} (0.785, 0.744)$ \\
\hline
3$\times$2pt, w/o LOWZ ${}^{\ast}$              &  $0.373_{-0.062}^{+0.080} (0.376, 0.384)$ &  $0.667_{-0.059}^{+0.083} (0.702, 0.690)$ &  $0.767_{-0.038}^{+0.046} (0.785, 0.771)$ \\
3$\times$2pt, w/o CMASS1 ${}^{\ast}$            &  $0.360_{-0.061}^{+0.071} (0.419, 0.369)$ &  $0.674_{-0.061}^{+0.080} (0.651, 0.694)$ &  $0.758_{-0.039}^{+0.043} (0.769, 0.760)$ \\
3$\times$2pt, w/o CMASS2 ${}^{\ast}$            &  $0.424_{-0.077}^{+0.060} (0.446, 0.408)$ &  $0.653_{-0.058}^{+0.067} (0.666, 0.671)$ &  $0.772_{-0.036}^{+0.044} (0.812, 0.775)$ \\
\hline
2$\times$2pt, w/o LOWZ ${}^{\ast}$              &  $0.407_{-0.091}^{+0.069} (0.427, 0.389)$ &  $0.721_{-0.073}^{+0.102} (0.703, 0.752)$ &  $0.841_{-0.057}^{+0.062} (0.838, 0.844)$ \\
2$\times$2pt, w/o CMASS1 ${}^{\ast}$            &  $0.314_{-0.064}^{+0.087} (0.352, 0.332)$ &  $0.757_{-0.096}^{+0.132} (0.722, 0.795)$ &  $0.816_{-0.061}^{+0.071} (0.782, 0.820)$ \\
2$\times$2pt, w/o CMASS2 ${}^{\ast}$            &  $0.403_{-0.090}^{+0.067} (0.254, 0.387)$ &  $0.721_{-0.084}^{+0.129} (0.973, 0.766)$ &  $0.864_{-0.075}^{+0.062} (0.895, 0.856)$ \\
\hline
no photo-$z$ error                              &  $0.394_{-0.068}^{+0.056} (0.453, 0.387)$ &  $0.691_{-0.059}^{+0.060} (0.647, 0.704)$ &  $0.796_{-0.022}^{+0.021} (0.795, 0.793)$ \\
no shear error ${}^{\ast}$                      &  $0.394_{-0.065}^{+0.058} (0.466, 0.388)$ &  $0.678_{-0.057}^{+0.068} (0.632, 0.695)$ &  $0.785_{-0.042}^{+0.038} (0.787, 0.783)$ \\
no magnification bias error ${}^{\ast}$         &  $0.404_{-0.064}^{+0.059} (0.400, 0.398)$ &  $0.669_{-0.055}^{+0.067} (0.661, 0.683)$ &  $0.781_{-0.043}^{+0.040} (0.764, 0.780)$ \\
no PSF error ${}^{\ast}$                        &  $0.407_{-0.061}^{+0.057} (0.394, 0.401)$ &  $0.667_{-0.054}^{+0.062} (0.685, 0.680)$ &  $0.781_{-0.039}^{+0.039} (0.786, 0.780)$ \\
no IA ${}^{\ast}$                               &  $0.396_{-0.063}^{+0.063} (0.392, 0.393)$ &  $0.653_{-0.047}^{+0.076} (0.669, 0.680)$ &  $0.774_{-0.038}^{+0.034} (0.765, 0.771)$ \\
extreme IA ${}^{\ast}$                          &  $0.402_{-0.062}^{+0.057} (0.380, 0.395)$ &  $0.668_{-0.056}^{+0.068} (0.707, 0.688)$ &  $0.781_{-0.039}^{+0.041} (0.796, 0.782)$ \\
\hline
$R_{\rm max}=30~h^{-1}{\rm Mpc}$ ${}^{\ast}$    &  $0.384_{-0.063}^{+0.062} (0.417, 0.385)$ &  $0.676_{-0.060}^{+0.072} (0.660, 0.692)$ &  $0.775_{-0.039}^{+0.040} (0.779, 0.776)$ \\
\hline
3$\times$2pt, 2 cosmo ${}^{\ast}$               &  $0.316_{-0.036}^{+0.038} (0.304, 0.323)$ &  $0.757_{-0.057}^{+0.058} (0.808, 0.758)$ &  $0.783_{-0.042}^{+0.039} (0.813, 0.782)$ \\
2$\times$2pt, 2 cosmo ${}^{\ast}$               &  $0.316_{-0.037}^{+0.037} (0.290, 0.318)$ &  $0.830_{-0.080}^{+0.068} (0.874, 0.830)$ &  $0.850_{-0.058}^{+0.056} (0.859, 0.850)$ \\
\hline
$\Delta z_{\rm ph}\sim {\cal U}(-1,1)$          &  $0.399_{-0.064}^{+0.061} (0.473, 0.396)$ &  $0.687_{-0.064}^{+0.079} (0.679, 0.703)$ &  $0.802_{-0.061}^{+0.059} (0.852, 0.800)$ \\
\hline
3$\times$2pt ${}^{\dagger}$                     &  $0.406_{-0.067}^{+0.053} (0.490, 0.397)$ &  $0.686_{-0.054}^{+0.070} (0.640, 0.703)$ &  $0.802_{-0.043}^{+0.042} (0.819, 0.801)$ \\
2$\times$2pt ${}^{\dagger}$                     &  $0.398_{-0.083}^{+0.067} (0.287, 0.380)$ &  $0.732_{-0.079}^{+0.115} (0.912, 0.776)$ &  $0.856_{-0.057}^{+0.066} (0.892, 0.860)$ \\
cosmic shear ${}^{\dagger}$                     &  $0.419_{-0.088}^{+0.096} (0.531, 0.393)$ &  $0.626_{-0.058}^{+0.111} (0.569, 0.677)$ &  $0.757_{-0.047}^{+0.046} (0.757, 0.759)$ \\
\hline
XMM $(\sim 33~\mathrm{deg}^2)$ ${}^{\ast}$      &  $0.387_{-0.073}^{+0.062} (0.357, 0.378)$ &  $0.618_{-0.071}^{+0.085} (0.678, 0.640)$ &  $0.711_{-0.074}^{+0.072} (0.739, 0.712)$ \\
GAMA15H $(\sim 41~\mathrm{deg}^2)$ ${}^{\ast}$  &  $0.338_{-0.071}^{+0.073} (0.356, 0.348)$ &  $0.719_{-0.091}^{+0.118} (0.752, 0.749)$ &  $0.793_{-0.066}^{+0.075} (0.819, 0.794)$ \\
HECTOMAP $(\sim 43~\mathrm{deg}^2)$ ${}^{\ast}$ &  $0.409_{-0.076}^{+0.060} (0.426, 0.395)$ &  $0.673_{-0.067}^{+0.095} (0.678, 0.706)$ &  $0.798_{-0.066}^{+0.067} (0.807, 0.800)$ \\
GAMA09H $(\sim 78~\mathrm{deg}^2)$ ${}^{\ast}$  &  $0.414_{-0.065}^{+0.060} (0.443, 0.404)$ &  $0.650_{-0.058}^{+0.070} (0.650, 0.668)$ &  $0.764_{-0.051}^{+0.057} (0.790, 0.769)$ \\
VVDS $(\sim96~\mathrm{deg}^2)$ ${}^{\ast}$      &  $0.409_{-0.076}^{+0.060} (0.426, 0.395)$ &  $0.673_{-0.067}^{+0.095} (0.678, 0.706)$ &  $0.798_{-0.066}^{+0.067} (0.807, 0.800)$ \\
WIDE12H $(\sim 121~\mathrm{deg}^2)$ ${}^{\ast}$ &  $0.389_{-0.072}^{+0.059} (0.404, 0.378)$ &  $0.631_{-0.060}^{+0.084} (0.662, 0.659)$ &  $0.727_{-0.051}^{+0.053} (0.768, 0.731)$ \\
\hline
$\textsc{DEmPz}$ \& WX ${}^{\ast}$              &  $0.404_{-0.069}^{+0.052} (0.422, 0.390)$ &  $0.653_{-0.048}^{+0.071} (0.641, 0.677)$ &  $0.763_{-0.035}^{+0.042} (0.760, 0.765)$ \\
$\textsc{Mizuki}$ ${}^{\ast}$                   &  $0.415_{-0.058}^{+0.057} (0.407, 0.409)$ &  $0.655_{-0.050}^{+0.057} (0.667, 0.667)$ &  $0.772_{-0.032}^{+0.036} (0.776, 0.774)$ \\
$\textsc{DNNz}$ ${}^{\ast}$                     &  $0.403_{-0.071}^{+0.059} (0.400, 0.390)$ &  $0.678_{-0.057}^{+0.088} (0.677, 0.709)$ &  $0.796_{-0.045}^{+0.050} (0.782, 0.799)$ \\
\bottomrule
\end{tabular}
    \label{tab:cosmo-constraint}
    \setlength{\tabcolsep}{6pt} 
    \renewcommand{\arraystretch}{1}
\end{table*}

\begin{figure}
    \includegraphics[width=\columnwidth]{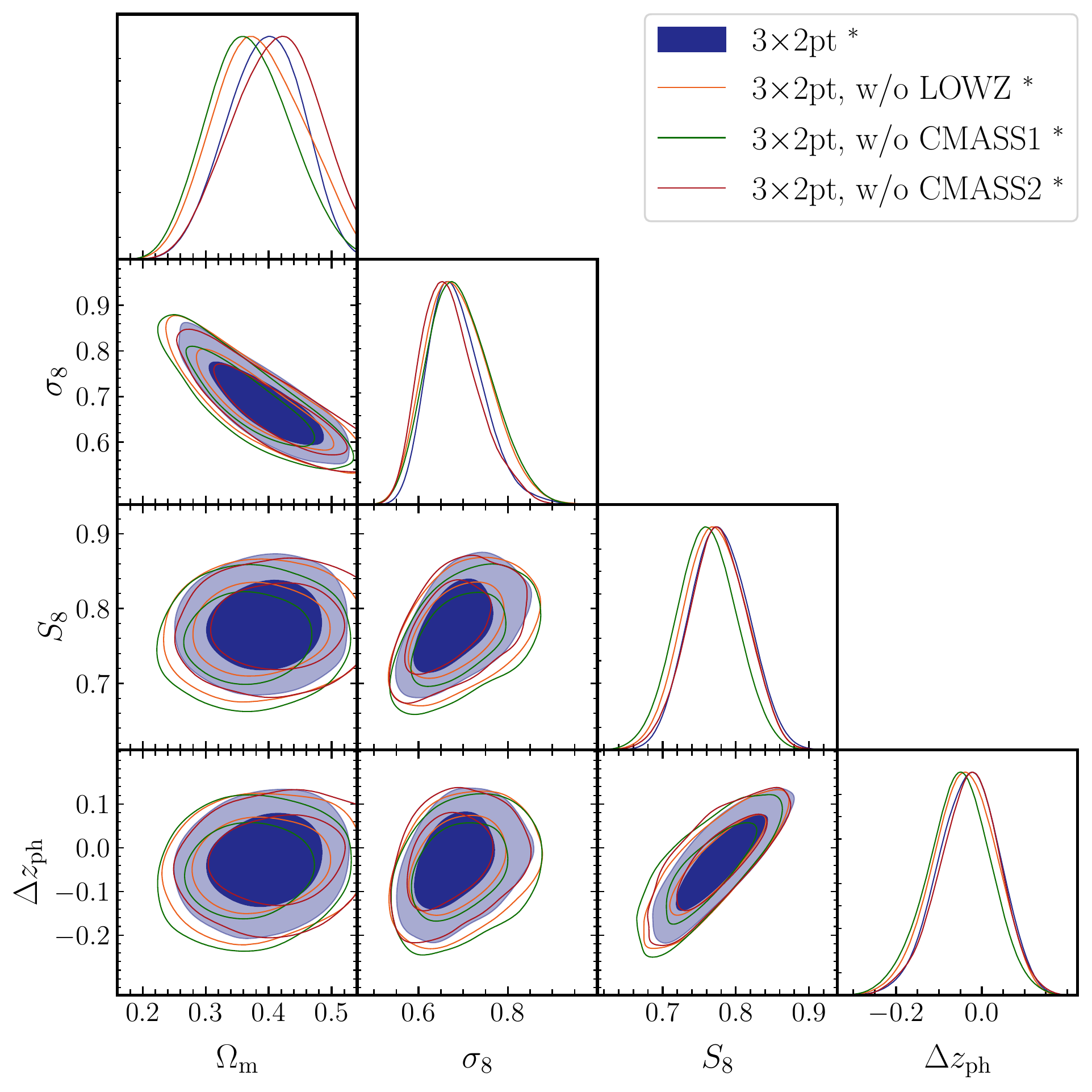}
    \caption{Cosmological constraints with 3$\times$2pt but removing a single lens redshift bin from each analysis.}
    \label{fig:3x2pt-wo-lens-sample}
\end{figure}

\begin{figure}
    \includegraphics[width=\columnwidth]{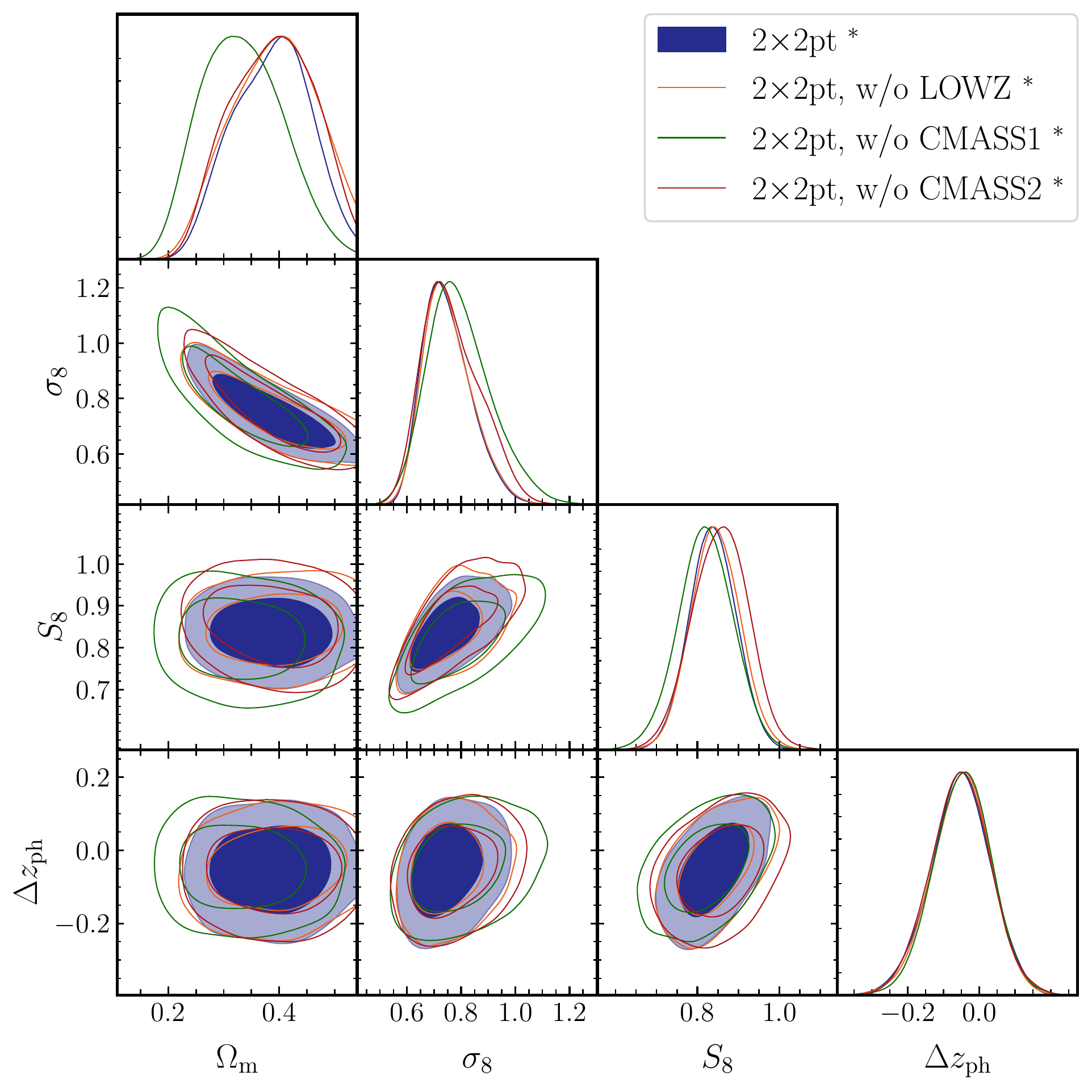}
    \caption{Cosmological constraints with 2$\times$2pt but removing a single lens redshift bin from each analysis.}
    \label{fig:2x2pt-wo-lens-sample}
\end{figure}

\begin{figure}
    \includegraphics[width=\columnwidth]{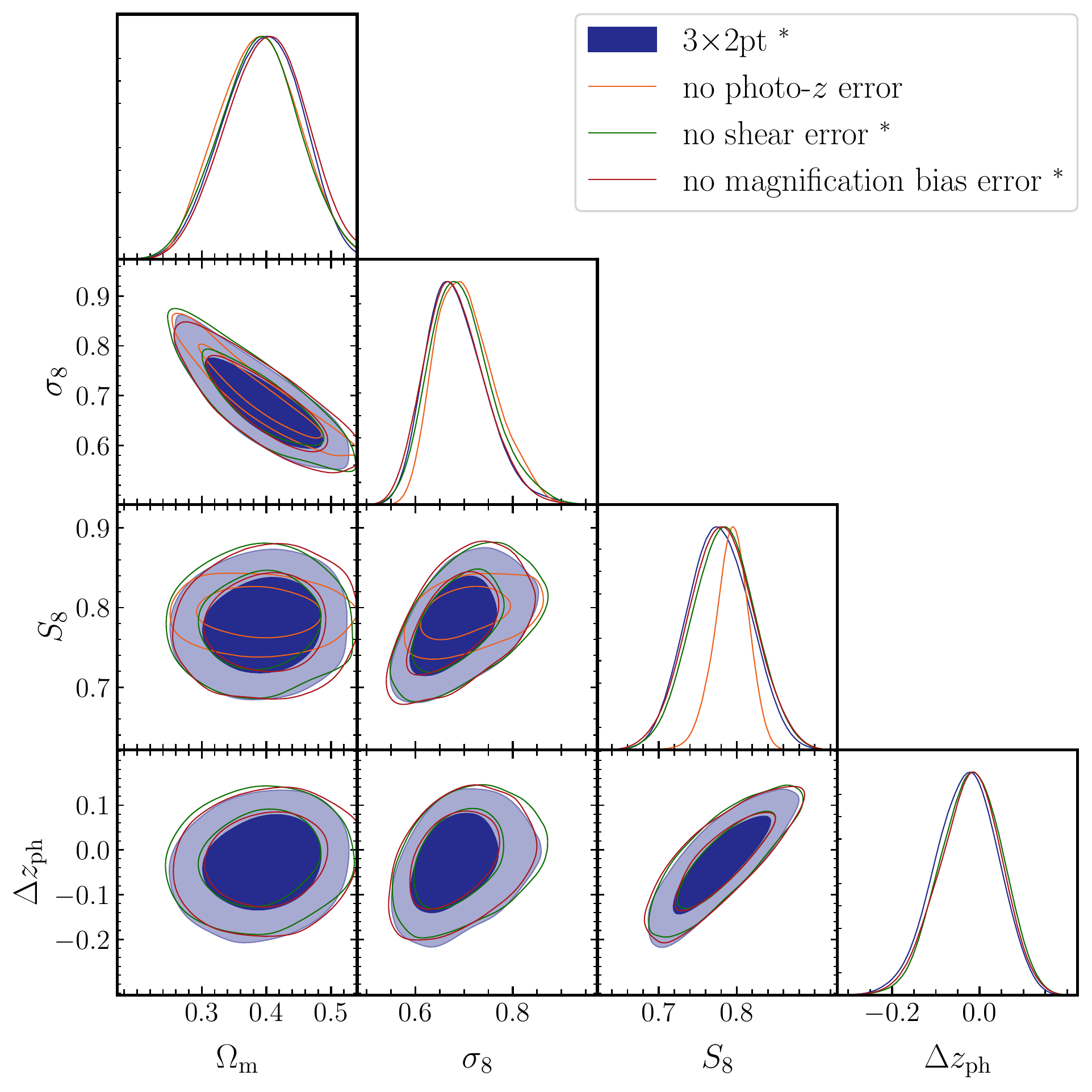}
    \caption{Cosmological constraints with 3$\times$2pt but fixing one of the nuisance parameters to zero.}
    \label{fig:3x2pt-nuisance1}
\end{figure}

\begin{figure}
    \includegraphics[width=\columnwidth]{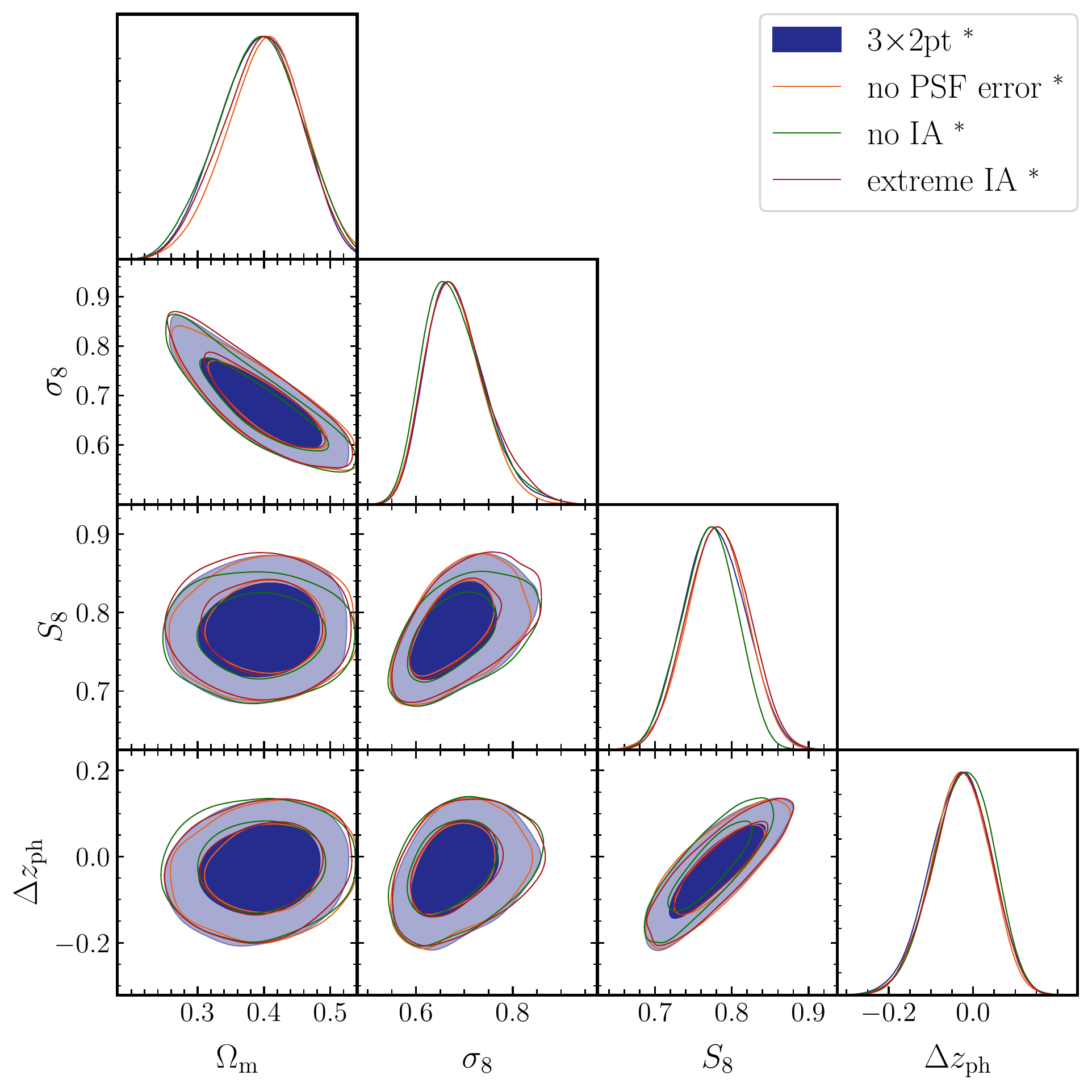}
    \caption{Cosmological constraints with 2$\times$2pt but fixing the cosmic shear related nuisance parameters to some fiducial values. In "no PSF error", we fix $\alpha_{\rm psf}$ and $\beta_{\rm psf}$ to the center of the prior. In the "no IA" case, we fix $A_{\rm IA}=0$, while we set $A_{\rm IA}=5$ in the "extreme IA" case.}
    \label{fig:3x2pt-nuisance2}
\end{figure}

\begin{figure}
    \includegraphics[width=\columnwidth]{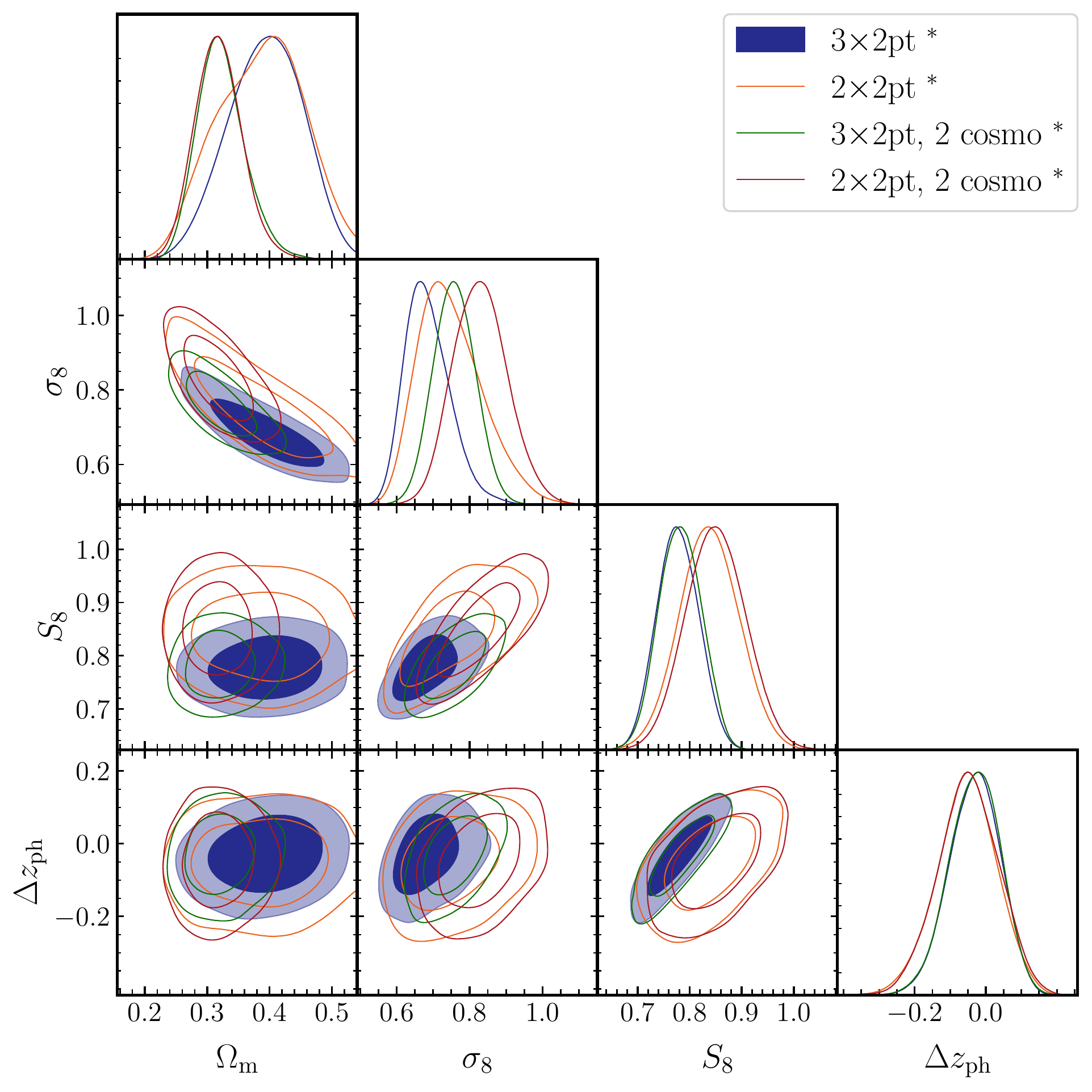}
    \caption{Cosmological constraints with 2$\times$2pt but fixing $n_{\rm s}$ and $\omega_{\rm cdm}\equiv\Omega_{\rm cdm}h^2$ to the best fit value of \cite{2020A&A...641A...6P}, and $\omega_{\rm b}\equiv\Omega_{\rm b}h^2$ to the best fit of BBN \citep{Aver:2015iza,Cooke:2017cwo,Schoneberg:2019wmt}.}
    \label{fig:3x2pt-2cosmo}
\end{figure}

In this section, we detail the internal consistency tests performed in Section~\ref{subsec:internal-consistency}. Table~\ref{tab:cosmo-constraint} shows the statistics of model parameters, $\Omega_{\rm m}$, $\sigma_8$, and $S_8$, obtained from each analysis setup of the internal consistency tests. In this table, we report the estimate of parameters as
\begin{align}
    \text{mode}^{+34\%~\text{upper}}_{-34\%~\text{lower}}~(\text{MAP},~\text{mean}),
\end{align}
for ease of comparison with other papers. As noted in Section~\ref{subsec:bayes-likelihood-prior}, the MAP value is estimated from the MC chain. The MAP obtained from the MC chain can be noisy and thus we should {\it not} take it as the robust estimate of MAP, 
but the difference between the MAP and the mode value gives an indication of how significant projection effects are in each case.  

Figs.~\ref{fig:3x2pt-wo-lens-sample}--\ref{fig:3x2pt-2cosmo} give contour diagrams of parameter constraints in the various internal consistency tests. The results of similar internal consistency tests are grouped and overplotted in each figure.

\section{Robustness of parameter sampling}
\subsection{Nestcheck}
\begin{figure}
    \includegraphics[width=\columnwidth]{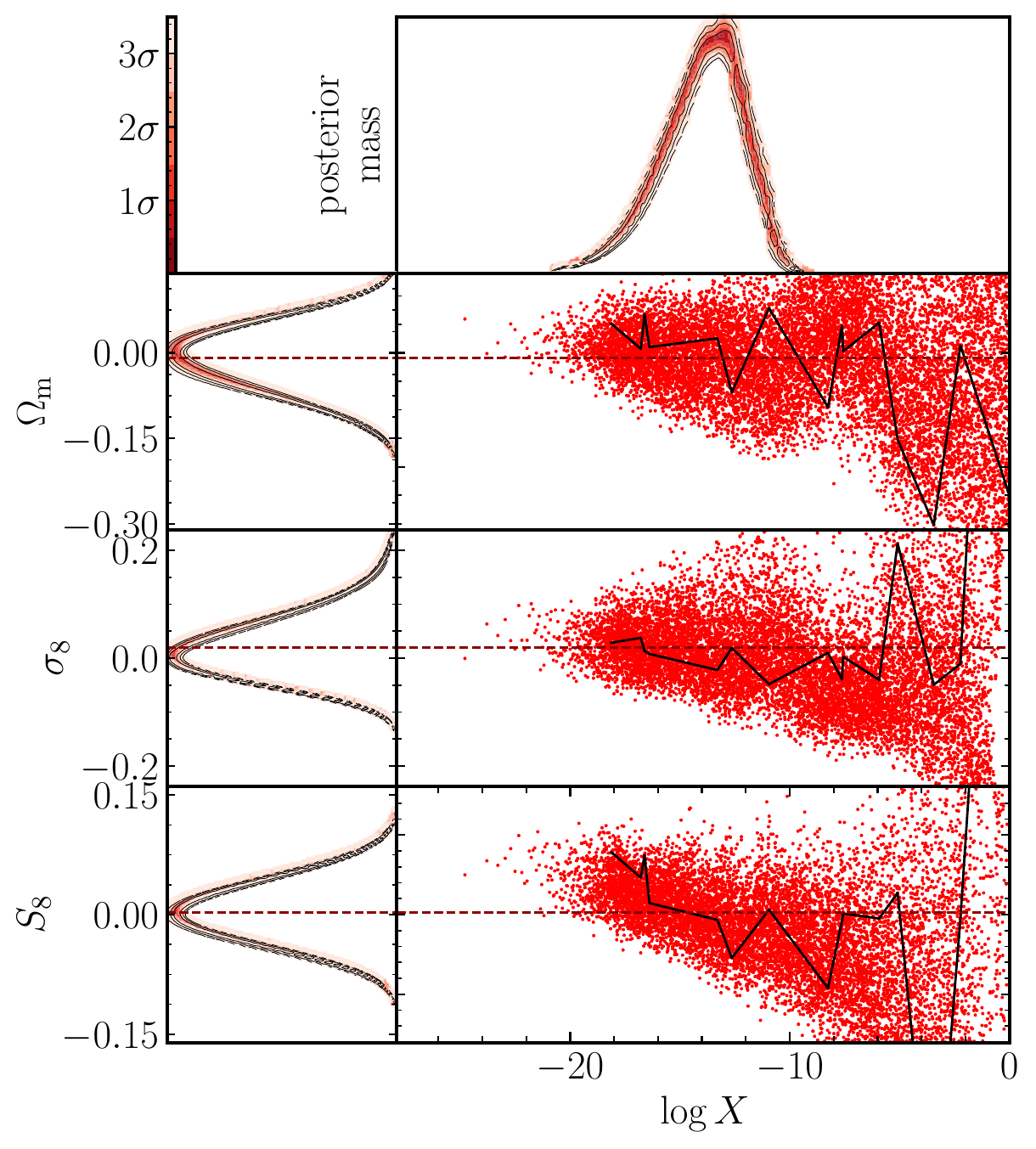}
    \caption{The result of {\tt nestcheck}\citep{higson2018nestcheck} for the baseline chain, sampled in the real data analysis. The top panel shows the posterior volume as a function of the prior volume, $X$. The left panels show the uncertainty of the posterior distribution from an input nested sampling chain, where the uncertainty is estimated by bootstrapping the chain.}
    \label{fig:nestcheck}
\end{figure}

\begin{figure}
    \includegraphics[width=\columnwidth]{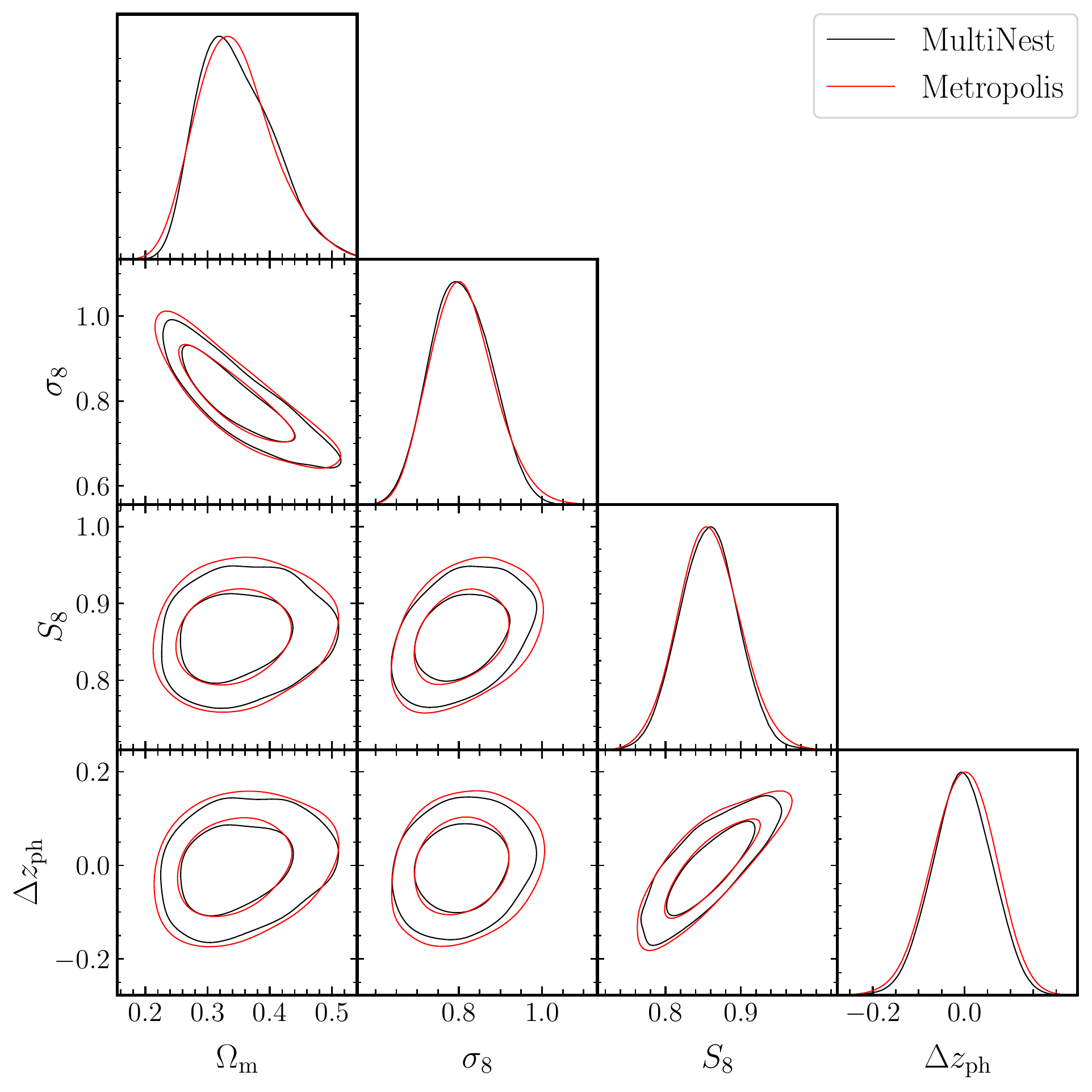}
    \caption{The comparison of posterior estimates from the nested sampling by {\tt MultiNest} and the Markov-Chain Monte Carlo sampling by the standard Metropolis algorithm.}
    \label{fig:mn-vs-mp}
\end{figure}

In this section, we test the robustness of the parameter sampling by {\tt MultiNest}. We use the {\tt nestcheck} diagnostic \citep{higson2019diagnostic} to test the convergence of the {\tt MultiNest} chain. Fig.~\ref{fig:nestcheck} shows the result of the convergence test for the main cosmological parameters, $\Omega_{\rm m}, \sigma_8$, and $S_8$. In the top right panel, we can see that the chain covers sufficient posterior volume. The left panels show the uncertainty of the posterior distributions, estimated by bootstrapping the original MultiNest chain, and indicating that our estimate of the posterior distributions is robust.

As an additional test of convergence of our parameter estimate, Fig.~\ref{fig:mn-vs-mp} compares the result of the nested sampling by {\tt MultiNest} \citep{Feroz:2009} to the result of the Markov Chain Monte Carlo sampling method of the standard Metropolis algorithm \citep{metropolis_1949}. The difference between the posterior estimates is almost negligible, and thus we conclude that our parameter inference by {\tt MultiNest} is robust.

\bibliography{refs-short}

\begin{thebibliography}{82}%
\makeatletter
\providecommand \@ifxundefined [1]{%
 \@ifx{#1\undefined}
}%
\providecommand \@ifnum [1]{%
 \ifnum #1\expandafter \@firstoftwo
 \else \expandafter \@secondoftwo
 \fi
}%
\providecommand \@ifx [1]{%
 \ifx #1\expandafter \@firstoftwo
 \else \expandafter \@secondoftwo
 \fi
}%
\providecommand \natexlab [1]{#1}%
\providecommand \enquote  [1]{``#1''}%
\providecommand \bibnamefont  [1]{#1}%
\providecommand \bibfnamefont [1]{#1}%
\providecommand \citenamefont [1]{#1}%
\providecommand \href@noop [0]{\@secondoftwo}%
\providecommand \href [0]{\begingroup \@sanitize@url \@href}%
\providecommand \@href[1]{\@@startlink{#1}\@@href}%
\providecommand \@@href[1]{\endgroup#1\@@endlink}%
\providecommand \@sanitize@url [0]{\catcode `\\12\catcode `\$12\catcode
  `\&12\catcode `\#12\catcode `\^12\catcode `\_12\catcode `\%12\relax}%
\providecommand \@@startlink[1]{}%
\providecommand \@@endlink[0]{}%
\providecommand \url  [0]{\begingroup\@sanitize@url \@url }%
\providecommand \@url [1]{\endgroup\@href {#1}{\urlprefix }}%
\providecommand \urlprefix  [0]{URL }%
\providecommand \Eprint [0]{\href }%
\providecommand \doibase [0]{http://dx.doi.org/}%
\providecommand \selectlanguage [0]{\@gobble}%
\providecommand \bibinfo  [0]{\@secondoftwo}%
\providecommand \bibfield  [0]{\@secondoftwo}%
\providecommand \translation [1]{[#1]}%
\providecommand \BibitemOpen [0]{}%
\providecommand \bibitemStop [0]{}%
\providecommand \bibitemNoStop [0]{.\EOS\space}%
\providecommand \EOS [0]{\spacefactor3000\relax}%
\providecommand \BibitemShut  [1]{\csname bibitem#1\endcsname}%
\let\auto@bib@innerbib\@empty
\bibitem [{\citenamefont {{Miyazaki}}\ \emph {et~al.}(2018)\citenamefont
  {{Miyazaki}}, \citenamefont {{Komiyama}}, \citenamefont {{Kawanomoto}},
  \citenamefont {{Doi}}, \citenamefont {{Furusawa}}, \citenamefont {{Hamana}},
  \citenamefont {{Hayashi}} \emph {et~al.}}]{2018PASJ...70S...1M}%
  \BibitemOpen
  \bibfield  {author} {\bibinfo {author} {\bibfnamefont {S.}~\bibnamefont
  {{Miyazaki}}}, \bibinfo {author} {\bibfnamefont {Y.}~\bibnamefont
  {{Komiyama}}}, \bibinfo {author} {\bibfnamefont {S.}~\bibnamefont
  {{Kawanomoto}}}, \bibinfo {author} {\bibfnamefont {Y.}~\bibnamefont {{Doi}}},
  \bibinfo {author} {\bibfnamefont {H.}~\bibnamefont {{Furusawa}}}, \bibinfo
  {author} {\bibfnamefont {T.}~\bibnamefont {{Hamana}}}, \bibinfo {author}
  {\bibfnamefont {Y.}~\bibnamefont {{Hayashi}}},  \emph {et~al.},\ }\href
  {\doibase 10.1093/pasj/psx063} {\bibfield  {journal} {\bibinfo  {journal}
  {\pasj}\ }\textbf {\bibinfo {volume} {70}},\ \bibinfo {eid} {S1} (\bibinfo
  {year} {2018})}\BibitemShut {NoStop}%
\bibitem [{\citenamefont {{Aihara}}\ \emph {et~al.}(2018)\citenamefont
  {{Aihara}}, \citenamefont {{Arimoto}}, \citenamefont {{Armstrong}},
  \citenamefont {{Arnouts}}, \citenamefont {{Bahcall}} \emph
  {et~al.}}]{HSCoverview:17}%
  \BibitemOpen
  \bibfield  {author} {\bibinfo {author} {\bibfnamefont {H.}~\bibnamefont
  {{Aihara}}}, \bibinfo {author} {\bibfnamefont {N.}~\bibnamefont {{Arimoto}}},
  \bibinfo {author} {\bibfnamefont {R.}~\bibnamefont {{Armstrong}}}, \bibinfo
  {author} {\bibfnamefont {S.}~\bibnamefont {{Arnouts}}}, \bibinfo {author}
  {\bibfnamefont {N.~A.}\ \bibnamefont {{Bahcall}}},  \emph {et~al.},\ }\href
  {\doibase 10.1093/pasj/psx066} {\bibfield  {journal} {\bibinfo  {journal}
  {\pasj}\ }\textbf {\bibinfo {volume} {70}},\ \bibinfo {eid} {S4} (\bibinfo
  {year} {2018})},\ \Eprint {http://arxiv.org/abs/1704.05858} {arXiv:1704.05858
  [astro-ph.IM]} \BibitemShut {NoStop}%
\bibitem [{\citenamefont {{Hikage}}\ \emph {et~al.}(2019)\citenamefont
  {{Hikage}}, \citenamefont {{Oguri}}, \citenamefont {{Hamana}}, \citenamefont
  {{More}}, \citenamefont {{Mandelbaum}}, \citenamefont {{Takada}},
  \citenamefont {{K{\"o}hlinger}} \emph {et~al.}}]{2019PASJ...71...43H}%
  \BibitemOpen
  \bibfield  {author} {\bibinfo {author} {\bibfnamefont {C.}~\bibnamefont
  {{Hikage}}}, \bibinfo {author} {\bibfnamefont {M.}~\bibnamefont {{Oguri}}},
  \bibinfo {author} {\bibfnamefont {T.}~\bibnamefont {{Hamana}}}, \bibinfo
  {author} {\bibfnamefont {S.}~\bibnamefont {{More}}}, \bibinfo {author}
  {\bibfnamefont {R.}~\bibnamefont {{Mandelbaum}}}, \bibinfo {author}
  {\bibfnamefont {M.}~\bibnamefont {{Takada}}}, \bibinfo {author}
  {\bibfnamefont {F.}~\bibnamefont {{K{\"o}hlinger}}},  \emph {et~al.},\ }\href
  {\doibase 10.1093/pasj/psz010} {\bibfield  {journal} {\bibinfo  {journal}
  {\pasj}\ }\textbf {\bibinfo {volume} {71}},\ \bibinfo {eid} {43} (\bibinfo
  {year} {2019})},\ \Eprint {http://arxiv.org/abs/1809.09148} {arXiv:1809.09148
  [astro-ph.CO]} \BibitemShut {NoStop}%
\bibitem [{\citenamefont {{Hamana}}\ \emph {et~al.}(2020)\citenamefont
  {{Hamana}}, \citenamefont {{Shirasaki}}, \citenamefont {{Miyazaki}},
  \citenamefont {{Hikage}}, \citenamefont {{Oguri}}, \citenamefont {{More}},
  \citenamefont {{Armstrong}} \emph {et~al.}}]{2020PASJ...72...16H}%
  \BibitemOpen
  \bibfield  {author} {\bibinfo {author} {\bibfnamefont {T.}~\bibnamefont
  {{Hamana}}}, \bibinfo {author} {\bibfnamefont {M.}~\bibnamefont
  {{Shirasaki}}}, \bibinfo {author} {\bibfnamefont {S.}~\bibnamefont
  {{Miyazaki}}}, \bibinfo {author} {\bibfnamefont {C.}~\bibnamefont
  {{Hikage}}}, \bibinfo {author} {\bibfnamefont {M.}~\bibnamefont {{Oguri}}},
  \bibinfo {author} {\bibfnamefont {S.}~\bibnamefont {{More}}}, \bibinfo
  {author} {\bibfnamefont {R.}~\bibnamefont {{Armstrong}}},  \emph {et~al.},\
  }\href {\doibase 10.1093/pasj/psz138} {\bibfield  {journal} {\bibinfo
  {journal} {\pasj}\ }\textbf {\bibinfo {volume} {72}},\ \bibinfo {eid} {16}
  (\bibinfo {year} {2020})},\ \Eprint {http://arxiv.org/abs/1906.06041}
  {arXiv:1906.06041 [astro-ph.CO]} \BibitemShut {NoStop}%
\bibitem [{\citenamefont {{Miyatake}}\ \emph
  {et~al.}(2022{\natexlab{a}})\citenamefont {{Miyatake}}, \citenamefont
  {{Sugiyama}}, \citenamefont {{Takada}}, \citenamefont {{Nishimichi}},
  \citenamefont {{Shirasaki}} \emph {et~al.}}]{Miyatake:2021sdd}%
  \BibitemOpen
  \bibfield  {author} {\bibinfo {author} {\bibfnamefont {H.}~\bibnamefont
  {{Miyatake}}}, \bibinfo {author} {\bibfnamefont {S.}~\bibnamefont
  {{Sugiyama}}}, \bibinfo {author} {\bibfnamefont {M.}~\bibnamefont
  {{Takada}}}, \bibinfo {author} {\bibfnamefont {T.}~\bibnamefont
  {{Nishimichi}}}, \bibinfo {author} {\bibfnamefont {M.}~\bibnamefont
  {{Shirasaki}}},  \emph {et~al.},\ }\href {\doibase
  10.1103/PhysRevD.106.083520} {\bibfield  {journal} {\bibinfo  {journal}
  {\prd}\ }\textbf {\bibinfo {volume} {106}},\ \bibinfo {eid} {083520}
  (\bibinfo {year} {2022}{\natexlab{a}})},\ \Eprint
  {http://arxiv.org/abs/2111.02419} {arXiv:2111.02419 [astro-ph.CO]}
  \BibitemShut {NoStop}%
\bibitem [{\citenamefont {{Sugiyama}}\ \emph {et~al.}(2022)\citenamefont
  {{Sugiyama}}, \citenamefont {{Takada}}, \citenamefont {{Miyatake}},
  \citenamefont {{Nishimichi}}, \citenamefont {{Shirasaki}} \emph
  {et~al.}}]{Sugiyama:2021}%
  \BibitemOpen
  \bibfield  {author} {\bibinfo {author} {\bibfnamefont {S.}~\bibnamefont
  {{Sugiyama}}}, \bibinfo {author} {\bibfnamefont {M.}~\bibnamefont
  {{Takada}}}, \bibinfo {author} {\bibfnamefont {H.}~\bibnamefont
  {{Miyatake}}}, \bibinfo {author} {\bibfnamefont {T.}~\bibnamefont
  {{Nishimichi}}}, \bibinfo {author} {\bibfnamefont {M.}~\bibnamefont
  {{Shirasaki}}},  \emph {et~al.},\ }\href {\doibase
  10.1103/PhysRevD.105.123537} {\bibfield  {journal} {\bibinfo  {journal}
  {\prd}\ }\textbf {\bibinfo {volume} {105}},\ \bibinfo {eid} {123537}
  (\bibinfo {year} {2022})},\ \Eprint {http://arxiv.org/abs/2111.10966}
  {arXiv:2111.10966 [astro-ph.CO]} \BibitemShut {NoStop}%
\bibitem [{\citenamefont {{Porredon}}\ \emph {et~al.}(2022)\citenamefont
  {{Porredon}}, \citenamefont {{Crocce}}, \citenamefont {{Elvin-Poole}},
  \citenamefont {{Cawthon}}, \citenamefont {{Giannini}} \emph
  {et~al.}}]{Porredon:2021}%
  \BibitemOpen
  \bibfield  {author} {\bibinfo {author} {\bibfnamefont {A.}~\bibnamefont
  {{Porredon}}}, \bibinfo {author} {\bibfnamefont {M.}~\bibnamefont
  {{Crocce}}}, \bibinfo {author} {\bibfnamefont {J.}~\bibnamefont
  {{Elvin-Poole}}}, \bibinfo {author} {\bibfnamefont {R.}~\bibnamefont
  {{Cawthon}}}, \bibinfo {author} {\bibfnamefont {G.}~\bibnamefont
  {{Giannini}}},  \emph {et~al.},\ }\href {\doibase
  10.1103/PhysRevD.106.103530} {\bibfield  {journal} {\bibinfo  {journal}
  {\prd}\ }\textbf {\bibinfo {volume} {106}},\ \bibinfo {eid} {103530}
  (\bibinfo {year} {2022})},\ \Eprint {http://arxiv.org/abs/2105.13546}
  {arXiv:2105.13546 [astro-ph.CO]} \BibitemShut {NoStop}%
\bibitem [{\citenamefont {{Heymans}}\ \emph {et~al.}(2021)\citenamefont
  {{Heymans}}, \citenamefont {{Tr{\"o}ster}}, \citenamefont {{Asgari}},
  \citenamefont {{Blake}}, \citenamefont {{Hildebrandt}} \emph
  {et~al.}}]{Heymansetal:2021}%
  \BibitemOpen
  \bibfield  {author} {\bibinfo {author} {\bibfnamefont {C.}~\bibnamefont
  {{Heymans}}}, \bibinfo {author} {\bibfnamefont {T.}~\bibnamefont
  {{Tr{\"o}ster}}}, \bibinfo {author} {\bibfnamefont {M.}~\bibnamefont
  {{Asgari}}}, \bibinfo {author} {\bibfnamefont {C.}~\bibnamefont {{Blake}}},
  \bibinfo {author} {\bibfnamefont {H.}~\bibnamefont {{Hildebrandt}}},  \emph
  {et~al.},\ }\href {\doibase 10.1051/0004-6361/202039063} {\bibfield
  {journal} {\bibinfo  {journal} {\aap}\ }\textbf {\bibinfo {volume} {646}},\
  \bibinfo {eid} {A140} (\bibinfo {year} {2021})},\ \Eprint
  {http://arxiv.org/abs/2007.15632} {arXiv:2007.15632 [astro-ph.CO]}
  \BibitemShut {NoStop}%
\bibitem [{\citenamefont {{Planck Collaboration}}\ \emph
  {et~al.}(2020)\citenamefont {{Planck Collaboration}}, \citenamefont
  {{Aghanim}}, \citenamefont {{Akrami}}, \citenamefont {{Ashdown}},
  \citenamefont {{Aumont}}, \citenamefont {{Baccigalupi}}, \citenamefont
  {{Ballardini}} \emph {et~al.}}]{2020A&A...641A...6P}%
  \BibitemOpen
  \bibfield  {author} {\bibinfo {author} {\bibnamefont {{Planck
  Collaboration}}}, \bibinfo {author} {\bibfnamefont {N.}~\bibnamefont
  {{Aghanim}}}, \bibinfo {author} {\bibfnamefont {Y.}~\bibnamefont {{Akrami}}},
  \bibinfo {author} {\bibfnamefont {M.}~\bibnamefont {{Ashdown}}}, \bibinfo
  {author} {\bibfnamefont {J.}~\bibnamefont {{Aumont}}}, \bibinfo {author}
  {\bibfnamefont {C.}~\bibnamefont {{Baccigalupi}}}, \bibinfo {author}
  {\bibfnamefont {M.}~\bibnamefont {{Ballardini}}},  \emph {et~al.},\ }\href
  {\doibase 10.1051/0004-6361/201833910} {\bibfield  {journal} {\bibinfo
  {journal} {\aap}\ }\textbf {\bibinfo {volume} {641}},\ \bibinfo {eid} {A6}
  (\bibinfo {year} {2020})},\ \Eprint {http://arxiv.org/abs/1807.06209}
  {arXiv:1807.06209 [astro-ph.CO]} \BibitemShut {NoStop}%
\bibitem [{\citenamefont {{Adhikari}}\ \emph {et~al.}(2022)\citenamefont
  {{Adhikari}}, \citenamefont {{Anchordoqui}}, \citenamefont {{Fang}},
  \citenamefont {{Sathyaprakash}}, \citenamefont {{Tollefson}} \emph
  {et~al.}}]{2022arXiv220911726A}%
  \BibitemOpen
  \bibfield  {author} {\bibinfo {author} {\bibfnamefont {R.~X.}\ \bibnamefont
  {{Adhikari}}}, \bibinfo {author} {\bibfnamefont {L.~A.}\ \bibnamefont
  {{Anchordoqui}}}, \bibinfo {author} {\bibfnamefont {K.}~\bibnamefont
  {{Fang}}}, \bibinfo {author} {\bibfnamefont {B.~S.}\ \bibnamefont
  {{Sathyaprakash}}}, \bibinfo {author} {\bibfnamefont {K.}~\bibnamefont
  {{Tollefson}}},  \emph {et~al.},\ }\href {\doibase 10.48550/arXiv.2209.11726}
  {\bibfield  {journal} {\bibinfo  {journal} {arXiv e-prints}\ ,\ \bibinfo
  {eid} {arXiv:2209.11726}} (\bibinfo {year} {2022})},\ \Eprint
  {http://arxiv.org/abs/2209.11726} {arXiv:2209.11726 [hep-ph]} \BibitemShut
  {NoStop}%
\bibitem [{\citenamefont {{Kaiser}}(1984)}]{Kaiser:1984}%
  \BibitemOpen
  \bibfield  {author} {\bibinfo {author} {\bibfnamefont {N.}~\bibnamefont
  {{Kaiser}}},\ }\href {\doibase 10.1086/184341} {\bibfield  {journal}
  {\bibinfo  {journal} {\apjl}\ }\textbf {\bibinfo {volume} {284}},\ \bibinfo
  {pages} {L9} (\bibinfo {year} {1984})}\BibitemShut {NoStop}%
\bibitem [{\citenamefont {{Desjacques}}\ \emph {et~al.}(2018)\citenamefont
  {{Desjacques}}, \citenamefont {{Jeong}},\ and\ \citenamefont
  {{Schmidt}}}]{Desjacquesetal:16}%
  \BibitemOpen
  \bibfield  {author} {\bibinfo {author} {\bibfnamefont {V.}~\bibnamefont
  {{Desjacques}}}, \bibinfo {author} {\bibfnamefont {D.}~\bibnamefont
  {{Jeong}}}, \ and\ \bibinfo {author} {\bibfnamefont {F.}~\bibnamefont
  {{Schmidt}}},\ }\href {\doibase 10.1016/j.physrep.2017.12.002} {\bibfield
  {journal} {\bibinfo  {journal} {\physrep}\ }\textbf {\bibinfo {volume}
  {733}},\ \bibinfo {pages} {1} (\bibinfo {year} {2018})},\ \Eprint
  {http://arxiv.org/abs/1611.09787} {arXiv:1611.09787 [astro-ph.CO]}
  \BibitemShut {NoStop}%
\bibitem [{\citenamefont {{Sugiyama}}\ \emph {et~al.}(2020)\citenamefont
  {{Sugiyama}}, \citenamefont {{Takada}}, \citenamefont {{Kobayashi}},
  \citenamefont {{Miyatake}}, \citenamefont {{Shirasaki}} \emph
  {et~al.}}]{Sugiyama:2020}%
  \BibitemOpen
  \bibfield  {author} {\bibinfo {author} {\bibfnamefont {S.}~\bibnamefont
  {{Sugiyama}}}, \bibinfo {author} {\bibfnamefont {M.}~\bibnamefont
  {{Takada}}}, \bibinfo {author} {\bibfnamefont {Y.}~\bibnamefont
  {{Kobayashi}}}, \bibinfo {author} {\bibfnamefont {H.}~\bibnamefont
  {{Miyatake}}}, \bibinfo {author} {\bibfnamefont {M.}~\bibnamefont
  {{Shirasaki}}},  \emph {et~al.},\ }\href {\doibase
  10.1103/PhysRevD.102.083520} {\bibfield  {journal} {\bibinfo  {journal}
  {\prd}\ }\textbf {\bibinfo {volume} {102}},\ \bibinfo {eid} {083520}
  (\bibinfo {year} {2020})},\ \Eprint {http://arxiv.org/abs/2008.06873}
  {arXiv:2008.06873 [astro-ph.CO]} \BibitemShut {NoStop}%
\bibitem [{\citenamefont {{Miyatake}}\ \emph
  {et~al.}(2022{\natexlab{b}})\citenamefont {{Miyatake}}, \citenamefont
  {{Kobayashi}}, \citenamefont {{Takada}}, \citenamefont {{Nishimichi}},
  \citenamefont {{Shirasaki}} \emph {et~al.}}]{2021arXiv210100113M}%
  \BibitemOpen
  \bibfield  {author} {\bibinfo {author} {\bibfnamefont {H.}~\bibnamefont
  {{Miyatake}}}, \bibinfo {author} {\bibfnamefont {Y.}~\bibnamefont
  {{Kobayashi}}}, \bibinfo {author} {\bibfnamefont {M.}~\bibnamefont
  {{Takada}}}, \bibinfo {author} {\bibfnamefont {T.}~\bibnamefont
  {{Nishimichi}}}, \bibinfo {author} {\bibfnamefont {M.}~\bibnamefont
  {{Shirasaki}}},  \emph {et~al.},\ }\href {\doibase
  10.1103/PhysRevD.106.083519} {\bibfield  {journal} {\bibinfo  {journal}
  {\prd}\ }\textbf {\bibinfo {volume} {106}},\ \bibinfo {eid} {083519}
  (\bibinfo {year} {2022}{\natexlab{b}})},\ \Eprint
  {http://arxiv.org/abs/2101.00113} {arXiv:2101.00113 [astro-ph.CO]}
  \BibitemShut {NoStop}%
\bibitem [{\citenamefont {{Bernardeau}}\ \emph {et~al.}(2002)\citenamefont
  {{Bernardeau}}, \citenamefont {{Colombi}}, \citenamefont {{Gazta{\~n}aga}},\
  and\ \citenamefont {{Scoccimarro}}}]{Bernardeauetal:02}%
  \BibitemOpen
  \bibfield  {author} {\bibinfo {author} {\bibfnamefont {F.}~\bibnamefont
  {{Bernardeau}}}, \bibinfo {author} {\bibfnamefont {S.}~\bibnamefont
  {{Colombi}}}, \bibinfo {author} {\bibfnamefont {E.}~\bibnamefont
  {{Gazta{\~n}aga}}}, \ and\ \bibinfo {author} {\bibfnamefont {R.}~\bibnamefont
  {{Scoccimarro}}},\ }\href {\doibase 10.1016/S0370-1573(02)00135-7} {\bibfield
   {journal} {\bibinfo  {journal} {\physrep}\ }\textbf {\bibinfo {volume}
  {367}},\ \bibinfo {pages} {1} (\bibinfo {year} {2002})},\ \Eprint
  {http://arxiv.org/abs/arXiv:astro-ph/0112551} {arXiv:astro-ph/0112551}
  \BibitemShut {NoStop}%
\bibitem [{\citenamefont {{Miyatake}}\ \emph {et~al.}(2023)\citenamefont
  {{Miyatake}}, \citenamefont {{Sugiyama}}, \citenamefont {{Takada}},
  \citenamefont {{Nishimichi}}, \citenamefont {{Li}} \emph
  {et~al.}}]{miyatake2023}%
  \BibitemOpen
  \bibfield  {author} {\bibinfo {author} {\bibfnamefont {H.}~\bibnamefont
  {{Miyatake}}}, \bibinfo {author} {\bibfnamefont {S.}~\bibnamefont
  {{Sugiyama}}}, \bibinfo {author} {\bibfnamefont {M.}~\bibnamefont
  {{Takada}}}, \bibinfo {author} {\bibfnamefont {T.}~\bibnamefont
  {{Nishimichi}}}, \bibinfo {author} {\bibfnamefont {X.}~\bibnamefont {{Li}}},
  \emph {et~al.},\ }\href {\doibase 10.1103/PhysRevD.108.123517} {\bibfield
  {journal} {\bibinfo  {journal} {\prd}\ }\textbf {\bibinfo {volume} {108}},\
  \bibinfo {eid} {123517} (\bibinfo {year} {2023})},\ \Eprint
  {http://arxiv.org/abs/2304.00704} {arXiv:2304.00704 [astro-ph.CO]}
  \BibitemShut {NoStop}%
\bibitem [{\citenamefont {{More}}\ \emph {et~al.}(2023)\citenamefont {{More}},
  \citenamefont {{Sugiyama}}, \citenamefont {{Miyatake}}, \citenamefont
  {{Rau}}, \citenamefont {{Shirasaki}} \emph {et~al.}}]{more2023}%
  \BibitemOpen
  \bibfield  {author} {\bibinfo {author} {\bibfnamefont {S.}~\bibnamefont
  {{More}}}, \bibinfo {author} {\bibfnamefont {S.}~\bibnamefont {{Sugiyama}}},
  \bibinfo {author} {\bibfnamefont {H.}~\bibnamefont {{Miyatake}}}, \bibinfo
  {author} {\bibfnamefont {M.~M.}\ \bibnamefont {{Rau}}}, \bibinfo {author}
  {\bibfnamefont {M.}~\bibnamefont {{Shirasaki}}},  \emph {et~al.},\ }\href
  {\doibase 10.1103/PhysRevD.108.123520} {\bibfield  {journal} {\bibinfo
  {journal} {\prd}\ }\textbf {\bibinfo {volume} {108}},\ \bibinfo {eid}
  {123520} (\bibinfo {year} {2023})},\ \Eprint
  {http://arxiv.org/abs/2304.00703} {arXiv:2304.00703 [astro-ph.CO]}
  \BibitemShut {NoStop}%
\bibitem [{\citenamefont {{Li}}\ \emph {et~al.}(2023)\citenamefont {{Li}},
  \citenamefont {{Zhang}}, \citenamefont {{Sugiyama}}, \citenamefont {{Dalal}},
  \citenamefont {{Terasawa}} \emph {et~al.}}]{li2023}%
  \BibitemOpen
  \bibfield  {author} {\bibinfo {author} {\bibfnamefont {X.}~\bibnamefont
  {{Li}}}, \bibinfo {author} {\bibfnamefont {T.}~\bibnamefont {{Zhang}}},
  \bibinfo {author} {\bibfnamefont {S.}~\bibnamefont {{Sugiyama}}}, \bibinfo
  {author} {\bibfnamefont {R.}~\bibnamefont {{Dalal}}}, \bibinfo {author}
  {\bibfnamefont {R.}~\bibnamefont {{Terasawa}}},  \emph {et~al.},\ }\href
  {\doibase 10.1103/PhysRevD.108.123518} {\bibfield  {journal} {\bibinfo
  {journal} {\prd}\ }\textbf {\bibinfo {volume} {108}},\ \bibinfo {eid}
  {123518} (\bibinfo {year} {2023})},\ \Eprint
  {http://arxiv.org/abs/2304.00702} {arXiv:2304.00702 [astro-ph.CO]}
  \BibitemShut {NoStop}%
\bibitem [{\citenamefont {{Dalal}}\ \emph {et~al.}(2023)\citenamefont
  {{Dalal}}, \citenamefont {{Li}}, \citenamefont {{Nicola}}, \citenamefont
  {{Zuntz}}, \citenamefont {{Strauss}} \emph {et~al.}}]{dalal2023}%
  \BibitemOpen
  \bibfield  {author} {\bibinfo {author} {\bibfnamefont {R.}~\bibnamefont
  {{Dalal}}}, \bibinfo {author} {\bibfnamefont {X.}~\bibnamefont {{Li}}},
  \bibinfo {author} {\bibfnamefont {A.}~\bibnamefont {{Nicola}}}, \bibinfo
  {author} {\bibfnamefont {J.}~\bibnamefont {{Zuntz}}}, \bibinfo {author}
  {\bibfnamefont {M.~A.}\ \bibnamefont {{Strauss}}},  \emph {et~al.},\ }\href
  {\doibase 10.1103/PhysRevD.108.123519} {\bibfield  {journal} {\bibinfo
  {journal} {\prd}\ }\textbf {\bibinfo {volume} {108}},\ \bibinfo {eid}
  {123519} (\bibinfo {year} {2023})},\ \Eprint
  {http://arxiv.org/abs/2304.00701} {arXiv:2304.00701 [astro-ph.CO]}
  \BibitemShut {NoStop}%
\bibitem [{\citenamefont {{Komiyama}}\ \emph {et~al.}(2018)\citenamefont
  {{Komiyama}}, \citenamefont {{Obuchi}}, \citenamefont {{Nakaya}},
  \citenamefont {{Kamata}}, \citenamefont {{Kawanomoto}}, \citenamefont
  {{Utsumi}}, \citenamefont {{Miyazaki}} \emph {et~al.}}]{2018PASJ...70S...2K}%
  \BibitemOpen
  \bibfield  {author} {\bibinfo {author} {\bibfnamefont {Y.}~\bibnamefont
  {{Komiyama}}}, \bibinfo {author} {\bibfnamefont {Y.}~\bibnamefont
  {{Obuchi}}}, \bibinfo {author} {\bibfnamefont {H.}~\bibnamefont {{Nakaya}}},
  \bibinfo {author} {\bibfnamefont {Y.}~\bibnamefont {{Kamata}}}, \bibinfo
  {author} {\bibfnamefont {S.}~\bibnamefont {{Kawanomoto}}}, \bibinfo {author}
  {\bibfnamefont {Y.}~\bibnamefont {{Utsumi}}}, \bibinfo {author}
  {\bibfnamefont {S.}~\bibnamefont {{Miyazaki}}},  \emph {et~al.},\ }\href
  {\doibase 10.1093/pasj/psx069} {\bibfield  {journal} {\bibinfo  {journal}
  {\pasj}\ }\textbf {\bibinfo {volume} {70}},\ \bibinfo {eid} {S2} (\bibinfo
  {year} {2018})}\BibitemShut {NoStop}%
\bibitem [{\citenamefont {{Furusawa}}\ \emph {et~al.}(2018)\citenamefont
  {{Furusawa}}, \citenamefont {{Koike}}, \citenamefont {{Takata}},
  \citenamefont {{Okura}}, \citenamefont {{Miyatake}}, \citenamefont
  {{Lupton}}, \citenamefont {{Bickerton}} \emph
  {et~al.}}]{2018PASJ...70S...3F}%
  \BibitemOpen
  \bibfield  {author} {\bibinfo {author} {\bibfnamefont {H.}~\bibnamefont
  {{Furusawa}}}, \bibinfo {author} {\bibfnamefont {M.}~\bibnamefont {{Koike}}},
  \bibinfo {author} {\bibfnamefont {T.}~\bibnamefont {{Takata}}}, \bibinfo
  {author} {\bibfnamefont {Y.}~\bibnamefont {{Okura}}}, \bibinfo {author}
  {\bibfnamefont {H.}~\bibnamefont {{Miyatake}}}, \bibinfo {author}
  {\bibfnamefont {R.~H.}\ \bibnamefont {{Lupton}}}, \bibinfo {author}
  {\bibfnamefont {S.}~\bibnamefont {{Bickerton}}},  \emph {et~al.},\ }\href
  {\doibase 10.1093/pasj/psx079} {\bibfield  {journal} {\bibinfo  {journal}
  {\pasj}\ }\textbf {\bibinfo {volume} {70}},\ \bibinfo {eid} {S3} (\bibinfo
  {year} {2018})}\BibitemShut {NoStop}%
\bibitem [{\citenamefont {{Kawanomoto}}\ \emph {et~al.}(2018)\citenamefont
  {{Kawanomoto}}, \citenamefont {{Uraguchi}}, \citenamefont {{Komiyama}},
  \citenamefont {{Miyazaki}}, \citenamefont {{Furusawa}}, \citenamefont
  {{Finet}}, \citenamefont {{Hattori}} \emph {et~al.}}]{2018PASJ...70...66K}%
  \BibitemOpen
  \bibfield  {author} {\bibinfo {author} {\bibfnamefont {S.}~\bibnamefont
  {{Kawanomoto}}}, \bibinfo {author} {\bibfnamefont {F.}~\bibnamefont
  {{Uraguchi}}}, \bibinfo {author} {\bibfnamefont {Y.}~\bibnamefont
  {{Komiyama}}}, \bibinfo {author} {\bibfnamefont {S.}~\bibnamefont
  {{Miyazaki}}}, \bibinfo {author} {\bibfnamefont {H.}~\bibnamefont
  {{Furusawa}}}, \bibinfo {author} {\bibfnamefont {F.}~\bibnamefont {{Finet}}},
  \bibinfo {author} {\bibfnamefont {T.}~\bibnamefont {{Hattori}}},  \emph
  {et~al.},\ }\href {\doibase 10.1093/pasj/psy056} {\bibfield  {journal}
  {\bibinfo  {journal} {\pasj}\ }\textbf {\bibinfo {volume} {70}},\ \bibinfo
  {eid} {66} (\bibinfo {year} {2018})}\BibitemShut {NoStop}%
\bibitem [{\citenamefont {{Mandelbaum}}\ \emph
  {et~al.}(2018{\natexlab{a}})\citenamefont {{Mandelbaum}}, \citenamefont
  {{Miyatake}}, \citenamefont {{Hamana}}, \citenamefont {{Oguri}},
  \citenamefont {{Simet}} \emph {et~al.}}]{HSCDR1_shear:17}%
  \BibitemOpen
  \bibfield  {author} {\bibinfo {author} {\bibfnamefont {R.}~\bibnamefont
  {{Mandelbaum}}}, \bibinfo {author} {\bibfnamefont {H.}~\bibnamefont
  {{Miyatake}}}, \bibinfo {author} {\bibfnamefont {T.}~\bibnamefont
  {{Hamana}}}, \bibinfo {author} {\bibfnamefont {M.}~\bibnamefont {{Oguri}}},
  \bibinfo {author} {\bibfnamefont {M.}~\bibnamefont {{Simet}}},  \emph
  {et~al.},\ }\href {\doibase 10.1093/pasj/psx130} {\bibfield  {journal}
  {\bibinfo  {journal} {\pasj}\ }\textbf {\bibinfo {volume} {70}},\ \bibinfo
  {eid} {S25} (\bibinfo {year} {2018}{\natexlab{a}})},\ \Eprint
  {http://arxiv.org/abs/1705.06745} {arXiv:1705.06745 [astro-ph.CO]}
  \BibitemShut {NoStop}%
\bibitem [{\citenamefont {{Mandelbaum}}\ \emph
  {et~al.}(2018{\natexlab{b}})\citenamefont {{Mandelbaum}}, \citenamefont
  {{Lanusse}}, \citenamefont {{Leauthaud}}, \citenamefont {{Armstrong}},
  \citenamefont {{Simet}} \emph {et~al.}}]{2018MNRAS.481.3170M}%
  \BibitemOpen
  \bibfield  {author} {\bibinfo {author} {\bibfnamefont {R.}~\bibnamefont
  {{Mandelbaum}}}, \bibinfo {author} {\bibfnamefont {F.}~\bibnamefont
  {{Lanusse}}}, \bibinfo {author} {\bibfnamefont {A.}~\bibnamefont
  {{Leauthaud}}}, \bibinfo {author} {\bibfnamefont {R.}~\bibnamefont
  {{Armstrong}}}, \bibinfo {author} {\bibfnamefont {M.}~\bibnamefont
  {{Simet}}},  \emph {et~al.},\ }\href {\doibase 10.1093/mnras/sty2420}
  {\bibfield  {journal} {\bibinfo  {journal} {\mnras}\ }\textbf {\bibinfo
  {volume} {481}},\ \bibinfo {pages} {3170} (\bibinfo {year}
  {2018}{\natexlab{b}})},\ \Eprint {http://arxiv.org/abs/1710.00885}
  {arXiv:1710.00885 [astro-ph.CO]} \BibitemShut {NoStop}%
\bibitem [{\citenamefont {{Bosch}}\ \emph {et~al.}(2018)\citenamefont
  {{Bosch}}, \citenamefont {{Armstrong}}, \citenamefont {{Bickerton}},
  \citenamefont {{Furusawa}}, \citenamefont {{Ikeda}} \emph
  {et~al.}}]{2018PASJ...70S...5B}%
  \BibitemOpen
  \bibfield  {author} {\bibinfo {author} {\bibfnamefont {J.}~\bibnamefont
  {{Bosch}}}, \bibinfo {author} {\bibfnamefont {R.}~\bibnamefont
  {{Armstrong}}}, \bibinfo {author} {\bibfnamefont {S.}~\bibnamefont
  {{Bickerton}}}, \bibinfo {author} {\bibfnamefont {H.}~\bibnamefont
  {{Furusawa}}}, \bibinfo {author} {\bibfnamefont {H.}~\bibnamefont {{Ikeda}}},
   \emph {et~al.},\ }\href {\doibase 10.1093/pasj/psx080} {\bibfield  {journal}
  {\bibinfo  {journal} {\pasj}\ }\textbf {\bibinfo {volume} {70}},\ \bibinfo
  {eid} {S5} (\bibinfo {year} {2018})},\ \Eprint
  {http://arxiv.org/abs/1705.06766} {arXiv:1705.06766 [astro-ph.IM]}
  \BibitemShut {NoStop}%
\bibitem [{\citenamefont {{Li}}\ \emph {et~al.}(2022)\citenamefont {{Li}},
  \citenamefont {{Miyatake}}, \citenamefont {{Luo}}, \citenamefont {{More}},
  \citenamefont {{Oguri}} \emph {et~al.}}]{Li2021}%
  \BibitemOpen
  \bibfield  {author} {\bibinfo {author} {\bibfnamefont {X.}~\bibnamefont
  {{Li}}}, \bibinfo {author} {\bibfnamefont {H.}~\bibnamefont {{Miyatake}}},
  \bibinfo {author} {\bibfnamefont {W.}~\bibnamefont {{Luo}}}, \bibinfo
  {author} {\bibfnamefont {S.}~\bibnamefont {{More}}}, \bibinfo {author}
  {\bibfnamefont {M.}~\bibnamefont {{Oguri}}},  \emph {et~al.},\ }\href
  {\doibase 10.1093/pasj/psac006} {\bibfield  {journal} {\bibinfo  {journal}
  {\pasj}\ }\textbf {\bibinfo {volume} {74}},\ \bibinfo {pages} {421} (\bibinfo
  {year} {2022})},\ \Eprint {http://arxiv.org/abs/2107.00136} {arXiv:2107.00136
  [astro-ph.CO]} \BibitemShut {NoStop}%
\bibitem [{\citenamefont {{Nishizawa}}\ \emph {et~al.}(2020)\citenamefont
  {{Nishizawa}}, \citenamefont {{Hsieh}}, \citenamefont {{Tanaka}},\ and\
  \citenamefont {{Takata}}}]{2020arXiv200301511N}%
  \BibitemOpen
  \bibfield  {author} {\bibinfo {author} {\bibfnamefont {A.~J.}\ \bibnamefont
  {{Nishizawa}}}, \bibinfo {author} {\bibfnamefont {B.-C.}\ \bibnamefont
  {{Hsieh}}}, \bibinfo {author} {\bibfnamefont {M.}~\bibnamefont {{Tanaka}}}, \
  and\ \bibinfo {author} {\bibfnamefont {T.}~\bibnamefont {{Takata}}},\
  }\href@noop {} {\bibfield  {journal} {\bibinfo  {journal} {arXiv e-prints}\
  ,\ \bibinfo {eid} {arXiv:2003.01511}} (\bibinfo {year} {2020})},\ \Eprint
  {http://arxiv.org/abs/2003.01511} {arXiv:2003.01511 [astro-ph.GA]}
  \BibitemShut {NoStop}%
\bibitem [{\citenamefont {{Oguri}}(2014)}]{2014MNRAS.444..147O}%
  \BibitemOpen
  \bibfield  {author} {\bibinfo {author} {\bibfnamefont {M.}~\bibnamefont
  {{Oguri}}},\ }\href {\doibase 10.1093/mnras/stu1446} {\bibfield  {journal}
  {\bibinfo  {journal} {\mnras}\ }\textbf {\bibinfo {volume} {444}},\ \bibinfo
  {pages} {147} (\bibinfo {year} {2014})},\ \Eprint
  {http://arxiv.org/abs/1407.4693} {arXiv:1407.4693 [astro-ph.CO]} \BibitemShut
  {NoStop}%
\bibitem [{\citenamefont {{Medezinski}}\ \emph {et~al.}(2018)\citenamefont
  {{Medezinski}}, \citenamefont {{Oguri}}, \citenamefont {{Nishizawa}},
  \citenamefont {{Speagle}}, \citenamefont {{Miyatake}} \emph
  {et~al.}}]{2018PASJ...70...30M}%
  \BibitemOpen
  \bibfield  {author} {\bibinfo {author} {\bibfnamefont {E.}~\bibnamefont
  {{Medezinski}}}, \bibinfo {author} {\bibfnamefont {M.}~\bibnamefont
  {{Oguri}}}, \bibinfo {author} {\bibfnamefont {A.~J.}\ \bibnamefont
  {{Nishizawa}}}, \bibinfo {author} {\bibfnamefont {J.~S.}\ \bibnamefont
  {{Speagle}}}, \bibinfo {author} {\bibfnamefont {H.}~\bibnamefont
  {{Miyatake}}},  \emph {et~al.},\ }\href {\doibase 10.1093/pasj/psy009}
  {\bibfield  {journal} {\bibinfo  {journal} {\pasj}\ }\textbf {\bibinfo
  {volume} {70}},\ \bibinfo {eid} {30} (\bibinfo {year} {2018})},\ \Eprint
  {http://arxiv.org/abs/1706.00427} {arXiv:1706.00427 [astro-ph.CO]}
  \BibitemShut {NoStop}%
\bibitem [{\citenamefont {{Miyatake}}\ \emph {et~al.}(2019)\citenamefont
  {{Miyatake}}, \citenamefont {{Battaglia}}, \citenamefont {{Hilton}},
  \citenamefont {{Medezinski}}, \citenamefont {{Nishizawa}}, \citenamefont
  {{More}}, \citenamefont {{Aiola}} \emph {et~al.}}]{2019ApJ...875...63M}%
  \BibitemOpen
  \bibfield  {author} {\bibinfo {author} {\bibfnamefont {H.}~\bibnamefont
  {{Miyatake}}}, \bibinfo {author} {\bibfnamefont {N.}~\bibnamefont
  {{Battaglia}}}, \bibinfo {author} {\bibfnamefont {M.}~\bibnamefont
  {{Hilton}}}, \bibinfo {author} {\bibfnamefont {E.}~\bibnamefont
  {{Medezinski}}}, \bibinfo {author} {\bibfnamefont {A.~J.}\ \bibnamefont
  {{Nishizawa}}}, \bibinfo {author} {\bibfnamefont {S.}~\bibnamefont {{More}}},
  \bibinfo {author} {\bibfnamefont {S.}~\bibnamefont {{Aiola}}},  \emph
  {et~al.},\ }\href {\doibase 10.3847/1538-4357/ab0af0} {\bibfield  {journal}
  {\bibinfo  {journal} {\apj}\ }\textbf {\bibinfo {volume} {875}},\ \bibinfo
  {eid} {63} (\bibinfo {year} {2019})},\ \Eprint
  {http://arxiv.org/abs/1804.05873} {arXiv:1804.05873 [astro-ph.CO]}
  \BibitemShut {NoStop}%
\bibitem [{\citenamefont {{Alam}}\ \emph {et~al.}(2015)\citenamefont {{Alam}},
  \citenamefont {{Albareti}}, \citenamefont {{Allende Prieto}}, \citenamefont
  {{Anders}}, \citenamefont {{Anderson}} \emph {et~al.}}]{Alam:2015}%
  \BibitemOpen
  \bibfield  {author} {\bibinfo {author} {\bibfnamefont {S.}~\bibnamefont
  {{Alam}}}, \bibinfo {author} {\bibfnamefont {F.~D.}\ \bibnamefont
  {{Albareti}}}, \bibinfo {author} {\bibfnamefont {C.}~\bibnamefont {{Allende
  Prieto}}}, \bibinfo {author} {\bibfnamefont {F.}~\bibnamefont {{Anders}}},
  \bibinfo {author} {\bibfnamefont {S.~F.}\ \bibnamefont {{Anderson}}},  \emph
  {et~al.},\ }\href {\doibase 10.1088/0067-0049/219/1/12} {\bibfield  {journal}
  {\bibinfo  {journal} {\apjs}\ }\textbf {\bibinfo {volume} {219}},\ \bibinfo
  {eid} {12} (\bibinfo {year} {2015})},\ \Eprint
  {http://arxiv.org/abs/1501.00963} {arXiv:1501.00963 [astro-ph.IM]}
  \BibitemShut {NoStop}%
\bibitem [{\citenamefont {{Dawson}}\ \emph {et~al.}(2013)\citenamefont
  {{Dawson}}, \citenamefont {{Schlegel}}, \citenamefont {{Ahn}}, \citenamefont
  {{Anderson}}, \citenamefont {{Aubourg}}, \citenamefont {{Bailey}},
  \citenamefont {{Barkhouser}} \emph {et~al.}}]{2013AJ....145...10D}%
  \BibitemOpen
  \bibfield  {author} {\bibinfo {author} {\bibfnamefont {K.~S.}\ \bibnamefont
  {{Dawson}}}, \bibinfo {author} {\bibfnamefont {D.~J.}\ \bibnamefont
  {{Schlegel}}}, \bibinfo {author} {\bibfnamefont {C.~P.}\ \bibnamefont
  {{Ahn}}}, \bibinfo {author} {\bibfnamefont {S.~F.}\ \bibnamefont
  {{Anderson}}}, \bibinfo {author} {\bibfnamefont {{\'E}.}~\bibnamefont
  {{Aubourg}}}, \bibinfo {author} {\bibfnamefont {S.}~\bibnamefont {{Bailey}}},
  \bibinfo {author} {\bibfnamefont {R.~H.}\ \bibnamefont {{Barkhouser}}},
  \emph {et~al.},\ }\href {\doibase 10.1088/0004-6256/145/1/10} {\bibfield
  {journal} {\bibinfo  {journal} {\aj}\ }\textbf {\bibinfo {volume} {145}},\
  \bibinfo {eid} {10} (\bibinfo {year} {2013})},\ \Eprint
  {http://arxiv.org/abs/1208.0022} {arXiv:1208.0022 [astro-ph.CO]} \BibitemShut
  {NoStop}%
\bibitem [{\citenamefont {{Abazajian}}\ \emph {et~al.}(2009)\citenamefont
  {{Abazajian}}, \citenamefont {{Adelman-McCarthy}}, \citenamefont
  {{Ag{\"u}eros}}, \citenamefont {{Allam}}, \citenamefont {{Allende Prieto}}
  \emph {et~al.}}]{2009ApJS..182..543A}%
  \BibitemOpen
  \bibfield  {author} {\bibinfo {author} {\bibfnamefont {K.~N.}\ \bibnamefont
  {{Abazajian}}}, \bibinfo {author} {\bibfnamefont {J.~K.}\ \bibnamefont
  {{Adelman-McCarthy}}}, \bibinfo {author} {\bibfnamefont {M.~A.}\ \bibnamefont
  {{Ag{\"u}eros}}}, \bibinfo {author} {\bibfnamefont {S.~S.}\ \bibnamefont
  {{Allam}}}, \bibinfo {author} {\bibfnamefont {C.}~\bibnamefont {{Allende
  Prieto}}},  \emph {et~al.},\ }\href {\doibase 10.1088/0067-0049/182/2/543}
  {\bibfield  {journal} {\bibinfo  {journal} {\apjs}\ }\textbf {\bibinfo
  {volume} {182}},\ \bibinfo {pages} {543} (\bibinfo {year} {2009})},\ \Eprint
  {http://arxiv.org/abs/0812.0649} {arXiv:0812.0649 [astro-ph]} \BibitemShut
  {NoStop}%
\bibitem [{\citenamefont {{Gunn}}\ \emph {et~al.}(2006)\citenamefont {{Gunn}},
  \citenamefont {{Siegmund}}, \citenamefont {{Mannery}}, \citenamefont
  {{Owen}}, \citenamefont {{Hull}} \emph {et~al.}}]{2006AJ....131.2332G}%
  \BibitemOpen
  \bibfield  {author} {\bibinfo {author} {\bibfnamefont {J.~E.}\ \bibnamefont
  {{Gunn}}}, \bibinfo {author} {\bibfnamefont {W.~A.}\ \bibnamefont
  {{Siegmund}}}, \bibinfo {author} {\bibfnamefont {E.~J.}\ \bibnamefont
  {{Mannery}}}, \bibinfo {author} {\bibfnamefont {R.~E.}\ \bibnamefont
  {{Owen}}}, \bibinfo {author} {\bibfnamefont {C.~L.}\ \bibnamefont {{Hull}}},
  \emph {et~al.},\ }\href {\doibase 10.1086/500975} {\bibfield  {journal}
  {\bibinfo  {journal} {\aj}\ }\textbf {\bibinfo {volume} {131}},\ \bibinfo
  {pages} {2332} (\bibinfo {year} {2006})},\ \Eprint
  {http://arxiv.org/abs/astro-ph/0602326} {arXiv:astro-ph/0602326 [astro-ph]}
  \BibitemShut {NoStop}%
\bibitem [{\citenamefont {{Fukugita}}\ \emph {et~al.}(1996)\citenamefont
  {{Fukugita}}, \citenamefont {{Ichikawa}}, \citenamefont {{Gunn}},
  \citenamefont {{Doi}}, \citenamefont {{Shimasaku}} \emph
  {et~al.}}]{1996AJ....111.1748F}%
  \BibitemOpen
  \bibfield  {author} {\bibinfo {author} {\bibfnamefont {M.}~\bibnamefont
  {{Fukugita}}}, \bibinfo {author} {\bibfnamefont {T.}~\bibnamefont
  {{Ichikawa}}}, \bibinfo {author} {\bibfnamefont {J.~E.}\ \bibnamefont
  {{Gunn}}}, \bibinfo {author} {\bibfnamefont {M.}~\bibnamefont {{Doi}}},
  \bibinfo {author} {\bibfnamefont {K.}~\bibnamefont {{Shimasaku}}},  \emph
  {et~al.},\ }\href {\doibase 10.1086/117915} {\bibfield  {journal} {\bibinfo
  {journal} {\aj}\ }\textbf {\bibinfo {volume} {111}},\ \bibinfo {pages} {1748}
  (\bibinfo {year} {1996})}\BibitemShut {NoStop}%
\bibitem [{\citenamefont {{Smith}}\ \emph {et~al.}(2002)\citenamefont
  {{Smith}}, \citenamefont {{Tucker}}, \citenamefont {{Kent}}, \citenamefont
  {{Richmond}}, \citenamefont {{Fukugita}} \emph
  {et~al.}}]{2002AJ....123.2121S}%
  \BibitemOpen
  \bibfield  {author} {\bibinfo {author} {\bibfnamefont {J.~A.}\ \bibnamefont
  {{Smith}}}, \bibinfo {author} {\bibfnamefont {D.~L.}\ \bibnamefont
  {{Tucker}}}, \bibinfo {author} {\bibfnamefont {S.}~\bibnamefont {{Kent}}},
  \bibinfo {author} {\bibfnamefont {M.~W.}\ \bibnamefont {{Richmond}}},
  \bibinfo {author} {\bibfnamefont {M.}~\bibnamefont {{Fukugita}}},  \emph
  {et~al.},\ }\href {\doibase 10.1086/339311} {\bibfield  {journal} {\bibinfo
  {journal} {\aj}\ }\textbf {\bibinfo {volume} {123}},\ \bibinfo {pages} {2121}
  (\bibinfo {year} {2002})},\ \Eprint {http://arxiv.org/abs/astro-ph/0201143}
  {arXiv:astro-ph/0201143 [astro-ph]} \BibitemShut {NoStop}%
\bibitem [{\citenamefont {{Doi}}\ \emph {et~al.}(2010)\citenamefont {{Doi}},
  \citenamefont {{Tanaka}}, \citenamefont {{Fukugita}}, \citenamefont {{Gunn}},
  \citenamefont {{Yasuda}} \emph {et~al.}}]{2010AJ....139.1628D}%
  \BibitemOpen
  \bibfield  {author} {\bibinfo {author} {\bibfnamefont {M.}~\bibnamefont
  {{Doi}}}, \bibinfo {author} {\bibfnamefont {M.}~\bibnamefont {{Tanaka}}},
  \bibinfo {author} {\bibfnamefont {M.}~\bibnamefont {{Fukugita}}}, \bibinfo
  {author} {\bibfnamefont {J.~E.}\ \bibnamefont {{Gunn}}}, \bibinfo {author}
  {\bibfnamefont {N.}~\bibnamefont {{Yasuda}}},  \emph {et~al.},\ }\href
  {\doibase 10.1088/0004-6256/139/4/1628} {\bibfield  {journal} {\bibinfo
  {journal} {\aj}\ }\textbf {\bibinfo {volume} {139}},\ \bibinfo {pages} {1628}
  (\bibinfo {year} {2010})},\ \Eprint {http://arxiv.org/abs/1002.3701}
  {arXiv:1002.3701 [astro-ph.IM]} \BibitemShut {NoStop}%
\bibitem [{\citenamefont {{Eisenstein}}\ \emph {et~al.}(2011)\citenamefont
  {{Eisenstein}}, \citenamefont {{Weinberg}}, \citenamefont {{Agol}},
  \citenamefont {{Aihara}}, \citenamefont {{Allende Prieto}} \emph
  {et~al.}}]{2011AJ....142...72E}%
  \BibitemOpen
  \bibfield  {author} {\bibinfo {author} {\bibfnamefont {D.~J.}\ \bibnamefont
  {{Eisenstein}}}, \bibinfo {author} {\bibfnamefont {D.~H.}\ \bibnamefont
  {{Weinberg}}}, \bibinfo {author} {\bibfnamefont {E.}~\bibnamefont {{Agol}}},
  \bibinfo {author} {\bibfnamefont {H.}~\bibnamefont {{Aihara}}}, \bibinfo
  {author} {\bibfnamefont {C.}~\bibnamefont {{Allende Prieto}}},  \emph
  {et~al.},\ }\href {\doibase 10.1088/0004-6256/142/3/72} {\bibfield  {journal}
  {\bibinfo  {journal} {\aj}\ }\textbf {\bibinfo {volume} {142}},\ \bibinfo
  {eid} {72} (\bibinfo {year} {2011})},\ \Eprint
  {http://arxiv.org/abs/1101.1529} {arXiv:1101.1529 [astro-ph.IM]} \BibitemShut
  {NoStop}%
\bibitem [{\citenamefont {{Ahn}}\ \emph {et~al.}(2012)\citenamefont {{Ahn}},
  \citenamefont {{Alexandroff}}, \citenamefont {{Allende Prieto}},
  \citenamefont {{Anderson}}, \citenamefont {{Anderton}} \emph
  {et~al.}}]{2012ApJS..203...21A}%
  \BibitemOpen
  \bibfield  {author} {\bibinfo {author} {\bibfnamefont {C.~P.}\ \bibnamefont
  {{Ahn}}}, \bibinfo {author} {\bibfnamefont {R.}~\bibnamefont
  {{Alexandroff}}}, \bibinfo {author} {\bibfnamefont {C.}~\bibnamefont
  {{Allende Prieto}}}, \bibinfo {author} {\bibfnamefont {S.~F.}\ \bibnamefont
  {{Anderson}}}, \bibinfo {author} {\bibfnamefont {T.}~\bibnamefont
  {{Anderton}}},  \emph {et~al.},\ }\href {\doibase 10.1088/0067-0049/203/2/21}
  {\bibfield  {journal} {\bibinfo  {journal} {\apjs}\ }\textbf {\bibinfo
  {volume} {203}},\ \bibinfo {eid} {21} (\bibinfo {year} {2012})},\ \Eprint
  {http://arxiv.org/abs/1207.7137} {arXiv:1207.7137 [astro-ph.IM]} \BibitemShut
  {NoStop}%
\bibitem [{\citenamefont {{Aihara}}\ \emph {et~al.}(2011)\citenamefont
  {{Aihara}}, \citenamefont {{Allende Prieto}}, \citenamefont {{An}},
  \citenamefont {{Anderson}}, \citenamefont {{Aubourg}} \emph
  {et~al.}}]{2011ApJS..193...29A}%
  \BibitemOpen
  \bibfield  {author} {\bibinfo {author} {\bibfnamefont {H.}~\bibnamefont
  {{Aihara}}}, \bibinfo {author} {\bibfnamefont {C.}~\bibnamefont {{Allende
  Prieto}}}, \bibinfo {author} {\bibfnamefont {D.}~\bibnamefont {{An}}},
  \bibinfo {author} {\bibfnamefont {S.~F.}\ \bibnamefont {{Anderson}}},
  \bibinfo {author} {\bibfnamefont {{\'E}.}~\bibnamefont {{Aubourg}}},  \emph
  {et~al.},\ }\href {\doibase 10.1088/0067-0049/193/2/29} {\bibfield  {journal}
  {\bibinfo  {journal} {\apjs}\ }\textbf {\bibinfo {volume} {193}},\ \bibinfo
  {eid} {29} (\bibinfo {year} {2011})},\ \Eprint
  {http://arxiv.org/abs/1101.1559} {arXiv:1101.1559 [astro-ph.IM]} \BibitemShut
  {NoStop}%
\bibitem [{\citenamefont {{Lupton}}\ \emph {et~al.}(2001)\citenamefont
  {{Lupton}}, \citenamefont {{Gunn}}, \citenamefont {{Ivezi{\'c}}},
  \citenamefont {{Knapp}},\ and\ \citenamefont {{Kent}}}]{2001ASPC..238..269L}%
  \BibitemOpen
  \bibfield  {author} {\bibinfo {author} {\bibfnamefont {R.}~\bibnamefont
  {{Lupton}}}, \bibinfo {author} {\bibfnamefont {J.~E.}\ \bibnamefont
  {{Gunn}}}, \bibinfo {author} {\bibfnamefont {Z.}~\bibnamefont
  {{Ivezi{\'c}}}}, \bibinfo {author} {\bibfnamefont {G.~R.}\ \bibnamefont
  {{Knapp}}}, \ and\ \bibinfo {author} {\bibfnamefont {S.~o.}\ \bibnamefont
  {{Kent}}},\ }in\ \href@noop {} {\emph {\bibinfo {booktitle} {Astronomical
  Data Analysis Software and Systems X}}},\ \bibinfo {series} {Astronomical
  Society of the Pacific Conference Series}, Vol.\ \bibinfo {volume} {238},\
  \bibinfo {editor} {edited by\ \bibinfo {editor} {\bibfnamefont
  {J.}~\bibnamefont {{Harnden}}, \bibfnamefont {F.~R.}}, \bibinfo {editor}
  {\bibfnamefont {F.~A.}\ \bibnamefont {{Primini}}}, \ and\ \bibinfo {editor}
  {\bibfnamefont {H.~E.}\ \bibnamefont {{Payne}}}}\ (\bibinfo {year} {2001})\
  p.\ \bibinfo {pages} {269},\ \Eprint {http://arxiv.org/abs/astro-ph/0101420}
  {arXiv:astro-ph/0101420 [astro-ph]} \BibitemShut {NoStop}%
\bibitem [{\citenamefont {{Pier}}\ \emph {et~al.}(2003)\citenamefont {{Pier}},
  \citenamefont {{Munn}}, \citenamefont {{Hindsley}}, \citenamefont
  {{Hennessy}}, \citenamefont {{Kent}} \emph {et~al.}}]{2003AJ....125.1559P}%
  \BibitemOpen
  \bibfield  {author} {\bibinfo {author} {\bibfnamefont {J.~R.}\ \bibnamefont
  {{Pier}}}, \bibinfo {author} {\bibfnamefont {J.~A.}\ \bibnamefont {{Munn}}},
  \bibinfo {author} {\bibfnamefont {R.~B.}\ \bibnamefont {{Hindsley}}},
  \bibinfo {author} {\bibfnamefont {G.~S.}\ \bibnamefont {{Hennessy}}},
  \bibinfo {author} {\bibfnamefont {S.~M.}\ \bibnamefont {{Kent}}},  \emph
  {et~al.},\ }\href {\doibase 10.1086/346138} {\bibfield  {journal} {\bibinfo
  {journal} {\aj}\ }\textbf {\bibinfo {volume} {125}},\ \bibinfo {pages} {1559}
  (\bibinfo {year} {2003})},\ \Eprint {http://arxiv.org/abs/astro-ph/0211375}
  {arXiv:astro-ph/0211375 [astro-ph]} \BibitemShut {NoStop}%
\bibitem [{\citenamefont {{Padmanabhan}}\ \emph {et~al.}(2008)\citenamefont
  {{Padmanabhan}}, \citenamefont {{Schlegel}}, \citenamefont {{Finkbeiner}},
  \citenamefont {{Barentine}}, \citenamefont {{Blanton}} \emph
  {et~al.}}]{2008ApJ...674.1217P}%
  \BibitemOpen
  \bibfield  {author} {\bibinfo {author} {\bibfnamefont {N.}~\bibnamefont
  {{Padmanabhan}}}, \bibinfo {author} {\bibfnamefont {D.~J.}\ \bibnamefont
  {{Schlegel}}}, \bibinfo {author} {\bibfnamefont {D.~P.}\ \bibnamefont
  {{Finkbeiner}}}, \bibinfo {author} {\bibfnamefont {J.~C.}\ \bibnamefont
  {{Barentine}}}, \bibinfo {author} {\bibfnamefont {M.~R.}\ \bibnamefont
  {{Blanton}}},  \emph {et~al.},\ }\href {\doibase 10.1086/524677} {\bibfield
  {journal} {\bibinfo  {journal} {\apj}\ }\textbf {\bibinfo {volume} {674}},\
  \bibinfo {pages} {1217} (\bibinfo {year} {2008})},\ \Eprint
  {http://arxiv.org/abs/astro-ph/0703454} {arXiv:astro-ph/0703454 [astro-ph]}
  \BibitemShut {NoStop}%
\bibitem [{\citenamefont {{Schlegel}}\ \emph {et~al.}(1998)\citenamefont
  {{Schlegel}}, \citenamefont {{Finkbeiner}},\ and\ \citenamefont
  {{Davis}}}]{1998ApJ...500..525S}%
  \BibitemOpen
  \bibfield  {author} {\bibinfo {author} {\bibfnamefont {D.~J.}\ \bibnamefont
  {{Schlegel}}}, \bibinfo {author} {\bibfnamefont {D.~P.}\ \bibnamefont
  {{Finkbeiner}}}, \ and\ \bibinfo {author} {\bibfnamefont {M.}~\bibnamefont
  {{Davis}}},\ }\href {\doibase 10.1086/305772} {\bibfield  {journal} {\bibinfo
   {journal} {\apj}\ }\textbf {\bibinfo {volume} {500}},\ \bibinfo {pages}
  {525} (\bibinfo {year} {1998})},\ \Eprint
  {http://arxiv.org/abs/astro-ph/9710327} {arXiv:astro-ph/9710327 [astro-ph]}
  \BibitemShut {NoStop}%
\bibitem [{\citenamefont {{Bolton}}\ \emph {et~al.}(2012)\citenamefont
  {{Bolton}}, \citenamefont {{Schlegel}}, \citenamefont {{Aubourg}},
  \citenamefont {{Bailey}}, \citenamefont {{Bhardwaj}} \emph
  {et~al.}}]{2012AJ....144..144B}%
  \BibitemOpen
  \bibfield  {author} {\bibinfo {author} {\bibfnamefont {A.~S.}\ \bibnamefont
  {{Bolton}}}, \bibinfo {author} {\bibfnamefont {D.~J.}\ \bibnamefont
  {{Schlegel}}}, \bibinfo {author} {\bibfnamefont {{\'E}.}~\bibnamefont
  {{Aubourg}}}, \bibinfo {author} {\bibfnamefont {S.}~\bibnamefont {{Bailey}}},
  \bibinfo {author} {\bibfnamefont {V.}~\bibnamefont {{Bhardwaj}}},  \emph
  {et~al.},\ }\href {\doibase 10.1088/0004-6256/144/5/144} {\bibfield
  {journal} {\bibinfo  {journal} {\aj}\ }\textbf {\bibinfo {volume} {144}},\
  \bibinfo {eid} {144} (\bibinfo {year} {2012})},\ \Eprint
  {http://arxiv.org/abs/1207.7326} {arXiv:1207.7326 [astro-ph.CO]} \BibitemShut
  {NoStop}%
\bibitem [{\citenamefont {{van den Bosch}}\ \emph {et~al.}(2013)\citenamefont
  {{van den Bosch}}, \citenamefont {{More}}, \citenamefont {{Cacciato}},
  \citenamefont {{Mo}},\ and\ \citenamefont {{Yang}}}]{vandenBosch:2013}%
  \BibitemOpen
  \bibfield  {author} {\bibinfo {author} {\bibfnamefont {F.~C.}\ \bibnamefont
  {{van den Bosch}}}, \bibinfo {author} {\bibfnamefont {S.}~\bibnamefont
  {{More}}}, \bibinfo {author} {\bibfnamefont {M.}~\bibnamefont {{Cacciato}}},
  \bibinfo {author} {\bibfnamefont {H.}~\bibnamefont {{Mo}}}, \ and\ \bibinfo
  {author} {\bibfnamefont {X.~o.}\ \bibnamefont {{Yang}}},\ }\href {\doibase
  10.1093/mnras/sts006} {\bibfield  {journal} {\bibinfo  {journal} {\mnras}\
  }\textbf {\bibinfo {volume} {430}},\ \bibinfo {pages} {725} (\bibinfo {year}
  {2013})},\ \Eprint {http://arxiv.org/abs/1206.6890} {arXiv:1206.6890
  [astro-ph.CO]} \BibitemShut {NoStop}%
\bibitem [{\citenamefont {{Smith}}\ \emph {et~al.}(2003)\citenamefont
  {{Smith}}, \citenamefont {{Peacock}}, \citenamefont {{Jenkins}},
  \citenamefont {{White}}, \citenamefont {{Frenk}} \emph
  {et~al.}}]{Smithetal:03}%
  \BibitemOpen
  \bibfield  {author} {\bibinfo {author} {\bibfnamefont {R.~E.}\ \bibnamefont
  {{Smith}}}, \bibinfo {author} {\bibfnamefont {J.~A.}\ \bibnamefont
  {{Peacock}}}, \bibinfo {author} {\bibfnamefont {A.}~\bibnamefont
  {{Jenkins}}}, \bibinfo {author} {\bibfnamefont {S.~D.~M.}\ \bibnamefont
  {{White}}}, \bibinfo {author} {\bibfnamefont {C.~S.}\ \bibnamefont
  {{Frenk}}},  \emph {et~al.},\ }\href {\doibase
  10.1046/j.1365-8711.2003.06503.x} {\bibfield  {journal} {\bibinfo  {journal}
  {\mnras}\ }\textbf {\bibinfo {volume} {341}},\ \bibinfo {pages} {1311}
  (\bibinfo {year} {2003})},\ \Eprint
  {http://arxiv.org/abs/arXiv:astro-ph/0207664} {arXiv:astro-ph/0207664}
  \BibitemShut {NoStop}%
\bibitem [{\citenamefont {Takahashi}\ \emph {et~al.}(2012)\citenamefont
  {Takahashi}, \citenamefont {Sato}, \citenamefont {Nishimichi}, \citenamefont
  {Taruya},\ and\ \citenamefont {Oguri}}]{Takahashi_2012}%
  \BibitemOpen
  \bibfield  {author} {\bibinfo {author} {\bibfnamefont {R.}~\bibnamefont
  {Takahashi}}, \bibinfo {author} {\bibfnamefont {M.}~\bibnamefont {Sato}},
  \bibinfo {author} {\bibfnamefont {T.}~\bibnamefont {Nishimichi}}, \bibinfo
  {author} {\bibfnamefont {A.}~\bibnamefont {Taruya}}, \ and\ \bibinfo {author}
  {\bibfnamefont {M.~o.}\ \bibnamefont {Oguri}},\ }\href {\doibase
  10.1088/0004-637x/761/2/152} {\bibfield  {journal} {\bibinfo  {journal} {The
  Astrophysical Journal}\ }\textbf {\bibinfo {volume} {761}},\ \bibinfo {pages}
  {152} (\bibinfo {year} {2012})}\BibitemShut {NoStop}%
\bibitem [{\citenamefont {{Hirata}}\ \emph {et~al.}(2004)\citenamefont
  {{Hirata}}, \citenamefont {{Mandelbaum}}, \citenamefont {{Seljak}},
  \citenamefont {{Guzik}}, \citenamefont {{Padmanabhan}} \emph
  {et~al.}}]{Hirata:2004}%
  \BibitemOpen
  \bibfield  {author} {\bibinfo {author} {\bibfnamefont {C.~M.}\ \bibnamefont
  {{Hirata}}}, \bibinfo {author} {\bibfnamefont {R.}~\bibnamefont
  {{Mandelbaum}}}, \bibinfo {author} {\bibfnamefont {U.}~\bibnamefont
  {{Seljak}}}, \bibinfo {author} {\bibfnamefont {J.}~\bibnamefont {{Guzik}}},
  \bibinfo {author} {\bibfnamefont {N.}~\bibnamefont {{Padmanabhan}}},  \emph
  {et~al.},\ }\href {\doibase 10.1111/j.1365-2966.2004.08090.x} {\bibfield
  {journal} {\bibinfo  {journal} {\mnras}\ }\textbf {\bibinfo {volume} {353}},\
  \bibinfo {pages} {529} (\bibinfo {year} {2004})},\ \Eprint
  {http://arxiv.org/abs/astro-ph/0403255} {arXiv:astro-ph/0403255 [astro-ph]}
  \BibitemShut {NoStop}%
\bibitem [{\citenamefont {{Heavens}}\ \emph {et~al.}(2000)\citenamefont
  {{Heavens}}, \citenamefont {{Refregier}},\ and\ \citenamefont
  {{Heymans}}}]{Heavensetal:00}%
  \BibitemOpen
  \bibfield  {author} {\bibinfo {author} {\bibfnamefont {A.}~\bibnamefont
  {{Heavens}}}, \bibinfo {author} {\bibfnamefont {A.}~\bibnamefont
  {{Refregier}}}, \ and\ \bibinfo {author} {\bibfnamefont {C.}~\bibnamefont
  {{Heymans}}},\ }\href {\doibase 10.1046/j.1365-8711.2000.03907.x} {\bibfield
  {journal} {\bibinfo  {journal} {\mnras}\ }\textbf {\bibinfo {volume} {319}},\
  \bibinfo {pages} {649} (\bibinfo {year} {2000})},\ \Eprint
  {http://arxiv.org/abs/astro-ph/0005269} {astro-ph/0005269} \BibitemShut
  {NoStop}%
\bibitem [{\citenamefont {{Bridle}}\ and\ \citenamefont
  {{King}}(2007)}]{2007NJPh....9..444B}%
  \BibitemOpen
  \bibfield  {author} {\bibinfo {author} {\bibfnamefont {S.}~\bibnamefont
  {{Bridle}}}\ and\ \bibinfo {author} {\bibfnamefont {L.}~\bibnamefont
  {{King}}},\ }\href {\doibase 10.1088/1367-2630/9/12/444} {\bibfield
  {journal} {\bibinfo  {journal} {New Journal of Physics}\ }\textbf {\bibinfo
  {volume} {9}},\ \bibinfo {pages} {444} (\bibinfo {year} {2007})},\ \Eprint
  {http://arxiv.org/abs/0705.0166} {arXiv:0705.0166 [astro-ph]} \BibitemShut
  {NoStop}%
\bibitem [{\citenamefont {{Huterer}}\ \emph {et~al.}(2006)\citenamefont
  {{Huterer}}, \citenamefont {{Takada}}, \citenamefont {{Bernstein}},\ and\
  \citenamefont {{Jain}}}]{Hutereretal:06}%
  \BibitemOpen
  \bibfield  {author} {\bibinfo {author} {\bibfnamefont {D.}~\bibnamefont
  {{Huterer}}}, \bibinfo {author} {\bibfnamefont {M.}~\bibnamefont {{Takada}}},
  \bibinfo {author} {\bibfnamefont {G.}~\bibnamefont {{Bernstein}}}, \ and\
  \bibinfo {author} {\bibfnamefont {B.}~\bibnamefont {{Jain}}},\ }\href
  {\doibase 10.1111/j.1365-2966.2005.09782.x} {\bibfield  {journal} {\bibinfo
  {journal} {\mnras}\ }\textbf {\bibinfo {volume} {366}},\ \bibinfo {pages}
  {101} (\bibinfo {year} {2006})},\ \Eprint
  {http://arxiv.org/abs/arXiv:astro-ph/0506030} {arXiv:astro-ph/0506030}
  \BibitemShut {NoStop}%
\bibitem [{\citenamefont {{Oguri}}\ and\ \citenamefont
  {{Takada}}(2011)}]{OguriTakada:11}%
  \BibitemOpen
  \bibfield  {author} {\bibinfo {author} {\bibfnamefont {M.}~\bibnamefont
  {{Oguri}}}\ and\ \bibinfo {author} {\bibfnamefont {M.}~\bibnamefont
  {{Takada}}},\ }\href {\doibase 10.1103/PhysRevD.83.023008} {\bibfield
  {journal} {\bibinfo  {journal} {\prd}\ }\textbf {\bibinfo {volume} {83}},\
  \bibinfo {pages} {023008} (\bibinfo {year} {2011})},\ \Eprint
  {http://arxiv.org/abs/1010.0744} {arXiv:1010.0744 [astro-ph.CO]} \BibitemShut
  {NoStop}%
\bibitem [{\citenamefont {Zhang}\ \emph {et~al.}(2022)\citenamefont {Zhang},
  \citenamefont {Li}, \citenamefont {Dalal}, \citenamefont {Mandelbaum},
  \citenamefont {Strauss} \emph {et~al.}}]{zhang:2022dvs}%
  \BibitemOpen
  \bibfield  {author} {\bibinfo {author} {\bibfnamefont {T.}~\bibnamefont
  {Zhang}}, \bibinfo {author} {\bibfnamefont {X.}~\bibnamefont {Li}}, \bibinfo
  {author} {\bibfnamefont {R.}~\bibnamefont {Dalal}}, \bibinfo {author}
  {\bibfnamefont {R.}~\bibnamefont {Mandelbaum}}, \bibinfo {author}
  {\bibfnamefont {M.~A.}\ \bibnamefont {Strauss}},  \emph {et~al.},\
  }\href@noop {} {\  (\bibinfo {year} {2022})},\ \Eprint
  {http://arxiv.org/abs/2212.03257} {arXiv:2212.03257 [astro-ph.CO]}
  \BibitemShut {NoStop}%
\bibitem [{\citenamefont {{More}}\ \emph {et~al.}(2015)\citenamefont {{More}},
  \citenamefont {{Miyatake}}, \citenamefont {{Mandelbaum}}, \citenamefont
  {{Takada}}, \citenamefont {{Spergel}} \emph {et~al.}}]{2015ApJ...806....2M}%
  \BibitemOpen
  \bibfield  {author} {\bibinfo {author} {\bibfnamefont {S.}~\bibnamefont
  {{More}}}, \bibinfo {author} {\bibfnamefont {H.}~\bibnamefont {{Miyatake}}},
  \bibinfo {author} {\bibfnamefont {R.}~\bibnamefont {{Mandelbaum}}}, \bibinfo
  {author} {\bibfnamefont {M.}~\bibnamefont {{Takada}}}, \bibinfo {author}
  {\bibfnamefont {D.~N.}\ \bibnamefont {{Spergel}}},  \emph {et~al.},\ }\href
  {\doibase 10.1088/0004-637X/806/1/2} {\bibfield  {journal} {\bibinfo
  {journal} {\apj}\ }\textbf {\bibinfo {volume} {806}},\ \bibinfo {eid} {2}
  (\bibinfo {year} {2015})},\ \Eprint {http://arxiv.org/abs/1407.1856}
  {arXiv:1407.1856} \BibitemShut {NoStop}%
\bibitem [{\citenamefont {{Troxel}}\ \emph {et~al.}(2018)\citenamefont
  {{Troxel}}, \citenamefont {{MacCrann}}, \citenamefont {{Zuntz}},
  \citenamefont {{Eifler}}, \citenamefont {{Krause}} \emph
  {et~al.}}]{2018PhRvD..98d3528T}%
  \BibitemOpen
  \bibfield  {author} {\bibinfo {author} {\bibfnamefont {M.~A.}\ \bibnamefont
  {{Troxel}}}, \bibinfo {author} {\bibfnamefont {N.}~\bibnamefont
  {{MacCrann}}}, \bibinfo {author} {\bibfnamefont {J.}~\bibnamefont {{Zuntz}}},
  \bibinfo {author} {\bibfnamefont {T.~F.}\ \bibnamefont {{Eifler}}}, \bibinfo
  {author} {\bibfnamefont {E.}~\bibnamefont {{Krause}}},  \emph {et~al.},\
  }\href {\doibase 10.1103/PhysRevD.98.043528} {\bibfield  {journal} {\bibinfo
  {journal} {\prd}\ }\textbf {\bibinfo {volume} {98}},\ \bibinfo {eid} {043528}
  (\bibinfo {year} {2018})},\ \Eprint {http://arxiv.org/abs/1708.01538}
  {arXiv:1708.01538 [astro-ph.CO]} \BibitemShut {NoStop}%
\bibitem [{\citenamefont {{Mandelbaum}}\ \emph {et~al.}(2005)\citenamefont
  {{Mandelbaum}}, \citenamefont {{Hirata}}, \citenamefont {{Seljak}},
  \citenamefont {{Guzik}}, \citenamefont {{Padmanabhan}} \emph
  {et~al.}}]{Mandelbaum:05b}%
  \BibitemOpen
  \bibfield  {author} {\bibinfo {author} {\bibfnamefont {R.}~\bibnamefont
  {{Mandelbaum}}}, \bibinfo {author} {\bibfnamefont {C.~M.}\ \bibnamefont
  {{Hirata}}}, \bibinfo {author} {\bibfnamefont {U.}~\bibnamefont {{Seljak}}},
  \bibinfo {author} {\bibfnamefont {J.}~\bibnamefont {{Guzik}}}, \bibinfo
  {author} {\bibfnamefont {N.}~\bibnamefont {{Padmanabhan}}},  \emph {et~al.},\
  }\href {\doibase 10.1111/j.1365-2966.2005.09282.x} {\bibfield  {journal}
  {\bibinfo  {journal} {\mnras}\ }\textbf {\bibinfo {volume} {361}},\ \bibinfo
  {pages} {1287} (\bibinfo {year} {2005})},\ \Eprint
  {http://arxiv.org/abs/astro-ph/0501201} {arXiv:astro-ph/0501201 [astro-ph]}
  \BibitemShut {NoStop}%
\bibitem [{\citenamefont {Fang}\ \emph {et~al.}(2020)\citenamefont {Fang},
  \citenamefont {Krause}, \citenamefont {Eifler},\ and\ \citenamefont
  {MacCrann}}]{Fang:2019xat}%
  \BibitemOpen
  \bibfield  {author} {\bibinfo {author} {\bibfnamefont {X.}~\bibnamefont
  {Fang}}, \bibinfo {author} {\bibfnamefont {E.}~\bibnamefont {Krause}},
  \bibinfo {author} {\bibfnamefont {T.}~\bibnamefont {Eifler}}, \ and\ \bibinfo
  {author} {\bibfnamefont {N.}~\bibnamefont {MacCrann}},\ }\href {\doibase
  10.1088/1475-7516/2020/05/010} {\bibfield  {journal} {\bibinfo  {journal}
  {JCAP}\ }\textbf {\bibinfo {volume} {05}},\ \bibinfo {pages} {010} (\bibinfo
  {year} {2020})},\ \Eprint {http://arxiv.org/abs/1911.11947} {arXiv:1911.11947
  [astro-ph.CO]} \BibitemShut {NoStop}%
\bibitem [{\citenamefont {{Hamilton}}(2000)}]{Hamilton00}%
  \BibitemOpen
  \bibfield  {author} {\bibinfo {author} {\bibfnamefont {A.~J.~S.}\
  \bibnamefont {{Hamilton}}},\ }\href {\doibase
  10.1046/j.1365-8711.2000.03071.x} {\bibfield  {journal} {\bibinfo  {journal}
  {\mnras}\ }\textbf {\bibinfo {volume} {312}},\ \bibinfo {pages} {257}
  (\bibinfo {year} {2000})},\ \Eprint {http://arxiv.org/abs/astro-ph/9905191}
  {astro-ph/9905191} \BibitemShut {NoStop}%
\bibitem [{\citenamefont {{Shirasaki}}\ \emph {et~al.}(2019)\citenamefont
  {{Shirasaki}}, \citenamefont {{Hamana}}, \citenamefont {{Takada}},
  \citenamefont {{Takahashi}},\ and\ \citenamefont
  {{Miyatake}}}]{2019MNRAS.486...52S}%
  \BibitemOpen
  \bibfield  {author} {\bibinfo {author} {\bibfnamefont {M.}~\bibnamefont
  {{Shirasaki}}}, \bibinfo {author} {\bibfnamefont {T.}~\bibnamefont
  {{Hamana}}}, \bibinfo {author} {\bibfnamefont {M.}~\bibnamefont {{Takada}}},
  \bibinfo {author} {\bibfnamefont {R.}~\bibnamefont {{Takahashi}}}, \ and\
  \bibinfo {author} {\bibfnamefont {H.}~\bibnamefont {{Miyatake}}},\ }\href
  {\doibase 10.1093/mnras/stz791} {\bibfield  {journal} {\bibinfo  {journal}
  {\mnras}\ }\textbf {\bibinfo {volume} {486}},\ \bibinfo {pages} {52}
  (\bibinfo {year} {2019})},\ \Eprint {http://arxiv.org/abs/1901.09488}
  {arXiv:1901.09488 [astro-ph.CO]} \BibitemShut {NoStop}%
\bibitem [{\citenamefont {Shirasaki}\ \emph {et~al.}(2017)\citenamefont
  {Shirasaki}, \citenamefont {Takada}, \citenamefont {Miyatake}, \citenamefont
  {Takahashi}, \citenamefont {Hamana}, \citenamefont {Nishimichi} \emph
  {et~al.}}]{Shirasakietal:17}%
  \BibitemOpen
  \bibfield  {author} {\bibinfo {author} {\bibfnamefont {M.}~\bibnamefont
  {Shirasaki}}, \bibinfo {author} {\bibfnamefont {M.}~\bibnamefont {Takada}},
  \bibinfo {author} {\bibfnamefont {H.}~\bibnamefont {Miyatake}}, \bibinfo
  {author} {\bibfnamefont {R.}~\bibnamefont {Takahashi}}, \bibinfo {author}
  {\bibfnamefont {T.}~\bibnamefont {Hamana}}, \bibinfo {author} {\bibfnamefont
  {T.}~\bibnamefont {Nishimichi}},  \emph {et~al.},\ }\href {\doibase
  10.1093/mnras/stx1477} {\bibfield  {journal} {\bibinfo  {journal} {Mon. Not.
  Roy. Astron. Soc.}\ }\textbf {\bibinfo {volume} {470}},\ \bibinfo {pages}
  {3476} (\bibinfo {year} {2017})},\ \Eprint {http://arxiv.org/abs/1607.08679}
  {arXiv:1607.08679 [astro-ph.CO]} \BibitemShut {NoStop}%
\bibitem [{\citenamefont {{Hartlap}}\ \emph {et~al.}(2007)\citenamefont
  {{Hartlap}}, \citenamefont {{Simon}},\ and\ \citenamefont
  {{Schneider}}}]{2007A&A...464..399H}%
  \BibitemOpen
  \bibfield  {author} {\bibinfo {author} {\bibfnamefont {J.}~\bibnamefont
  {{Hartlap}}}, \bibinfo {author} {\bibfnamefont {P.}~\bibnamefont {{Simon}}},
  \ and\ \bibinfo {author} {\bibfnamefont {P.}~\bibnamefont {{Schneider}}},\
  }\href {\doibase 10.1051/0004-6361:20066170} {\bibfield  {journal} {\bibinfo
  {journal} {\aap}\ }\textbf {\bibinfo {volume} {464}},\ \bibinfo {pages} {399}
  (\bibinfo {year} {2007})},\ \Eprint {http://arxiv.org/abs/astro-ph/0608064}
  {arXiv:astro-ph/0608064 [astro-ph]} \BibitemShut {NoStop}%
\bibitem [{\citenamefont {{Takahashi}}\ \emph {et~al.}(2017)\citenamefont
  {{Takahashi}}, \citenamefont {{Hamana}}, \citenamefont {{Shirasaki}},
  \citenamefont {{Namikawa}}, \citenamefont {{Nishimichi}}, \citenamefont
  {{Osato}},\ and\ \citenamefont {{Shiroyama}}}]{2017ApJ...850...24T}%
  \BibitemOpen
  \bibfield  {author} {\bibinfo {author} {\bibfnamefont {R.}~\bibnamefont
  {{Takahashi}}}, \bibinfo {author} {\bibfnamefont {T.}~\bibnamefont
  {{Hamana}}}, \bibinfo {author} {\bibfnamefont {M.}~\bibnamefont
  {{Shirasaki}}}, \bibinfo {author} {\bibfnamefont {T.}~\bibnamefont
  {{Namikawa}}}, \bibinfo {author} {\bibfnamefont {T.}~\bibnamefont
  {{Nishimichi}}}, \bibinfo {author} {\bibfnamefont {K.}~\bibnamefont
  {{Osato}}}, \ and\ \bibinfo {author} {\bibfnamefont {K.~o.}\ \bibnamefont
  {{Shiroyama}}},\ }\href {\doibase 10.3847/1538-4357/aa943d} {\bibfield
  {journal} {\bibinfo  {journal} {\apj}\ }\textbf {\bibinfo {volume} {850}},\
  \bibinfo {eid} {24} (\bibinfo {year} {2017})},\ \Eprint
  {http://arxiv.org/abs/1706.01472} {arXiv:1706.01472 [astro-ph.CO]}
  \BibitemShut {NoStop}%
\bibitem [{\citenamefont {{Takada}}\ and\ \citenamefont
  {{Hu}}(2013)}]{TakadaHu:13}%
  \BibitemOpen
  \bibfield  {author} {\bibinfo {author} {\bibfnamefont {M.}~\bibnamefont
  {{Takada}}}\ and\ \bibinfo {author} {\bibfnamefont {W.}~\bibnamefont
  {{Hu}}},\ }\href {\doibase 10.1103/PhysRevD.87.123504} {\bibfield  {journal}
  {\bibinfo  {journal} {\prd}\ }\textbf {\bibinfo {volume} {87}},\ \bibinfo
  {eid} {123504} (\bibinfo {year} {2013})},\ \Eprint
  {http://arxiv.org/abs/1302.6994} {arXiv:1302.6994 [astro-ph.CO]} \BibitemShut
  {NoStop}%
\bibitem [{\citenamefont {Aver}\ \emph {et~al.}(2015)\citenamefont {Aver},
  \citenamefont {Olive},\ and\ \citenamefont {Skillman}}]{Aver:2015iza}%
  \BibitemOpen
  \bibfield  {author} {\bibinfo {author} {\bibfnamefont {E.}~\bibnamefont
  {Aver}}, \bibinfo {author} {\bibfnamefont {K.~A.}\ \bibnamefont {Olive}}, \
  and\ \bibinfo {author} {\bibfnamefont {E.~D.}\ \bibnamefont {Skillman}},\
  }\href {\doibase 10.1088/1475-7516/2015/07/011} {\bibfield  {journal}
  {\bibinfo  {journal} {JCAP}\ }\textbf {\bibinfo {volume} {07}},\ \bibinfo
  {pages} {011} (\bibinfo {year} {2015})},\ \Eprint
  {http://arxiv.org/abs/1503.08146} {arXiv:1503.08146 [astro-ph.CO]}
  \BibitemShut {NoStop}%
\bibitem [{\citenamefont {Cooke}\ \emph {et~al.}(2018)\citenamefont {Cooke},
  \citenamefont {Pettini},\ and\ \citenamefont {Steidel}}]{Cooke:2017cwo}%
  \BibitemOpen
  \bibfield  {author} {\bibinfo {author} {\bibfnamefont {R.~J.}\ \bibnamefont
  {Cooke}}, \bibinfo {author} {\bibfnamefont {M.}~\bibnamefont {Pettini}}, \
  and\ \bibinfo {author} {\bibfnamefont {C.~C.}\ \bibnamefont {Steidel}},\
  }\href {\doibase 10.3847/1538-4357/aaab53} {\bibfield  {journal} {\bibinfo
  {journal} {Astrophys. J.}\ }\textbf {\bibinfo {volume} {855}},\ \bibinfo
  {pages} {102} (\bibinfo {year} {2018})},\ \Eprint
  {http://arxiv.org/abs/1710.11129} {arXiv:1710.11129 [astro-ph.CO]}
  \BibitemShut {NoStop}%
\bibitem [{\citenamefont {Sch\"oneberg}\ \emph {et~al.}(2019)\citenamefont
  {Sch\"oneberg}, \citenamefont {Lesgourgues},\ and\ \citenamefont
  {Hooper}}]{Schoneberg:2019wmt}%
  \BibitemOpen
  \bibfield  {author} {\bibinfo {author} {\bibfnamefont {N.}~\bibnamefont
  {Sch\"oneberg}}, \bibinfo {author} {\bibfnamefont {J.}~\bibnamefont
  {Lesgourgues}}, \ and\ \bibinfo {author} {\bibfnamefont {D.~C.}\ \bibnamefont
  {Hooper}},\ }\href {\doibase 10.1088/1475-7516/2019/10/029} {\bibfield
  {journal} {\bibinfo  {journal} {JCAP}\ }\textbf {\bibinfo {volume} {10}},\
  \bibinfo {pages} {029} (\bibinfo {year} {2019})},\ \Eprint
  {http://arxiv.org/abs/1907.11594} {arXiv:1907.11594 [astro-ph.CO]}
  \BibitemShut {NoStop}%
\bibitem [{\citenamefont {{Feroz}}\ \emph {et~al.}(2009)\citenamefont
  {{Feroz}}, \citenamefont {{Hobson}},\ and\ \citenamefont
  {{Bridges}}}]{Feroz:2009}%
  \BibitemOpen
  \bibfield  {author} {\bibinfo {author} {\bibfnamefont {F.}~\bibnamefont
  {{Feroz}}}, \bibinfo {author} {\bibfnamefont {M.~P.}\ \bibnamefont
  {{Hobson}}}, \ and\ \bibinfo {author} {\bibfnamefont {M.}~\bibnamefont
  {{Bridges}}},\ }\href {\doibase 10.1111/j.1365-2966.2009.14548.x} {\bibfield
  {journal} {\bibinfo  {journal} {\mnras}\ }\textbf {\bibinfo {volume} {398}},\
  \bibinfo {pages} {1601} (\bibinfo {year} {2009})},\ \Eprint
  {http://arxiv.org/abs/0809.3437} {arXiv:0809.3437 [astro-ph]} \BibitemShut
  {NoStop}%
\bibitem [{\citenamefont {{Buchner}}\ \emph {et~al.}(2014)\citenamefont
  {{Buchner}}, \citenamefont {{Georgakakis}}, \citenamefont {{Nandra}},
  \citenamefont {{Hsu}}, \citenamefont {{Rangel}} \emph
  {et~al.}}]{2014A&A...564A.125B}%
  \BibitemOpen
  \bibfield  {author} {\bibinfo {author} {\bibfnamefont {J.}~\bibnamefont
  {{Buchner}}}, \bibinfo {author} {\bibfnamefont {A.}~\bibnamefont
  {{Georgakakis}}}, \bibinfo {author} {\bibfnamefont {K.}~\bibnamefont
  {{Nandra}}}, \bibinfo {author} {\bibfnamefont {L.}~\bibnamefont {{Hsu}}},
  \bibinfo {author} {\bibfnamefont {C.}~\bibnamefont {{Rangel}}},  \emph
  {et~al.},\ }\href {\doibase 10.1051/0004-6361/201322971} {\bibfield
  {journal} {\bibinfo  {journal} {\aap}\ }\textbf {\bibinfo {volume} {564}},\
  \bibinfo {eid} {A125} (\bibinfo {year} {2014})},\ \Eprint
  {http://arxiv.org/abs/1402.0004} {arXiv:1402.0004 [astro-ph.HE]} \BibitemShut
  {NoStop}%
\bibitem [{\citenamefont {{DES Collaboration}}\ \emph
  {et~al.}(2021)\citenamefont {{DES Collaboration}}, \citenamefont {{Abbott}},
  \citenamefont {{Aguena}}, \citenamefont {{Alarcon}}, \citenamefont {{Allam}}
  \emph {et~al.}}]{DES-Y3}%
  \BibitemOpen
  \bibfield  {author} {\bibinfo {author} {\bibnamefont {{DES Collaboration}}},
  \bibinfo {author} {\bibfnamefont {T.~M.~C.}\ \bibnamefont {{Abbott}}},
  \bibinfo {author} {\bibfnamefont {M.}~\bibnamefont {{Aguena}}}, \bibinfo
  {author} {\bibfnamefont {A.}~\bibnamefont {{Alarcon}}}, \bibinfo {author}
  {\bibfnamefont {S.}~\bibnamefont {{Allam}}},  \emph {et~al.},\ }\href@noop {}
  {\bibfield  {journal} {\bibinfo  {journal} {arXiv e-prints}\ ,\ \bibinfo
  {eid} {arXiv:2105.13549}} (\bibinfo {year} {2021})},\ \Eprint
  {http://arxiv.org/abs/2105.13549} {arXiv:2105.13549 [astro-ph.CO]}
  \BibitemShut {NoStop}%
\bibitem [{\citenamefont {{Park}}\ and\ \citenamefont
  {{Rozo}}(2020)}]{Park:2020}%
  \BibitemOpen
  \bibfield  {author} {\bibinfo {author} {\bibfnamefont {Y.}~\bibnamefont
  {{Park}}}\ and\ \bibinfo {author} {\bibfnamefont {E.}~\bibnamefont
  {{Rozo}}},\ }\href {\doibase 10.1093/mnras/staa2647} {\bibfield  {journal}
  {\bibinfo  {journal} {\mnras}\ }\textbf {\bibinfo {volume} {499}},\ \bibinfo
  {pages} {4638} (\bibinfo {year} {2020})},\ \Eprint
  {http://arxiv.org/abs/1907.05798} {arXiv:1907.05798 [astro-ph.CO]}
  \BibitemShut {NoStop}%
\bibitem [{\citenamefont {{Raveri}}\ and\ \citenamefont
  {{Hu}}(2019)}]{2019PhRvD..99d3506R}%
  \BibitemOpen
  \bibfield  {author} {\bibinfo {author} {\bibfnamefont {M.}~\bibnamefont
  {{Raveri}}}\ and\ \bibinfo {author} {\bibfnamefont {W.}~\bibnamefont
  {{Hu}}},\ }\href {\doibase 10.1103/PhysRevD.99.043506} {\bibfield  {journal}
  {\bibinfo  {journal} {\prd}\ }\textbf {\bibinfo {volume} {99}},\ \bibinfo
  {eid} {043506} (\bibinfo {year} {2019})},\ \Eprint
  {http://arxiv.org/abs/1806.04649} {arXiv:1806.04649 [astro-ph.CO]}
  \BibitemShut {NoStop}%
\bibitem [{\citenamefont {{Planck Collaboration}}\ \emph
  {et~al.}(2016)\citenamefont {{Planck Collaboration}}, \citenamefont {{Ade}},
  \citenamefont {{Aghanim}}, \citenamefont {{Arnaud}}, \citenamefont
  {{Ashdown}} \emph {et~al.}}]{PlanckCosmology:16}%
  \BibitemOpen
  \bibfield  {author} {\bibinfo {author} {\bibnamefont {{Planck
  Collaboration}}}, \bibinfo {author} {\bibfnamefont {P.~A.~R.}\ \bibnamefont
  {{Ade}}}, \bibinfo {author} {\bibfnamefont {N.}~\bibnamefont {{Aghanim}}},
  \bibinfo {author} {\bibfnamefont {M.}~\bibnamefont {{Arnaud}}}, \bibinfo
  {author} {\bibfnamefont {M.}~\bibnamefont {{Ashdown}}},  \emph {et~al.},\
  }\href {\doibase 10.1051/0004-6361/201525830} {\bibfield  {journal} {\bibinfo
   {journal} {\aap}\ }\textbf {\bibinfo {volume} {594}},\ \bibinfo {eid} {A13}
  (\bibinfo {year} {2016})},\ \Eprint {http://arxiv.org/abs/1502.01589}
  {arXiv:1502.01589} \BibitemShut {NoStop}%
\bibitem [{\citenamefont {{Vogelsberger}}\ \emph {et~al.}(2014)\citenamefont
  {{Vogelsberger}}, \citenamefont {{Genel}}, \citenamefont {{Springel}},
  \citenamefont {{Torrey}}, \citenamefont {{Sijacki}}, \citenamefont {{Xu}}
  \emph {et~al.}}]{2014Natur.509..177V}%
  \BibitemOpen
  \bibfield  {author} {\bibinfo {author} {\bibfnamefont {M.}~\bibnamefont
  {{Vogelsberger}}}, \bibinfo {author} {\bibfnamefont {S.}~\bibnamefont
  {{Genel}}}, \bibinfo {author} {\bibfnamefont {V.}~\bibnamefont {{Springel}}},
  \bibinfo {author} {\bibfnamefont {P.}~\bibnamefont {{Torrey}}}, \bibinfo
  {author} {\bibfnamefont {D.}~\bibnamefont {{Sijacki}}}, \bibinfo {author}
  {\bibfnamefont {D.}~\bibnamefont {{Xu}}},  \emph {et~al.},\ }\href {\doibase
  10.1038/nature13316} {\bibfield  {journal} {\bibinfo  {journal} {\nat}\
  }\textbf {\bibinfo {volume} {509}},\ \bibinfo {pages} {177} (\bibinfo {year}
  {2014})},\ \Eprint {http://arxiv.org/abs/1405.1418} {arXiv:1405.1418
  [astro-ph.CO]} \BibitemShut {NoStop}%
\bibitem [{\citenamefont {Mead}\ \emph {et~al.}(2020)\citenamefont {Mead},
  \citenamefont {Brieden}, \citenamefont {Tr\"oster},\ and\ \citenamefont
  {Heymans}}]{Mead:2020vgs}%
  \BibitemOpen
  \bibfield  {author} {\bibinfo {author} {\bibfnamefont {A.}~\bibnamefont
  {Mead}}, \bibinfo {author} {\bibfnamefont {S.}~\bibnamefont {Brieden}},
  \bibinfo {author} {\bibfnamefont {T.}~\bibnamefont {Tr\"oster}}, \ and\
  \bibinfo {author} {\bibfnamefont {C.}~\bibnamefont {Heymans}},\ }\href
  {\doibase 10.1093/mnras/stab082} {\  (\bibinfo {year} {2020}),\
  10.1093/mnras/stab082},\ \Eprint {http://arxiv.org/abs/2009.01858}
  {arXiv:2009.01858 [astro-ph.CO]} \BibitemShut {NoStop}%
\bibitem [{\citenamefont {Mead}\ \emph {et~al.}(2015)\citenamefont {Mead},
  \citenamefont {Peacock}, \citenamefont {Heymans}, \citenamefont {Joudaki},\
  and\ \citenamefont {Heavens}}]{Mead:2015yca}%
  \BibitemOpen
  \bibfield  {author} {\bibinfo {author} {\bibfnamefont {A.}~\bibnamefont
  {Mead}}, \bibinfo {author} {\bibfnamefont {J.}~\bibnamefont {Peacock}},
  \bibinfo {author} {\bibfnamefont {C.}~\bibnamefont {Heymans}}, \bibinfo
  {author} {\bibfnamefont {S.}~\bibnamefont {Joudaki}}, \ and\ \bibinfo
  {author} {\bibfnamefont {A.~o.}\ \bibnamefont {Heavens}},\ }\href {\doibase
  10.1093/mnras/stv2036} {\bibfield  {journal} {\bibinfo  {journal} {Mon. Not.
  Roy. Astron. Soc.}\ }\textbf {\bibinfo {volume} {454}},\ \bibinfo {pages}
  {1958} (\bibinfo {year} {2015})},\ \Eprint {http://arxiv.org/abs/1505.07833}
  {arXiv:1505.07833 [astro-ph.CO]} \BibitemShut {NoStop}%
\bibitem [{\citenamefont {{Schaye}}\ \emph {et~al.}(2010)\citenamefont
  {{Schaye}}, \citenamefont {{Dalla Vecchia}}, \citenamefont {{Booth}},
  \citenamefont {{Wiersma}}, \citenamefont {{Theuns}}, \citenamefont {{Haas}}
  \emph {et~al.}}]{2010mnras.402.1536s}%
  \BibitemOpen
  \bibfield  {author} {\bibinfo {author} {\bibfnamefont {J.}~\bibnamefont
  {{Schaye}}}, \bibinfo {author} {\bibfnamefont {C.}~\bibnamefont {{Dalla
  Vecchia}}}, \bibinfo {author} {\bibfnamefont {C.~M.}\ \bibnamefont
  {{Booth}}}, \bibinfo {author} {\bibfnamefont {R.~P.~C.}\ \bibnamefont
  {{Wiersma}}}, \bibinfo {author} {\bibfnamefont {T.}~\bibnamefont {{Theuns}}},
  \bibinfo {author} {\bibfnamefont {M.~R.}\ \bibnamefont {{Haas}}},  \emph
  {et~al.},\ }\href {\doibase 10.1111/j.1365-2966.2009.16029.x} {\bibfield
  {journal} {\bibinfo  {journal} {\mnras}\ }\textbf {\bibinfo {volume} {402}},\
  \bibinfo {pages} {1536} (\bibinfo {year} {2010})},\ \Eprint
  {http://arxiv.org/abs/0909.5196} {arXiv:0909.5196 [astro-ph.CO]} \BibitemShut
  {NoStop}%
\bibitem [{\citenamefont {{McCarthy}}\ \emph {et~al.}(2017)\citenamefont
  {{McCarthy}}, \citenamefont {{Schaye}}, \citenamefont {{Bird}},\ and\
  \citenamefont {{Le Brun}}}]{2017MNRAS.465.2936M}%
  \BibitemOpen
  \bibfield  {author} {\bibinfo {author} {\bibfnamefont {I.~G.}\ \bibnamefont
  {{McCarthy}}}, \bibinfo {author} {\bibfnamefont {J.}~\bibnamefont
  {{Schaye}}}, \bibinfo {author} {\bibfnamefont {S.}~\bibnamefont {{Bird}}}, \
  and\ \bibinfo {author} {\bibfnamefont {A.~M.~C.}\ \bibnamefont {{Le Brun}}},\
  }\href {\doibase 10.1093/mnras/stw2792} {\bibfield  {journal} {\bibinfo
  {journal} {\mnras}\ }\textbf {\bibinfo {volume} {465}},\ \bibinfo {pages}
  {2936} (\bibinfo {year} {2017})},\ \Eprint {http://arxiv.org/abs/1603.02702}
  {arXiv:1603.02702 [astro-ph.CO]} \BibitemShut {NoStop}%
\bibitem [{\citenamefont {Brun}\ \emph {et~al.}(2014)\citenamefont {Brun},
  \citenamefont {McCarthy}, \citenamefont {Schaye},\ and\ \citenamefont
  {Ponman}}]{brun:2013yva}%
  \BibitemOpen
  \bibfield  {author} {\bibinfo {author} {\bibfnamefont {A.~M. C.~L.}\
  \bibnamefont {Brun}}, \bibinfo {author} {\bibfnamefont {I.~G.}\ \bibnamefont
  {McCarthy}}, \bibinfo {author} {\bibfnamefont {J.}~\bibnamefont {Schaye}}, \
  and\ \bibinfo {author} {\bibfnamefont {T.~J.}\ \bibnamefont {Ponman}},\
  }\href {\doibase 10.1093/mnras/stu608} {\bibfield  {journal} {\bibinfo
  {journal} {Mon. Not. Roy. Astron. Soc.}\ }\textbf {\bibinfo {volume} {441}},\
  \bibinfo {pages} {1270} (\bibinfo {year} {2014})},\ \Eprint
  {http://arxiv.org/abs/1312.5462} {arXiv:1312.5462 [astro-ph.CO]} \BibitemShut
  {NoStop}%
\bibitem [{\citenamefont {Higson}(2018)}]{higson2018nestcheck}%
  \BibitemOpen
  \bibfield  {author} {\bibinfo {author} {\bibfnamefont {E.}~\bibnamefont
  {Higson}},\ }\href {\doibase 10.21105/joss.00916} {\bibfield  {journal}
  {\bibinfo  {journal} {Journal of Open Source Software}\ }\textbf {\bibinfo
  {volume} {3}},\ \bibinfo {pages} {916} (\bibinfo {year} {2018})}\BibitemShut
  {NoStop}%
\bibitem [{\citenamefont {Higson}\ \emph {et~al.}(2019)\citenamefont {Higson},
  \citenamefont {Handley}, \citenamefont {Hobson},\ and\ \citenamefont
  {Lasenby}}]{higson2019diagnostic}%
  \BibitemOpen
  \bibfield  {author} {\bibinfo {author} {\bibfnamefont {E.}~\bibnamefont
  {Higson}}, \bibinfo {author} {\bibfnamefont {W.}~\bibnamefont {Handley}},
  \bibinfo {author} {\bibfnamefont {M.}~\bibnamefont {Hobson}}, \ and\ \bibinfo
  {author} {\bibfnamefont {A.}~\bibnamefont {Lasenby}},\ }\href {\doibase
  10.1093/mnras/sty3090} {\bibfield  {journal} {\bibinfo  {journal} {Monthly
  Notices of the Royal Astronomical Society}\ }\textbf {\bibinfo {volume}
  {483}},\ \bibinfo {pages} {2044} (\bibinfo {year} {2019})}\BibitemShut
  {NoStop}%
\bibitem [{\citenamefont {Metropolis}\ and\ \citenamefont
  {Ulam}(1949)}]{metropolis_1949}%
  \BibitemOpen
  \bibfield  {author} {\bibinfo {author} {\bibfnamefont {N.}~\bibnamefont
  {Metropolis}}\ and\ \bibinfo {author} {\bibfnamefont {S.}~\bibnamefont
  {Ulam}},\ }\href {\doibase 10.1080/01621459.1949.10483310} {\bibfield
  {journal} {\bibinfo  {journal} {Journal of the American Statistical
  Association}\ }\textbf {\bibinfo {volume} {44}},\ \bibinfo {pages} {335}
  (\bibinfo {year} {1949})},\ \bibinfo {note} {pMID: 18139350},\ \Eprint
  {http://arxiv.org/abs/https://www.tandfonline.com/doi/pdf/10.1080/01621459.1949.10483310}
  {https://www.tandfonline.com/doi/pdf/10.1080/01621459.1949.10483310}
  \BibitemShut {NoStop}%
\end{thebibliography}%

\end{document}